\documentclass[11pt, onecolumn]{IEEEtran}
\makeatletter
\def\@begintheorem#1#2{\@IEEEtmpitemindent\itemindent\relax\topsep 0pt\rmfamily\trivlist%
    \item[]\textit{\indent #1\ #2:} \itemindent\@IEEEtmpitemindent\relax}
\def\@opargbegintheorem#1#2#3{\@IEEEtmpitemindent\itemindent\relax\topsep 0pt\rmfamily \trivlist%
    \item[]\textit{\indent #1\ #2\ (#3):} \itemindent\@IEEEtmpitemindent\relax}
\makeatother

\usepackage{algorithmic}
\usepackage{algorithm}

\usepackage{cite}
\usepackage{graphicx}

\usepackage{amsmath}   % From the American Mathematical Society
\usepackage{amssymb}
\usepackage{amsfonts}

\usepackage{multirow}

\usepackage{dsfont}
\usepackage[latin1]{inputenc}
\usepackage{times}

\newtheorem{thm}{Theorem}
\newtheorem{rem}{Remark}
\newtheorem{cor}{Corollary}
\newtheorem{lem}{Lemma}

\newtheorem{lemap}{Lemma}[subsection]
\newtheorem{prpap}[lemap]{Proposition}
\newtheorem{corap}[lemap]{Corollary}
\newtheorem{dfnap}[lemap]{Definition}
\newtheorem{remap}[lemap]{Remark}

\def\mb{\mathbf}
\def\opn{\operatorname*}
\def\bs{\boldsymbol}

\def\ds{\mathds}
\def\mc{\mathcal}
\def\ss#1{{\sf #1}}

\def\esc#1{\textrm{#1}}
\def\veci#1#2{\esc{#1}_{#2}}
\def\rveci#1#2{\uppercase{#1}_{#2}}
\def\rvecri#1#2{\lowercase{#1}_{#2}}

\def\vec#1{\mb{#1}}
\def\rvec#1{\bs{\uppercase{#1}}} % random vector
\def\rvecr#1{\bs{\lowercase{#1}}} % random vector realization
\def\vecs#1{\bs{#1}}

\def\mat#1{\mb{\uppercase{#1}}}
\def\mats#1{\bs{#1}}

\def\dim{n}
\def\dimp{m}
\def\dimpp{p}

\def\R{\ds{R}}
\def\C{\ds{C}}

\def\SymM#1{\ds{S}^{#1}}
\def\PSD#1{\SymM{#1}_+}
\def\PD#1{\SymM{#1}_{++}}

\def\N#1#2{\mc{N}\left(#1,#2\right)}
\def\NZ#1{\N{\vecs{0}}{#1}}

\def\pdfvec#1{P_{\rvec{#1}}(\rvecr{#1})}
\def\pdf#1#2{P_{\rvec{#1}}(#2)}

\def\EspOp{\ss{E}}

\newcommand{\Esp}[2][5]{%
  \ifcase#1
     \EspOp\{ #2 \}
     \or \EspOp \bigl\{ #2 \bigr\}
     \or \EspOp \Bigl\{ #2 \Bigr\}
     \or \EspOp \biggl\{ #2 \biggr\}
     \or \EspOp \Biggl\{ #2 \Biggr\}
  \else
     \EspOp \left\{ #2  \right\}
\fi}

\newcommand{\Earg}[3][5]{%
  \ifcase#1
     \EspOp_{#3} \{ #2 \}
     \or \EspOp_{#3} \bigl\{ #2 \bigr\}
     \or \EspOp_{#3} \Bigl\{ #2 \Bigr\}
     \or \EspOp_{#3} \biggl\{ #2 \biggr\}
     \or \EspOp_{#3} \Biggl\{ #2 \Biggr\}
  \else
     \EspOp_{#3} \left\{ #2  \right\}
\fi}

\newcommand{\CEsp}[3][5]{%
  \ifcase#1
     \EspOp\{ #2 \mid #3 \}
     \or \EspOp \bigl\{ #2 \bigm\vert #3 \bigr\}
     \or \EspOp \Bigl\{ #2 \Bigm\vert #3 \Bigr\}
     \or \EspOp \biggl\{ #2 \biggm\vert #3 \biggr\}
     \or \EspOp \Biggl\{ #2 \Biggm\vert #3 \Biggr\}
  \else
     \EspOp \left\{ #2  \,\middle\vert\, #3 \right\}
\fi}

\def\Jacob{{\ss D}}
\def\Hess{{\ss H}}
\def\Tr{\ss{Tr}}
\def\det{\ss{det}}
\def\log{\ss{log}}
\def\rank{\ss{rank}}
\def\vecop{\ss{vec}}
\def\vechop{\ss{vech}}
\def\pinv{\textrm{\footnotesize{+}}}

\def\d{\opn{d}\!}
\def\H{\dagger}
\def\T{\ss{T}}

\def\nume{\esc{e}}
\def\diag#1{\mb{diag}(#1)}
\def\Diag#1{\mb{Diag}(#1)}
\def\real{\Re \textrm{e}}
\def\imag{\Im \textrm{m}}

\def\I{I} % \mc{I}
\def\Ent{h} % \mc{H}

\def\EP{N} % \mc{N}

\def\req#1{(\ref{#1})}
\def\ie{i.e.}
\def\eg{e.g.}

\def\dilogP#1#2{\frac{\partial \log \pdf{\rvec{#1}}{\rvec{#1}}}{\partial \rvecri{#1}{#2}}}
\def\dijlogP#1#2#3{\frac{\partial^2 \log \pdf{\rvec{#1}}{\rvec{#1}}}{\partial \rvecri{#1}{#2} \partial \rvecri{#1}{#3}}}

\def\dijklogP#1#2#3#4{\frac{\partial^3 \log \pdf{\rvec{#1}}{\rvec{#1}}}{\partial \rvecri{#1}{#2} \partial \rvecri{#1}{#3} \partial \rvecri{#1}{#4}}}
\def\dijkllogP#1#2#3#4#5{\frac{\partial^4 \log \pdf{\rvec{#1}}{\rvec{#1}}}{\partial \rvecri{#1}{#2} \partial \rvecri{#1}{#3} \partial \rvecri{#1}{#4} \partial \rvecri{#1}{#5}}}

\def\dilogPz#1#2{\frac{\partial \log \pdfvec{#1}}{\partial \rvecri{#1}{#2}}}
\def\dijlogPz#1#2#3{\frac{\partial^2 \log \pdfvec{#1}}{\partial \rvecri{#1}{#2} \partial \rvecri{#1}{#3}}}

\def\dijklogPz#1#2#3#4{\frac{\partial^3 \log \pdfvec{#1}}{\partial \rvecri{#1}{#2} \partial \rvecri{#1}{#3} \partial \rvecri{#1}{#4}}}
\def\dijkllogPz#1#2#3#4#5{\frac{\partial^4 \log \pdfvec{#1}}{\partial \rvecri{#1}{#2} \partial \rvecri{#1}{#3} \partial \rvecri{#1}{#4} \partial \rvecri{#1}{#5}}}

\def\diPz#1#2{\frac{\partial \pdfvec{#1}}{\partial \rvecri{#1}{#2}}}
\def\dijPz#1#2#3{\frac{\partial^2 \pdfvec{#1}}{\partial \rvecri{#1}{#2} \partial \rvecri{#1}{#3}}}

\def\CM{\mats{\Phi}}
\def\EM{\mat{E}}
\def\MSE#1{\EM_{\rvec{#1}}}

\def\CMSE#1#2{\CM_{\rvec{#1}}(\rvec{#2})}
\def\CMSEr#1#2{\CM_{\rvec{#1}}(\rvecr{#2})}

\def\JM{\mat{J}}
\def\Cj{\mats{\Gamma}}
\def\J#1{\JM_{\rvec{#1}}}
\def\CJ#1{\Cj_{\rvec{#1}}(\rvec{#1})}
\def\CJr#1{\Cj_{\rvec{#1}}(\rvecr{#1})}

\def\preS{\mat{G}}
\def\Chan{\mat{H}}
\def\Prec{\mat{P}}
\def\preN{\mat{C}}

\def\Cov#1{\mat{R}_{\rvec{#1}}}
\def\Com#1{\mat{K}_{#1}}
\def\Sym#1{\mat{N}_{#1}}
\def\Dup#1{\mat{D}_{#1}}
\def\Rdx#1{\mat{S}_{#1}}

\def\snr{\ss{snr}}
\def\CostaT{t}

\begin{document}

\title{Hessian and concavity of mutual information, differential entropy, and entropy power in linear vector Gaussian channels}

\author{Miquel Payar\'o and Daniel P.~Palomar
\thanks{A shorter version of this paper is to appear in \emph{IEEE Transactions on Information Theory}.}%
\thanks{This work was supported by the RGC 618008 research grant.}%
\thanks{M.~Payaró conducted his part of this research while he was with
the Department of Electronic and Computer Engineering, Hong Kong
University of Science and Technology, Clear Water Bay, Kowloon, Hong
Kong. He is now with the Centre Tecnològic de Telecomunicacions de
Catalunya (CTTC), Barcelona, Spain (e-mail:
miquel.payaro@cttc.es).}\thanks{ D.~P.~Palomar is with the
Department of Electronic and Computer Engineering, Hong Kong
University of Science and Technology, Clear Water Bay, Kowloon, Hong
Kong (e-mail: palomar@ust.hk).}}

\maketitle

\begin{abstract} %PASSIVE VOICE
Within the framework of linear vector Gaussian channels with
arbitrary signaling, closed-form expressions for the Jacobian of the
minimum mean square error and Fisher information matrices with
respect to arbitrary parameters of the system are calculated in this
paper. Capitalizing on prior research where the minimum mean square
error and Fisher information matrices were linked to
information-theoretic quantities through differentiation,
closed-form expressions for the Hessian of the mutual information
and the differential entropy are derived. These expressions are then
used to assess the concavity properties of mutual information and
differential entropy under different channel conditions and also to
derive a multivariate version of the entropy power inequality due to
Costa.
\end{abstract}

\section{Introduction and motivation} \label{sec:intro}

Closed-form expressions for the Hessian matrix of the mutual
information with respect to arbitrary parameters of the system are
useful from a theoretical perspective but also from a practical
standpoint. In system design, if the mutual information is to be
optimized through a gradient algorithm as in \cite{palomar:06}, the
Hessian matrix may be used alongside the gradient in the Newton's
method to speed up the convergence of the algorithm. Additionally,
from a system analysis perspective, the Hessian matrix can also
complement the gradient in studying the sensitivity of the mutual
information to variations of the system parameters and, more
importantly, in the cases where the mutual information is concave
with respect to the system design parameters, it can also be used to
guarantee the global optimality of a given design.

In this sense and within the framework of linear vector Gaussian
channels with arbitrary signaling, the purpose of this work is
twofold. First, we find closed-form expressions for the Hessian
matrix of the mutual information, differential entropy and entropy
power with respect to arbitrary parameters of the system and,
second, we study the concavity properties of these quantities. Both
goals are intimately related since concavity can be assessed through
the negative definiteness of the Hessian matrix. As intermediate
results of our study, we derive closed-form expressions for the
Jacobian of the minimum mean-square error (MMSE) and Fisher
information matrices, which are interesting results in their own
right and contribute to the exploration of the fundamental links
between information theory and estimation theory.

Initial connections between information- and estimation-theoretic
quantities for linear channels with additive Gaussian noise date
back from the late fifties: in the proof of Shannon's entropy power
inequality \cite{stam:59}, Stam used the fact that the derivative of
the output differential entropy with respect to the added noise
power is equal to the Fisher information of the channel output and
attributed this identity to De Bruijn. More than a decade later, the
links between both worlds strengthened when Duncan \cite{duncan:70}
and Kadota, Zakai, and Ziv \cite{kadota:71} independently
represented mutual information as a function of the error in causal
filtering.

Much more recently, in \cite{guo:05}, Guo, Shamai, and Verd\'u
fruitfully explored further these connections and, as their main
result, proved that the derivative of the mutual information (and
differential entropy) with respect to the signal-to-noise ratio
(SNR) is equal to half the MMSE regardless of the input statistics.
The main result in \cite{guo:05} was generalized to the abstract
Wiener space by Zakai in \cite{zakai:05} and by Palomar and Verd\'u
in two different directions: in \cite{palomar:06} they calculated
the partial derivatives of the mutual information and differential
entropy with respect to arbitrary parameters of the system, rather
than with respect to the SNR alone, and in \cite{palomar:07} they
represented the derivative of mutual information as a function of
the conditional marginal input given the output for channels where
the noise is not constrained to be Gaussian.

In this paper we build upon the setting of \cite{palomar:06}, where
loosely speaking, it was proved that, for the linear vector Gaussian
channel
\begin{gather}
\rvec{y} = \preS \rvec{s} + \preN \rvec{n},
\end{gather}
i) the gradients of the differential entropy $\Ent(\rvec{y})$ and
the mutual information $\I(\rvec{s}; \rvec{y})$ with respect to
functions of the linear transformation undergone by the input,
$\preS$, are linear functions of the MMSE matrix $\MSE{s}$ and ii)
the gradient of the differential entropy $\Ent(\rvec{y})$ with
respect to the linear transformation undergone by the noise,
$\preN$, are linear functions of the Fisher information matrix,
$\J{y}$.
\begin{figure}[t]
\centering
\includegraphics[width=0.9\textwidth]{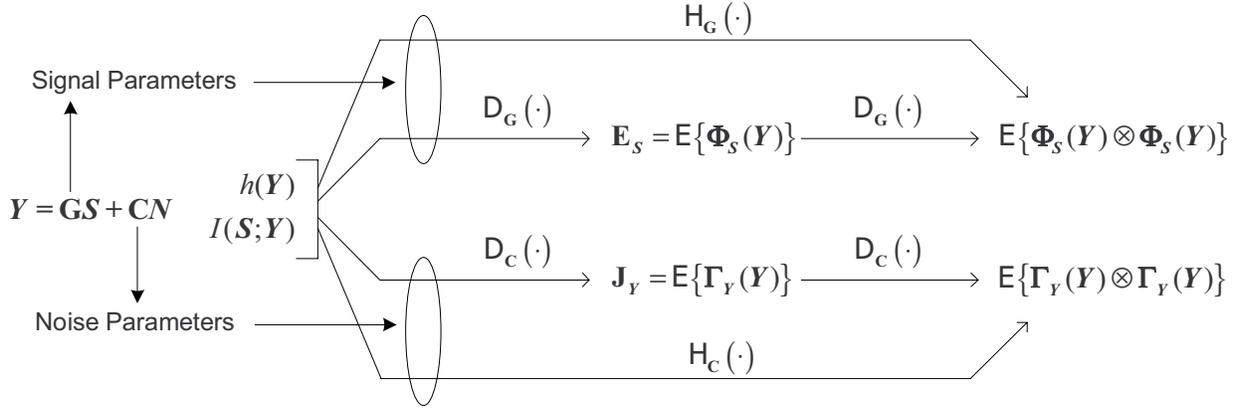}
\caption{Simplified representation of the relations between the quantities dealt with in this work. The Jacobian, $\Jacob$, and Hessian, $\Hess$, operators represent first and second order differentiation, respectively.}
\label{fig:relations}
\end{figure}

In this work, we show that the previous two key quantities $\MSE{s}$ and
$\J{y}$, which completely characterize the first-order derivatives,
are not enough to describe the second-order derivatives. For that
purpose, we introduce the more refined conditional MMSE matrix
$\CMSEr{s}{y}$ and conditional Fisher information matrix
$\CJr{y}$ (note that when these quantities are averaged with
respect to the distribution of the output $\rvecr{y}$, we recover
$\MSE{s} = \Esp{\CMSE{s}{y}}$ and $\J{y} = \Esp{\CJ{y}}$). In
particular, the second-order derivatives depend on $\CMSEr{s}{y}$
and $\CJr{y}$ through the following terms:
$\Esp{\CMSE{s}{y} \otimes \CMSE{s}{y}}$ and $\Esp{\CJ{y}
\otimes \CJ{y}}$. See Fig.~\ref{fig:relations} for a schematic representation of these relations.

Analogous results to some of the expressions presented in this paper
particularized to the scalar Gaussian channel were simultaneously
derived in \cite{guo:08, guo:08a}, where the second and third
derivatives of the mutual information with respect to the SNR were
calculated.

As an application of the obtained expressions, we show concavity
properties of the mutual information and the differential entropy,
derive a multivariate generalization of the entropy power inequality
(EPI) due to Costa in \cite{costa:85}. Our multivariate EPI has
already found an application in \cite{tandon:08} to derive outer
bounds on the capacity region in multiuser channels with feedback.

This paper is organized as follows. In Section \ref{sec:model}, the
model for the linear vector Gaussian channel is given and the
differential entropy, mutual information, minimum mean-square error,
and Fisher information quantities as well as the relationships among
them are introduced. The main results of the paper are given in
Section \ref{sec:Jacob_Hess} where we present the expressions for
the Jacobian matrix of the MMSE and Fisher information and also for
the Hessian matrix of the mutual information and differential
entropy. In Section \ref{sec:mi_concavity} the concavity properties
of the mutual information are studied and in Section
\ref{sec:Costa_EPI} a multivariate generalization of Costa's EPI in
\cite{costa:85} is given. Finally, an extension to the
complex-valued case of some of the obtained results is considered in
Section \ref{sec:complex}.

\emph{Notation:} Straight boldface denote multivariate quantities
such as vectors (lowercase) and matrices (uppercase). Uppercase
italics denote random variables, and their realizations are
represented by lowercase italics. The sets of $q$-dimensional
symmetric, positive semidefinite, and positive definite matrices are
denoted by $\SymM{q}$, $\PSD{q}$, and $\PD{q}$, respectively. The
elements of a matrix $\mat{A}$ are represented by $\mat{A}_{ij}$ or
$[\mat{A}]_{ij}$ interchangeably, whereas the elements of a vector
$\vec{a}$ are represented by $\veci{a}{i}$. The operator
$\mb{diag}(\mat{A})$ represents a column vector with the diagonal
entries of matrix $\mat{A}$, $\mb{Diag}(\mat{A})$ and
$\mb{Diag}(\vec{a})$ represent a diagonal matrix whose non-zero
elements are given by the diagonal elements of matrix $\mat{A}$ and
by the elements of vector $\vec{a}$, respectively, and
$\vecop\mat{A}$ represents the vector obtained by stacking the
columns of $\mat{A}$. For symmetric matrices, $\vechop \mat{A}$ is
obtained from $\vecop\mat{A}$ by eliminating the repeated elements
located above the main diagonal of $\mat{A}$. The Kronecker matrix
product is represented by $\mat{A}\otimes\mat{B}$ and the Schur (or
Hadamard) element-wise matrix product is denoted by $\mat{A}\circ
\mat{B}$. The superscripts $(\cdot)^\T$, $(\cdot)^\H$, and
$(\cdot)^\pinv$, denote transpose, Hermitian, and Moore-Penrose
pseudo-inverse operations, respectively. With a slight abuse of
notation, we consider that when square root or multiplicative
inverse are applied to a vector, they act upon the entries of the
vector, we thus have $\big[\sqrt{\vec{a}}\big]_i =
\sqrt{\veci{a}{i}}$ and $[1/\vec{a}]_i = 1/\veci{a}{i}$.

\section{Signal model} \label{sec:model}

We consider a general discrete-time linear vector Gaussian channel, whose output $\rvec{y} \in \R^{\dim}$ is represented by the following signal model
\begin{gather} \label{eq:MIMOio-simple}
\rvec{y} = \preS\rvec{s} + \rvec{z},
\end{gather}
where $\rvec{s} \in \R^{\dimp}$ is the zero-mean channel input
vector with covariance matrix $\Cov{s}$, the matrix $\preS \in
\R^{\dim\times \dimp}$ specifies the linear transformation undergone
by the input vector, and $\rvec{z} \in \R^{\dim}$ represents a
zero-mean Gaussian noise with non-singular covariance matrix
$\Cov{z}$.

The channel transition probability density function corresponding to the channel model in \req{eq:MIMOio-simple} is
\begin{gather} \label{eq:pdfYcondS}
\pdfvec{y|s} = \pdf{\rvec{z}}{\rvecr{y} - \preS \rvecr{s}} = \frac{1}{\sqrt{(2\pi)^{\dim} \det\left(\Cov{z}\right)}} \exp\left(-\frac{1}{2}(\rvecr{y} - \preS\rvecr{s})^\T
\Cov{z}^{-1} (\rvecr{y} - \preS\rvecr{s})\right)
\end{gather}
and the marginal probability density function of the output is given
by\footnote{We highlight that in every expression involving
integrals, expectation operators, or even a density we should
include the statement \emph{if it exists}.}
\begin{gather} \label{eq:Py}
\pdfvec{y} = \Esp{\pdf{\rvec{y}|\rvec{s}}{\rvecr{y}|\rvec{s}}},
\end{gather}
which is an infinitely differentiable continuous function of $\rvecr{y}$ regardless of the distribution of the input vector $\rvec{s}$ thanks to the smoothing properties of the added noise \cite[Section II]{costa:85}.

At some points, it may be convenient to define the random vector
$\rvec{x} = \preS\rvec{s}$ with covariance matrix given by $\Cov{x}
= \preS \Cov{s} \preS^\T$ and also express the noise vector as
$\rvec{z} = \preN\rvec{n}$, where $\preN \in \R^{\dim\times\dim'}$,
such that $\dim'\geq \dim$, and the noise covariance matrix
$\Cov{z} = \preN \Cov{n} \preN^\T$ has an inverse so that
\req{eq:pdfYcondS} is meaningful.

With this notation, $\pdfvec{y|x}$ can be obtained by replacing $\preS \rvecr{s}$ by $\rvecr{x}$ in \req{eq:pdfYcondS} and the channel model \req{eq:MIMOio-simple} can be alternatively rewritten as
\begin{gather} \label{eq:MIMOio}
%\begin{split}
\rvec{y} = \preS\rvec{s} + \preN\rvec{n}
= \rvec{x} + \preN\rvec{n} = \preS\rvec{s} + \rvec{z} = \rvec{x} + \rvec{z}.
%\end{split}
\end{gather}

In the following subsections we describe the information- and estimation-theoretic quantities whose relations we are interested in.

\subsection{Differential entropy and mutual information} \label{ssec:ent_mi}

The differential entropy\footnote{Throughout this
paper we work with natural logarithms and thus nats are used as information units.} of the continuous random vector $\rvec{y}$ is defined as \cite[Chapter 9]{cover:91}
\begin{gather} \label{eq:diff_entropy}
\Ent(\rvec{y}) = -\Esp{\log \pdf{\rvec{Y}}{\rvec{Y}}}.
\end{gather}
For the case where the distribution of $\rvec{y}$ assigns positive
mass to one or more singletons in $\R^{\dim}$, the above definition
is usually extended with $\Ent(\rvec{Y}) = -\infty$.

For the linear vector Gaussian channel in \req{eq:MIMOio}, the input-output mutual information is \cite[Chapter 10]{cover:91}
\begin{gather} \label{eq:mi-ent}
\begin{split}
\I(\rvec{s}; \rvec{y}) &= \Ent(\rvec{y}) - \Ent(\rvec{z}) \\
&= \Ent(\rvec{y}) - \frac{1}{2} \log\det( 2\pi\nume \Cov{z}) = \Ent(\rvec{y}) - \frac{1}{2} \log\det\big( 2\pi\nume \preN \Cov{n} \preN^\T\big).
\end{split}
\end{gather}

\subsection{MMSE matrix} \label{ssec:mmse}

We consider the estimation of the input signal $\rvec{s}$ based on the observation of a realization of the output $\rvec{y} = \rvecr{y}$. The mean square error (MSE) matrix of an estimate $\widehat{\rvec{s}}(\rvecr{y})$ of the input $\rvec{s}$ given the realization of the output $\rvec{y} = \rvecr{y}$ is defined as $\Esp[1]{ (\rvec{s} - \widehat{\rvec{s}}(\rvec{y})) (\rvec{s} - \widehat{\rvec{s}}(\rvec{y}))^\T}$ and it gives us a description of the performance of the estimator.

The estimator that simultaneously achieves the minimum MSE for all the
components of the estimation error vector is given by the
conditional mean estimator $\widehat{\rvec{s}}(\rvecr{y}) =
\CEsp{\rvec{s}}{\rvecr{y}}$ and the corresponding MSE matrix,
referred to as the MMSE matrix, is
\begin{gather} \label{eq:MSE_def}
\MSE{s} = \Esp[1]{ (\rvec{s} - \CEsp{\rvec{s}}{\rvec{y}}) (\rvec{s}
- \CEsp{\rvec{s}}{\rvec{y}})^\T}.
\end{gather}

An alternative and useful expression for the MMSE matrix can be
obtained by considering first the MMSE matrix conditioned on a
specific realization of the output $\rvec{y} = \rvecr{y}$, which is
denoted by $\CMSEr{s}{y}$ and defined as:
\begin{gather} \label{eq:CondE}
\CMSEr{s}{y} = \CEsp[1]{(\rvec{s} - \Esp{\rvec{s}|\rvecr{y}})(\rvec{s} - \Esp{\rvec{s}|\rvecr{y}})^\T}{\rvecr{y}}.
\end{gather}
Observe from \req{eq:CondE} that $\CMSEr{s}{y}$ is a positive semidefinite matrix. Finally, the MMSE matrix in \req{eq:MSE_def} can be obtained by taking
the expectation in \req{eq:CondE} with respect to the distribution
of the output:
\begin{gather} \label{eq:MMSE}
\MSE{s} = \Esp{\CMSE{s}{y}}.
\end{gather}

\subsection{Fisher information matrix} \label{ssec:FIM}

Besides the MMSE matrix, another quantity that is closely related to
the differential entropy is the Fisher information matrix with
respect to a translation parameter, which is a special case of the
Fisher information matrix \cite{dembo:91}. The Fisher information is
a measure of the minimum error in estimating a parameter of a
distribution and is closely related to the Cram\'er-Rao lower bound
\cite{kay:93}.

For an arbitrary random vector $\rvec{y}$, the Fisher information matrix with respect to a translation parameter is defined as
\begin{gather} \label{eq:FIM_general}
\J{y} = \Esp[1]{\Jacob_{\rvecr{y}}^\T \log \pdf{y}{\rvec{y}} \Jacob_{\rvecr{y}} \log \pdf{y}{\rvec{y}}},
\end{gather}
where $\Jacob$ is the Jacobian operator. This operator together with the Hessian operator, $\Hess$, and other definitions and conventions used for differentiation with respect to multidimensional parameters are described in Appendices \ref{ap:special_matrices} and \ref{ap:diff_conventions}.

The expression of the Fisher information in \req{eq:FIM_general} in
terms of the Jacobian of $\log \pdfvec{y}$ can be transformed into
an expression in terms of its Hessian matrix, thanks to the
logarithmic identity
\begin{gather} \label{eq:log_id}
\Hess_{\rvecr{y}} \log \pdfvec{y} = \frac{\Hess_{\rvecr{y}}
\pdfvec{y}}{\pdfvec{y}} - \Jacob_{\rvecr{y}}^\T \log
\pdf{y}{\rvecr{y}} \Jacob_{\rvecr{y}} \log \pdf{y}{\rvecr{y}}
\end{gather}
together with the fact that $\Esp[1]{\Hess_{\rvecr{y}}
\pdf{\rvec{y}}{\rvec{y}}/\pdf{\rvec{y}}{\rvec{y}}} = \int
\Hess_{\rvecr{y}} \pdfvec{y} \d \rvecr{y} = \mats{0}$, which follows
directly from the expression for $\Hess_{\rvecr{y}} \pdfvec{y}$ in
\req{eq:HessPz} in Appendix \ref{ap:partials_Py}. The alternative
expression for the Fisher information matrix in terms of the Hessian
is then
\begin{gather} \label{eq:FIM_Reg}
\J{y} = - \Esp{\Hess_{\rvecr{y}} \log \pdf{y}{\rvec{y}}}.
\end{gather}

Similarly to the previous section with the MMSE matrix, it will be useful to define a conditional form of the Fisher information matrix $\CJr{y}$, in such a way that $\J{y} = \Esp{\CJ{y}}$. At this point, it may not be clear which of the two forms \req{eq:FIM_general} or \req{eq:FIM_Reg} will be more useful for the rest of the paper; we advance that defining $\CJr{y}$ based on \req{eq:FIM_Reg} will prove more convenient:
\begin{gather}
\CJr{y} = - \Hess_{\rvecr{y}} \log \pdfvec{y} = \Cov{z}^{-1} - \Cov{z}^{-1}  \CMSEr{x}{y} \Cov{z}^{-1} \label{eq:CondJ},
\end{gather}
where the second equality is proved in Lemma \ref{lem:DlogP} in
Appendix \ref{ap:partials_Py}\footnote{Note that the lemmas placed
in the appendices have a prefix indicating the appendix where they
belong to ease its localization. From this point we will omit the
explicit reference to the appendix.} and where we have $\CMSEr{x}{y}
= \preS \CMSEr{s}{y} \preS^\T$.

\subsection{Prior known relations among information- and estimation-theoretic quantities} \label{ssec:relations}

The first known relation between the above described quantities is the De Bruijn identity \cite{stam:59} (see also the alternative derivation in \cite{guo:05}), which couples the Fisher information with the differential entropy according to
\begin{gather} \label{eq:DeBruijn}
\frac{\d}{\d t} \Ent\big(\rvec{x} + \sqrt{t} \rvec{z} \big) =
\frac{1}{2} \Tr \J{y},
\end{gather}
where, in this case $\rvec{y} = \rvec{x} + \sqrt{t} \rvec{z}$. A
multivariate extension of the De Bruijn identity was found in
\cite{palomar:06} as
\begin{gather} \label{eq:Multi-Bruijn}
\nabla_{\preN} \Ent(\rvec{x} + \preN \rvec{n}) = \J{y} \preN \Cov{n}.
\end{gather}

In \cite{guo:05}, the more canonical operational measures of mutual information and MMSE were coupled through the identity
\begin{gather} \label{eq:Guo}
\frac{\d}{\d \snr} \I\big(\rvec{s}; \sqrt{\snr}\rvec{s} + \rvec{z} \big) = \frac{1}{2} \Tr \MSE{s}.
\end{gather}
This result was generalized in \cite{palomar:06} to the multivariate case, yielding
\begin{gather} \label{eq:Multi-Guo}
\nabla_{\preS} \I(\rvec{s}; \preS\rvec{s} + \rvec{z}) = \Cov{z}^{-1} \preS \MSE{s}.
\end{gather}
Note that the simple dependence of mutual information on differential entropy established in \req{eq:mi-ent}, implies that $\nabla_{\preS} \I(\rvec{s}; \preS\rvec{s} + \rvec{z}) = \nabla_{\preS} \Ent(\preS\rvec{s} + \rvec{z})$.

From these previous existing results, we realize that the output
differential entropy function $\Ent(\preS\rvec{s} + \preN\rvec{n})$
is related to the MMSE matrix $\MSE{s}$ through differentiation with
respect to the transformation $\preS$ undergone by the signal
$\rvec{s}$ (see \req{eq:Multi-Guo}) and is related to the Fisher
information matrix $\J{y}$ through differentiation with respect to
the transformation $\preN$ undergone by the Gaussian noise
$\rvec{n}$ (see \req{eq:Multi-Bruijn}). This is illustrated in
Fig.~\ref{fig:relations}. A comprehensive account of other relations
can be found in \cite{guo:05}.

Since we are interested in calculating the Hessian matrix of
differential entropy and mutual information quantities, in the light
of the results in \req{eq:Multi-Bruijn} and \req{eq:Multi-Guo}, it
is instrumental to first calculate the Jacobian matrix of the MMSE
and Fisher information matrices, as considered in the next section.

\section{Jacobian and Hessian results} \label{sec:Jacob_Hess}

In order to derive the Hessian of the differential entropy and the
mutual information, we start by obtaining the Jacobians of the
Fisher information matrix and the MMSE matrix.

\subsection{Jacobian of the Fisher information matrix} \label{ssec:Jacob_J}

As a warm-up, consider first the signal model in \req{eq:MIMOio} with Gaussian signaling, $\rvec{y}_\mc{G} = \rvec{x}_\mc{G} + \preN \rvec{n}$. In this case, the conditional Fisher information matrix defined in \req{eq:CondJ} does not depend on the realization of the received vector $\rvecr{y}$ and is (\eg, \cite[Appendix 3C]{kay:93})
\begin{gather} \label{eq:Cj}
\Cj_{\rvec{y}_{\mc{G}}} = ( \Cov{x_{\mc{G}}} + \Cov{z} )^{-1} = \big( \Cov{x_{\mc{G}}} + \preN \Cov{n} \preN^{\T} \big)^{-1}.
\end{gather}
Consequently, we have that $\J{y_\mc{G}} = \Esp{\Cj_{\rvec{y}_{\mc{G}}}} = \Cj_{\rvec{y}_{\mc{G}}}$.

The Jacobian matrix of the Fisher information matrix with respect to the noise transformation $\preN$ can be readily obtained as
\begin{align}
\Jacob_{\preN} \J{y_\mc{G}} = \Jacob_{\Cov{z}} \J{y_\mc{G}} \cdot \Jacob_{\preN} \Cov{z} &= \Jacob_{\Cov{z}} ( \Cov{x_{\mc{G}}} + \Cov{z} )^{-1} \cdot \Jacob_{\preN} \preN \Cov{n} \preN^{\T} \label{eq:G_chainrule} \\ &= - \Dup{\dim}^\pinv \left( \J{y_\mc{G}} \otimes \J{y_\mc{G}}  \right) \Dup{\dim} \cdot 2 \Dup{\dim}^\pinv \left( \preN \Cov{n} \otimes \mat{I}_\dim \right) \label{eq:G_magnus} \\
&= -2 \Dup{\dim}^\pinv \left( \J{y_\mc{G}} \otimes \J{y_\mc{G}}
\right) \left( \preN \Cov{n} \otimes \mat{I}_\dim \right)
\label{eq:prevDcJg} \\ &= -2 \Dup{\dim}^\pinv \Esp{
\Cj_{\rvec{y}_{\mc{G}}} \otimes \Cj_{\rvec{y}_{\mc{G}}} } \left(
\preN \Cov{n} \otimes \mat{I}_\dim \right), \label{eq:DcJg}
\end{align}
where \req{eq:G_chainrule} follows from the Jacobian chain rule in Lemma \ref{lem:ChainRules}; in \req{eq:G_magnus} we have applied Lemmas \ref{lem:Jacob_chain_examples}.\ref{lem:id4.5} and \ref{lem:Jacob_chain_examples}.\ref{lem:id5} with $\Dup{\dim}^\pinv$ being the Moore-Penrose inverse of the duplication matrix $\Dup{\dim}$ defined in Appendix \ref{ap:special_matrices}\footnote{The matrix $\Dup{\dim}$ appears in \req{eq:DcJg} and in many successive expressions because we are explicitly taking into account the fact that $\J{y}$ is a symmetric matrix. The reader is referred to Appendices \ref{ap:special_matrices} and \ref{ap:diff_conventions} for more details on the conventions used in this paper.}; and finally
\req{eq:prevDcJg} follows from the facts that $\Dup{\dim} \Dup{\dim}^\pinv = \Sym{\dim}$, $\Dup{\dim}^\pinv \Sym{\dim} = \Dup{\dim}^\pinv$, and $(\mat{A}\otimes \mat{A})\Sym{\dim} = \Sym{\dim}(\mat{A}\otimes\mat{A})$, which are given in \req{eq:DupProp} and \req{eq:symm_AkronA} in Appendix \ref{ap:special_matrices}, respectively.

In the following theorem we generalize \req{eq:DcJg} for the case of arbitrary signaling.
\begin{thm}[Jacobian of the Fisher information matrix] \label{thm:DcJ}
Consider the signal model $\rvec{y} = \rvec{x} + \preN\rvec{n}$, where $\preN$ is an arbitrary deterministic matrix, the signaling $\rvec{x}$ is arbitrarily distributed, and the noise vector $\rvec{n}$ is Gaussian and independent of the input $\rvec{x}$. Then, the Jacobian of the Fisher information matrix of the $\dim$-dimensional output vector $\rvec{y}$ is
\begin{gather} \label{eq:DcJ}
\Jacob_{\preN} \J{y} = - 2 \Dup{\dim}^\pinv \Esp{ \CJ{y} \otimes
\CJ{y}} ( \preN \Cov{n} \otimes \mat{I}_\dim ),
\end{gather}
where $\CJr{y}$ is defined in \req{eq:CondJ}.
\end{thm}
\begin{IEEEproof}
Since $\J{y}$ is a symmetric matrix, its Jacobian can be written as
\begin{align}
\Jacob_{\preN} \J{y} &= \Jacob_{\preN} \vechop \J{y} \\ &=
\Jacob_{\preN} \Dup{\dim}^\pinv \vecop \J{y} \label{eq:DcJ1} \\ &=
\Dup{\dim}^\pinv \Jacob_{\preN} \vecop \J{y} \label{eq:DcJ2} \\ &=
\Dup{\dim}^\pinv (-2 \Sym{\dim} \Esp{ \CJ{y} \preN \Cov{n} \otimes
\CJ{y}}) \label{eq:DcJ3} \\ &= - 2 \Dup{\dim}^\pinv \Esp{ \CJ{y}
\otimes \CJ{y}} ( \preN \Cov{n} \otimes \mat{I}_\dim ),
\label{eq:DcJ4}
\end{align}
where \req{eq:DcJ1} follows from \req{eq:inv_duplication} in
Appendix \ref{ap:special_matrices} and \req{eq:DcJ2} follows from
Lemma \ref{lem:Jacob_chain_examples}.\ref{lem:id1.5}. The expression for $\Jacob_{\preN} \vecop
\J{y}$ is derived in Appendix \ref{ap:DcJ_proof}, which yields
\req{eq:DcJ3} and \req{eq:DcJ4} follows from Lemma \ref{lem:ACkronBD} and
$\Dup{\dim}^\pinv \Sym{\dim} = \Dup{\dim}^\pinv$ as detailed in
Appendix \ref{ap:special_matrices}.
\end{IEEEproof}

\begin{rem}
Due to the fact that, in general, the conditional Fisher information matrix $\CJr{y}$ does depend on the particular value of the observation $\rvecr{y}$, it is not possible to express the expectation of the Kronecker product as the Kronecker product of the expectations, as in \req{eq:prevDcJg} for the Gaussian signaling case, where $\Cj_{\rvec{y}_{\mc{G}}}$ does not depend on the particular value of the observation $\rvecr{y}$.
\end{rem}

\subsection{Jacobian of the MMSE matrix}  \label{ssec:Jacob_E}

Again, as a warm-up, before dealing with the arbitrary signaling case we consider first the signal model in \req{eq:MIMOio} with Gaussian signaling, $\rvec{y}_\mc{G} = \preS \rvec{s}_\mc{G} + \rvec{z}$, and study the properties of the conditional MMSE matrix, $\CMSEr{s}{y}$, which does not depend on the particular realization of the observed vector $\rvecr{y}$. Precisely, we have \cite[Chapter 11]{kay:93}
\begin{gather} \label{eq:Ce}
\CM_{\rvec{s}_{\mc{G}}} = \big( \Cov{s}^{-1} + \preS^\T \Cov{z}^{-1} \preS \big)^{-1}
\end{gather}
and thus $\MSE{s_{\mc{G}}} = \Esp{\CM_{\rvec{s}_{\mc{G}}}} = \CM_{\rvec{s}_{\mc{G}}}$.

Following similar steps as in \req{eq:G_chainrule}-\req{eq:DcJg} for the Fisher information matrix, the Jacobian matrix of the MMSE matrix with respect to the signal transformation $\preS$ can be readily obtained as
\begin{gather}
\Jacob_{\preS} \MSE{s_{\mc{G}}} = -2 \Dup{\dimp}^\pinv \Esp{
\CM_{\rvec{s}_{\mc{G}}} \otimes \CM_{\rvec{s}_{\mc{G}}} } \big(
\mat{I}_\dimp \otimes \preS^{\T} \Cov{z}^{-1}  \big),
\label{eq:DbEg}
\end{gather}
Note that the expression in \req{eq:DbEg} for the Jacobian of the MMSE matrix has a very similar structure as the Jacobian for the Fisher information matrix in \req{eq:DcJg}. The following theorem formalizes the fact that the Gaussian assumption is unnecessary for \req{eq:DbEg} to hold.
\begin{thm}[Jacobian of the MMSE matrix] \label{thm:DbE}
Consider the signal model $\rvec{y} = \preS \rvec{s} + \rvec{z}$, where $\preS$ is an arbitrary deterministic matrix, the $\dimp$-dimensional signaling $\rvec{s}$ is arbitrarily distributed, and the noise vector $\rvec{z}$ is Gaussian and independent of the input $\rvec{s}$. Then, the Jacobian of the MMSE matrix of the input vector $\rvec{s}$ is
\begin{gather} \label{eq:DbE}
\Jacob_{\preS} \MSE{s} = - 2 \Dup{\dimp}^\pinv \Esp{ \CMSE{s}{y} \otimes \CMSE{s}{y}} \big( \mat{I}_\dimp \otimes \preS^\T \Cov{z}^{-1} \big), %MAKE SURE \dimp IS CORRECT.
\end{gather}
where $\CMSEr{s}{y}$ is defined in \req{eq:CondE}.
\end{thm}
\begin{IEEEproof}
The proof is analogous to that of Theorem \ref{thm:DcJ} with the appropriate notation adaptation. The calculation of $\Jacob_{\preS} \vecop \MSE{s}$ can be found in Appendix \ref{ap:DbE_proof}.
\end{IEEEproof}

\begin{rem}
In light of the two results in Theorems \ref{thm:DcJ} and \ref{thm:DbE}, it is now apparent that $\CJr{y}$ plays an analogous role in the differentiation of the Fisher information matrix as the one played by the conditional MMSE matrix $\CMSEr{s}{y}$ when differentiating the MMSE matrix, which justifies the choice made in Section \ref{ssec:FIM} of identifying $\CJr{y}$ with the expression in \req{eq:FIM_Reg} and not with the expression in \req{eq:FIM_general}.
\end{rem}

\subsection{Jacobians with respect to arbitrary parameters} \label{ssec:Jacob_arb}

With the basic results for the Jacobian of the MMSE and Fisher information matrices in Theorems \ref{thm:DcJ} and \ref{thm:DbE}, one can easily find the Jacobian with respect to arbitrary parameters of the system through the chain rule for differentiation (see Lemma \ref{lem:ChainRules}). Precisely, we are interested in considering the case where the linear transformation undergone by the signal is decomposed as the product of two linear transformations, $\preS = \Chan \Prec$, where $\Chan$ represents the channel, which is externally determined by the propagation environment conditions, and $\Prec$ represents the linear precoder, which is specified by the system designer.

\begin{thm}[Jacobians with respect to arbitrary parameters] \label{thm:Jacob_arb}
Consider the signal model $\rvec{y} = \Chan \Prec \rvec{s} + \preN \rvec{n}$ , where $\Chan  \in \R^{\dim\times\dimpp}$, $\Prec  \in \R^{\dimpp\times\dimp}$, and $\preN  \in \R^{\dim\times\dim'}$, with $\dim'\geq\dim$, are arbitrary de\-ter\-mi\-nis\-tic matrices, the signaling $\rvec{s} \in \R^\dimp$ is arbitrarily distributed, the noise $\rvec{n} \in \R^\dim$ is Gaussian,
independent of the input $\rvec{s}$, and has covariance
matrix $\Cov{n}$, and the total noise, defined as $\rvec{z} = \preN\rvec{n} \in \R^{\dim}$, has a positive definite covariance matrix given by $\Cov{z} = \preN \Cov{n} \preN^\T$. Then, the MMSE and Fisher information matrices satisfy
\begin{align}
\Jacob_\Prec \MSE{s} &= - 2 \Dup{\dimp}^\pinv \Esp{ \CMSE{s}{y}
\otimes \CMSE{s}{y}} \big( \mat{I}_\dimp \otimes \Prec^\T \Chan^\T
\Cov{z}^{-1} \Chan \big) \\ \Jacob_\Chan \MSE{s} &= - 2
\Dup{\dimp}^\pinv \Esp{ \CMSE{s}{y} \otimes \CMSE{s}{y}} \big(
\Prec^\T \otimes \Prec^\T \Chan^\T \Cov{z}^{-1} \big) \\
\Jacob_{\Cov{z}} \J{y} &= -
\Dup{\dim}^\pinv \Esp{ \CJ{y} \otimes \CJ{y}} \Dup{\dim} \\ \Jacob_{\Cov{n}} \J{y} &= - \Dup{\dim}^\pinv \Esp{ \CJ{y} \otimes
\CJ{y}} (\preN \otimes \preN) \Dup{\dim'}.
\end{align}
\end{thm}
\begin{IEEEproof}
The Jacobians $\Jacob_{\Prec} \MSE{s}$ and $\Jacob_{\Chan} \MSE{s}$ follow from the Jacobian $\Jacob_{\preS} \MSE{s}$ calculated in Theorem \ref{thm:DbE} applying the following chain rules (from Lemma \ref{lem:ChainRules}):
\begin{align}
\Jacob_{\Prec} \MSE{s} &= \Jacob_{\preS} \MSE{s} \cdot \Jacob_{\Prec} \preS \\ \Jacob_{\Chan} \MSE{s} &= \Jacob_{\preS} \MSE{s} \cdot \Jacob_{\Chan} \preS,
\end{align}
where $\preS = \Chan\Prec$ and where $\Jacob_{\Prec} \preS  = \mat{I}_\dimp \otimes \Chan$ and $\Jacob_{\Chan} \preS = \Prec^\T \otimes \mat{I}_\dim$ can be found in Lemma \ref{lem:Jacob_chain_examples}.\ref{lem:id1}.

Similarly, the Jacobian $\Jacob_{\Cov{z}} \J{y}$ can be calculated by applying
\begin{align}
\Jacob_{\preN} \J{y} &= \Jacob_{\Cov{z}} \J{y} \cdot \Jacob_{\preN} \Cov{z},
\end{align}
where $\Jacob_{\preN} \Cov{z} = 2\Dup{\dim}^\pinv (\preN \Cov{n} \otimes \mat{I}_{\dim})$ as in Lemma \ref{lem:Jacob_chain_examples}.\ref{lem:id5}. Recalling that, in this case, the matrix $\preN$ is a dummy variable that is used only to obtain $\Jacob_{\Cov{z}} \J{y}$ through the chain rule, the factor $(\preN \Cov{n} \otimes \mat{I}_{\dim})$ can be eliminated from both sides of the equation. Using $\Dup{\dim}^\pinv \Dup{\dim} = \mat{I}_\dim$, the result follows.

Finally, the Jacobian $\Jacob_{\Cov{n}} \J{y}$ follows from the chain rule
\begin{gather}
\Jacob_{\Cov{n}} \J{y} = \Jacob_{\Cov{z}} \J{y} \cdot \Jacob_{\Cov{n}}
\Cov{z} = \Jacob_{\Cov{z}} \J{y} \cdot \Dup{\dim}^\pinv (\preN \otimes
\preN) \Dup{\dim'},
\end{gather}
where the expression for $\Jacob_{\Cov{n}}
\Cov{z}$ is obtained from Lemma
\ref{lem:Jacob_chain_examples}.\ref{lem:id2} and where we have used
that $\Dup{\dim}^\pinv (\mat{A}\otimes
\mat{A})\Dup{\dim}\Dup{\dim}^\pinv = \Dup{\dim}^\pinv
(\mat{A}\otimes \mat{A})\Sym{\dim} = \Dup{\dim}^\pinv \Sym{\dim}
(\mat{A}\otimes \mat{A}) = \Dup{\dim}^\pinv (\mat{A}\otimes
\mat{A})$.
\end{IEEEproof}

\subsection{Hessian of differential entropy and mutual information}
\label{ssec:Hessians}

Now that we have obtained the Jacobians of the MMSE and Fisher
matrices, we will capitalize on the results in \cite{palomar:06} to
obtain the Hessians of the mutual information $\I(\rvec{s};
\rvec{y})$ and the differential entropy $\Ent(\rvec{y})$. We start
by recalling the results that will be used.
\begin{lem}[Differential entropy Jacobians \cite{palomar:06}] \label{lem:Entropy_Jacobians}
Consider the setting of Theorem \ref{thm:Jacob_arb}. Then, the
differential entropy of the output vector $\rvec{y}$,
$\Ent(\rvec{y})$, satisfies
\begin{align}
\Jacob_\Prec \Ent(\rvec{y}) &= \vecop^\T \big(\Chan^\T \Cov{z}^{-1} \Chan \Prec \MSE{s}\big) \label{eq:JPh} \\ \Jacob_\Chan \Ent(\rvec{y}) &= \vecop^\T \big(\Cov{z}^{-1} \Chan \Prec \MSE{s} \Prec^\T\big) \label{eq:JHh} \\ \Jacob_\preN \Ent(\rvec{y}) &= \vecop^\T \big( \J{y} \preN \Cov{n} \big) \label{eq:JCh} \\ \Jacob_{\Cov{z}} \Ent(\rvec{y}) &= \frac{1}{2}\vecop^\T \big( \J{y}\big) \Dup{\dim} \label{eq:JRzh} \\ \Jacob_{\Cov{n}} \Ent(\rvec{y}) &= \frac{1}{2}\vecop^\T \big( \preN^\T \J{y} \preN \big) \Dup{\dim'}. \label{eq:JRnh}
\end{align}
\end{lem}
\begin{rem} \label{rem:Jacob_mi_ent_Signal}
Note that in \cite{palomar:06} the authors gave the expressions
\req{eq:JPh} and \req{eq:JHh} for the mutual information. Recalling
the simple relation \req{eq:mi-ent} between mutual information and
differential entropy for the linear vector Gaussian channel, it
becomes easy to see that \req{eq:JPh} and \req{eq:JHh} are also
valid by replacing the differential entropy by the mutual
information because the differential entropy of the noise vector is
independent of $\Prec$ and $\Chan$.
\end{rem}
\begin{rem} \label{rem:Jacob_mi_ent_Noise}
Alternatively, the expressions \req{eq:JCh}, \req{eq:JRzh}, and
\req{eq:JRnh} do not hold verbatim for the mutual information
because, in that case, the differential entropy of the noise vector
does depend on $\preN$, $\Cov{z}$, and $\Cov{n}$ and it has to be
taken into account. Then, from \req{eq:mi-ent} and applying basic
Jacobian results from \cite[Chapter 9]{magnus:88}, we have
\begin{align}
\Jacob_\preN \I(\rvec{s}; \rvec{y}) &= \Jacob_\preN \Ent(\rvec{y}) - \vecop^\T \Big( \big( \preN \Cov{n} \preN^\T \big)^{-1} \preN \Cov{n}\Big) \\ \Jacob_{\Cov{z}} \I(\rvec{s}; \rvec{y}) &= \Jacob_{\Cov{z}} \Ent(\rvec{y}) - \frac{1}{2} \vecop^\T \big( \Cov{z}^{-1} \big)\Dup{\dim} \\ \Jacob_{\Cov{n}} \I(\rvec{s}; \rvec{y}) &= \Jacob_{\Cov{n}} \Ent(\rvec{y}) - \frac{1}{2} \vecop^\T \big(\preN^\T (\preN \Cov{n} \preN^\T)^{-1} \preN \big) \Dup{\dim'}.
\end{align}
\end{rem}

With Lemma \ref{lem:Entropy_Jacobians} at hand, and the expressions obtained in the previous section for the Jacobian matrices of the Fisher information and the MMSE matrices, we are ready to calculate the Hessian matrix with respect to all the parameters of interest.

\begin{thm}[Differential entropy Hessians] \label{thm:entropy_Hessians}
Consider the setting of Theorem \ref{thm:Jacob_arb}. Then, the
differential entropy of the output vector $\rvec{y}$,
$\Ent(\rvec{y})$, satisfies
\begin{align}
\Hess_{\Prec} \Ent(\rvec{y}) &= \big(\MSE{s} \otimes \Chan^\T \Cov{z}^{-1} \Chan \big) - 2\big(\mat{I}_\dimp \otimes \Chan^\T \Cov{z}^{-1} \Chan \Prec\big) \Sym{\dimp} \Esp{ \CMSE{s}{y} \otimes \CMSE{s}{y}} \big( \mat{I}_\dimp \otimes \Prec^\T \Chan^\T \Cov{z}^{-1} \Chan \big) \nonumber \\ \Hess_{\Chan} \Ent(\rvec{y}) &= \big(\Prec \MSE{s} \Prec^\T \otimes \Cov{z}^{-1} \big) - 2\big(\Prec \otimes \Cov{z}^{-1} \Chan \Prec \big) \Sym{\dimp} \Esp{ \CMSE{s}{y} \otimes \CMSE{s}{y}} \big( \Prec^\T \otimes \Prec^\T \Chan^\T \Cov{z}^{-1} \big) \nonumber \\ &= \big(\MSE{\Prec s} \otimes \Cov{z}^{-1} \big) - 2\big(\mat{I}_\dimpp \otimes \Cov{z}^{-1} \Chan \big) \Sym{\dimpp} \Esp{ \CMSE{\Prec s}{y} \otimes \CMSE{\Prec s}{y}} \big( \mat{I}_\dimpp \otimes \Chan^\T \Cov{z}^{-1} \big) %Compte dimensions P
\\ \Hess_{\preN} \Ent(\rvec{y}) &= (\Cov{n} \otimes \J{y}) - 2 (\Cov{n}\preN^\T \otimes \mat{I}_\dim) \Sym{\dim} \Esp{ \CJ{y}  \otimes \CJ{y}}(\preN \Cov{n} \otimes \mat{I}_\dim) \\ \Hess_{\Cov{z}} \Ent(\rvec{y}) &= - \frac{1}{2} \Dup{\dim}^\T \Esp{ \CJ{y} \otimes \CJ{y}} \Dup{\dim} \label{eq:HRzh} \\ \Hess_{\Cov{n}} \Ent(\rvec{y}) &= - \frac{1}{2} \Dup{\dim'}^\T ( \preN^\T \otimes \preN^\T ) \Esp{ \CJ{y} \otimes \CJ{y}} ( \preN \otimes \preN ) \Dup{\dim'}. \label{eq:HRnh}
\end{align}
\end{thm}
\begin{IEEEproof}
See Appendix \ref{ap:entropy_Hessians}.
\end{IEEEproof}
\begin{rem} \label{rem:Hess_MI}
The Hessian results in Theorem \ref{thm:entropy_Hessians} are given
for the differential entropy. The Hessian matrices for the mutual
information can be found straightforwardly from \req{eq:mi-ent} and
Remarks \ref{rem:Jacob_mi_ent_Signal} and
\ref{rem:Jacob_mi_ent_Noise} as $\Hess_{\Prec} \I(\rvec{s};
\rvec{y}) = \Hess_{\Prec} \Ent(\rvec{y})$, $\Hess_{\Chan}
\I(\rvec{s}; \rvec{y}) = \Hess_{\Chan} \Ent(\rvec{y})$, and
\begin{align}
\Hess_{\preN} \I(\rvec{s}; \rvec{y}) &= \Hess_{\preN} \Ent(\rvec{y})
+ 2 (\Cov{n}\preN^\T \otimes \mat{I}_\dim) \Sym{\dim} \big( (\preN
\Cov{n} \preN^\T)^{-1} \otimes (\preN \Cov{n} \preN^\T)^{-1} \big)
(\preN \Cov{n} \otimes \mat{I}_\dim) \nonumber \\ & \quad
\quad - \Cov{n} \otimes (\preN \Cov{n} \preN^\T)^{-1} \\
\Hess_{\Cov{z}} \I(\rvec{s}; \rvec{y}) &= \Hess_{\Cov{z}}
\Ent(\rvec{y}) + \frac{1}{2} \Dup{\dim}^\T (\Cov{z}^{-1} \otimes
\Cov{z}^{-1}) \Dup{\dim} \\ \Hess_{\Cov{n}} \I(\rvec{s}; \rvec{y})
&= \Hess_{\Cov{n}} \Ent(\rvec{y}) + \frac{1}{2} \Dup{\dim'}^\T \big(
(\preN^\T(\preN \Cov{n} \preN^\T)^{-1} \preN) \otimes (
\preN^\T(\preN \Cov{n} \preN^\T)^{-1} \preN) \big) \Dup{\dim'}.
\end{align}
\end{rem}

\subsection{Hessian of mutual information with respect to the transmitted signal covariance} \label{ssec:mi_not_fQ}

While in the previous sections we have obtained expressions for the Jacobian of the MMSE and the Hessian of the mutual information and differential entropy with respect to the noise covariances $\Cov{z}$ and $\Cov{n}$ among others, we have purposely avoided calculating these Jacobian and Hessian matrices with respect to covariance matrices of the signal such as the squared precoder $\mat{Q}_{\Prec} = \Prec \Prec^\T$, the transmitted signal covariance $\mat{Q} = \Prec \Cov{s} \Prec^\T$, or the input signal covariance $\Cov{s}$.% as it was done in \cite[Theorem 2]{palomar:06}.

The reason is that, in general, the mutual information, the
differential entropy, and the MMSE are not functions of
$\mat{Q}_{\Prec}$, $\mat{Q}$, or $\Cov{s}$ alone. It can be seen,
for example, by noting that, given $\mat{Q}_{\Prec}$, the
corresponding precoder matrix $\Prec$ is specified up to an
arbitrary orthonormal transformation, as both $\Prec$ and $\Prec
\mat{V}$, with $\mat{V}$ being orthonormal, yield the same squared
precoder $\mat{Q}_{\Prec}$. Now, it is easy to see that the two
precoders $\Prec$ and $\Prec \mat{V}$ need not yield the same mutual
information, and, thus, the mutual information is not well defined
as a function of $\mat{Q}_\Prec$ alone because the mutual
information can not be uniquely determined from $\mat{Q}_\Prec$. The
same reasoning applies to the differential entropy and the MMSE
matrix.

There are, however, some particular cases where the quantities of
mutual information and differential entropy are indeed functions of
$\mat{Q}_\Prec$, $\mat{Q}$, or $\Cov{s}$. We have, for example, the
particular case where the signaling is Gaussian, $\rvec{s} =
\rvec{s}_{\mc{G}}$. In this case, the mutual information is given by
\begin{gather} \label{eq:miG}
\I(\rvec{s}_{\mc{G}};\rvec{y}_{\mc{G}}) = \frac{1}{2} \log \det( \mat{I}_{\dim} + \Cov{z}^{-1}\mat{H}\Prec\Cov{s}\Prec^\T\mat{H}^\T),
\end{gather}
which is, of course, a function of the transmitted signal covariance $\mat{Q} = \Prec\Cov{s}\Prec^\T$, a function of the input signal covariance $\Cov{s}$, and also a function of the squared precoder $\mat{Q}_\Prec = \Prec \Prec^\T$ when $\Cov{s} = \mat{I}_\dimp$.

Upon direct differentiation with respect to, \eg, $\mat{Q}$ we obtain \cite[Chapter 9]{magnus:88}
\begin{gather} \label{eq:DqIg}
\Jacob_{\mat{Q}} \I(\rvec{s}_{\mc{G}};\rvec{y}_{\mc{G}}) = \frac{1}{2} \vecop^\T \big( \Chan^\T \Cov{z}^{-1} \Chan ( \mat{I}_\dimpp  + \mat{Q} \Chan^\T \Cov{z}^{-1} \Chan )^{-1} \big) \Dup{\dimpp},
\end{gather}
which, after some algebra, agrees with the result in \cite[Theorem 2, Eq. (23)]{palomar:06} adapted to our notation,
\begin{gather}
\Jacob_{\mat{Q}} \I(\rvec{s}_{\mc{G}};\rvec{y}_{\mc{G}}) = \frac{1}{2} \vecop^\T \big( \Chan^\T \Cov{z}^{-1} \Chan \Prec \EM_{\rvec{s}_{\mc{G}}} \Cov{s}^{-1} \Prec^{-1}
\big) \Dup{\dimpp},
\end{gather}
where, for the sake of simplicity, we have assumed that the inverses of $\Prec$ and $\Cov{s}$ exist and where the MMSE is given by $\EM_{\rvec{s}_{\mc{G}}} = ( \Cov{s}^{-1} + \Prec^\T \Chan^\T \Cov{z}^{-1} \Chan \Prec )^{-1}$. Note now that the MMSE matrix is not a function of $\mat{Q}$ and, consequently, it cannot be used to derive the Hessian of the mutual information with respect to $\mat{Q}$ as we have done in Section \ref{ssec:Hessians} for other variables such as $\Prec$ or $\preN$. Therefore, the Hessian of the mutual information for the Gaussian signaling case has to be obtained by direct differentiation of the expression in \req{eq:DqIg} with respect to $\mat{Q}$, yielding \cite[Chapter 10]{magnus:88}
\begin{gather} \label{eq:HqI_Gauss}
\Hess_{\mat{Q}} \I(\rvec{s}_{\mc{G}}; \rvec{y}_{\mc{G}}) = \frac{1}{2} \Dup{\dimpp}^\T \big( \big((\mat{I}_{\dimpp} + \Chan^\T \Cov{z}^{-1} \Chan\mat{Q})^{-1} \Chan^\T \Cov{z}^{-1} \Chan\big) \otimes \big( \Chan^\T \Cov{z}^{-1} \Chan (\mat{I}_{\dimpp} + \Chan^\T \Cov{z}^{-1} \Chan\mat{Q})^{-1} \big) \big) \Dup{\dimpp}.
\end{gather}

Another particular case where the mutual information is a function of the transmit covariance matrices is in the low-SNR regime \cite{prelov:04}. Assuming that $\Cov{z} = N_0 \mat{I}$, Prelov and Verd\'u showed that \cite[Theorem 3]{prelov:04}
\begin{gather} \label{eq:lowSNR}
\I(\rvec{s}; \rvec{y}) = \frac{1}{2 N_0} \Tr\big( \mat{H} \Prec\Cov{s}\Prec^\T \mat{H}^\T\big) - \frac{1}{4N_0^2} \Tr\big( \big( \mat{H} \Prec\Cov{s}\Prec^\T \mat{H}^\T \big)^2 \big) + o\big(N_0^{-2}\big),
\end{gather}
where the dependence (up to terms $o(N_0^{-2})$) of the mutual information with respect to $\mat{Q} = \Prec\Cov{s}\Prec^\T$ is explicitly shown. The Jacobian and Hessian of the mutual information, for this particular case become \cite[Chapters 9 and 10]{magnus:88}:
\begin{align}
\Jacob_{\mat{Q}} \I(\rvec{s}; \rvec{y}) &= \frac{1}{2N_0}\vecop^\T\big( \Chan^\T \Chan \big)\Dup{\dimpp} - \frac{1}{2N_0^2} \vecop^\T\big( \Chan^\T \Chan \mat{Q} \Chan^\T \Chan \big)\Dup{\dimpp} + o\big(N_0^{-2}\big) \\ \Hess_{\mat{Q}} \I(\rvec{s}; \rvec{y}) &= -\frac{1}{2N_0^2}\Dup{\dimpp}^\T \big( \Chan^\T \Chan \otimes \Chan^\T \Chan \big)\Dup{\dimpp} + o\big(N_0^{-2}\big). \label{eq:HqI_lowSNR}
\end{align}

Even though we have shown two particular cases where the mutual
information is a function of the transmitted signal covariance
matrix $\mat{Q} = \Prec\Cov{s}\Prec^\T$, it is important to
highlight that care must be taken when calculating the Jacobian
matrix of the MMSE and the Hessian matrix of the mutual information
or differential entropy as, in general, these quantities are
\emph{not} functions of $\mat{Q}_\Prec$, $\mat{Q}$, nor $\Cov{s}$.
In this sense, the results in \cite[Theorem 2, Eqs. (23), (24),
(25); Corollary 2, Eq. (49); Theorem 4, Eq. (56)]{palomar:06} only
make sense when the mutual information is well defined as a function
of the signal covariance matrix (such as the cases seen above where
the signaling is Gaussian or the SNR is low).

\section{Mutual information concavity results}
\label{sec:mi_concavity}

As we have mentioned in the introduction, studying the concavity of
the mutual information with respect to design parameters of the
system is important from both analysis and design perspectives.

The first candidate as a system parameter of interest that naturally
arises is the precoder matrix $\Prec$ in the signal model $\rvec{y}
= \Chan \Prec \rvec{s} + \rvec{z}$. However, one realizes from the
expression $\Hess_{\Prec} \I(\rvec{s}; \rvec{y})$ in Remark
\ref{rem:Hess_MI} of Theorem \ref{thm:entropy_Hessians}, that for a
sufficiently small $\Prec$ the Hessian is approximately
$\Hess_{\Prec} \I(\rvec{s}; \rvec{y}) \approx \MSE{s} \otimes
\Chan^\T \Cov{z}^{-1} \Chan$, which, from Lemma
\ref{lem:kronecker_th} is positive definite and, consequently, the
mutual information is not concave in $\Prec$ (actually, it is
convex). Numerical computations show that the non-concavity of the
mutual information with respect to $\Prec$ also holds for non-small
$\Prec$.

The next candidate is the transmitted signal covariance matrix
$\mat{Q}$, which, at first sight, is better suited than the precoder
$\Prec$ as it is well known that, for the Gaussian signaling case,
the mutual information as in \req{eq:miG} is a concave function of
the transmitted signal covariance $\mat{Q}$. Similarly, in the low
SNR regime we have that, from \req{eq:HqI_lowSNR}, the mutual
information is also a concave function with respect to $\mat{Q}$.

Since in this work we are interested in the properties of the mutual
information for all the SNR range and for arbitrary signaling, we
wish to  study if the above results can be generalized.
Unfortunately, as discussed in the previous section, the first
difference of the general case with respect to the particular cases
of Gaussian signaling and low SNR is that the mutual information is
not well defined as a function of the transmitted signal covariance
$\mat{Q}$ only.

Having discarded the concavity of the mutual information with
respect to $\Prec$ and $\mat{Q}$, in the following subsections we
study the concavity of the mutual information with respect to other
parameters of the system.

For the sake of notation we define the channel covariance matrix as
$\Cov{\Chan} = \Chan^\T \Cov{z}^{-1} \Chan$, which will be used in
the remainder of the paper.

\subsection{The scalar case: concavity in the SNR}
\label{ssec:scalar_conc}

The concavity of the mutual information with respect to the SNR for
arbitrary input distributions can be derived as a corollary from
Costa's results in \cite{costa:85}, where he proved the concavity of
the entropy power of a random variable consisting of the sum of a
signal and Gaussian noise with respect to the power of the signal.
As a direct consequence, the concavity of the entropy power implies
the concavity of the mutual information in the signal power, or,
equivalently, in the SNR.

In this section, we give an explicit expression of the Hessian of
the mutual information with respect to the SNR, which was previously
unavailable for vector Gaussian channels.
\begin{cor}[Mutual information Hessian with respect to the SNR] \label{cor:snr_concav}
Consider the signal model $\rvec{y} = \sqrt{\snr} \Chan \rvec{s} + \rvec{z}$, with $\snr > 0$ and where all the terms are defined as in Theorem \ref{thm:Jacob_arb}. Then,
\begin{gather} \label{eq:HsnrI}
\Hess_{\snr} \I\left(\rvec{s}; \rvec{y}\right) =
\frac{\d^2\I\left(\rvec{s}; \rvec{y}\right)}{\d \snr^2} =
-\frac{1}{2} \Tr \Esp[1]{\left(\Cov{\Chan} \CMSE{s}{y} \right)^2}.
\end{gather}
Moreover, $\Hess_{\snr} \I\left(\rvec{s}; \rvec{y}\right) \leq 0$ for all $\snr$, which implies that the mutual information is a concave function with respect to $\snr$.
\end{cor}
\begin{IEEEproof}
First, we consider the result in \cite[Corollary 1]{palomar:06},
\begin{gather}
\Jacob_{\snr} \I\left(\rvec{s}; \rvec{y}\right) = \frac{1}{2} \Tr \: \Cov{\Chan} \MSE{s}.
\end{gather}
Now, we only need to choose $\Prec = \sqrt{\snr} \mat{I}_\dimpp$, which implies $\dimp = \dimpp$, and apply the results in Theorem \ref{thm:entropy_Hessians} and the chain rule in Lemma \ref{lem:ChainRules} to obtain
\begin{align}
\Hess_{\snr} \I\left(\rvec{s}; \rvec{y}\right) &= \frac{1}{2} \Jacob_{\snr} \Tr \: \Cov{\Chan} \MSE{s} \\ &= \frac{1}{2} \Jacob_{\MSE{s}} \Tr \: \Cov{\Chan} \MSE{s} \cdot \Jacob_{\Prec} \MSE{s} \cdot \Jacob_{\snr} \Prec \\ &= \frac{1}{2} \vecop^\T (\Cov{\Chan}) \Dup{\dimpp} (-2 \Dup{\dimpp}^\pinv \Esp{ \CMSE{s}{y} \otimes \CMSE{s}{y}} ( \mat{I}_\dimpp \otimes \sqrt{\snr} \Cov{\Chan})) \frac{1}{2\sqrt{\snr}} \vecop \mat{I}_\dimpp \\ &= -\frac{1}{2} \vecop^\T (\Cov{\Chan}) \Esp{ \CMSE{s}{y} \otimes \CMSE{s}{y}} \vecop \Cov{\Chan}, \label{eq:Hsnr_partial}
% \\ &= -\frac{1}{2} \Tr \Esp[1]{\left(\Cov{\Chan} \CMSE{s}{y} \right)^2},
\end{align}
where in last equality we have used Lemma \ref{lem:vecABC}, the equality $\Dup{\dimpp}\Dup{\dimpp}^\pinv = \Sym{\dimpp}$, and the fact that, for symmetric matrices, $\vecop^\T (\Cov{\Chan}) \Sym{\dimpp} = \vecop^\T \Cov{\Chan}$ as in \req{eq:symm_vecA} in Appendix \ref{ap:special_matrices}.

From the expression in \req{eq:Hsnr_partial}, it readily follows that the mutual information is a concave function of the $\snr$ parameter because, from Lemma \ref{lem:kronecker_th} we have that $\CMSEr{s}{y} \otimes \CMSEr{s}{y} \geq \mats{0}$, $\forall \rvecr{y}$, and, consequently, $\Hess_{\snr} \I\left(\rvec{s}; \rvec{y}\right) \leq 0$. Finally, applying again Lemma \ref{lem:vecABC} and $\vecop^\T \mat{A} \vecop \mat{B} = \Tr \mat{A}^{\! \T} \mat{B}$, the expression for the Hessian in the corollary follows.
\end{IEEEproof}
\begin{rem}
Observe that \req{eq:HsnrI} agrees with \cite[Prp.~5]{guo:08a} for
scalar Gaussian channels.
\end{rem}

We now wonder if the concavity result in Corollary \ref{cor:snr_concav} can be extended to more general quantities than the scalar SNR. In the following section we study the concavity of the mutual information with respect to the squared singular values of the precoder for the simple case where the left singular vectors of the precoder coincide with the eigenvectors of the channel covariance matrix $\Cov{\Chan}$, which is commonly referred to as the case where the precoder \emph{diagonalizes} the channel.

\subsection{Concavity in the squared singular values of the precoder when the precoder diagonalizes the channel}
\label{ssec:commute}

Consider the eigendecomposition of the $\dimpp \times \dimpp$ channel covariance matrix $\Cov{\Chan} = \mat{U}_\Chan \mb{Diag}(\vecs{\sigma}) \mat{U}_\Chan^\T$, where $\mat{U}_\Chan \in \R^{\dimpp \times \dimpp}$ is an orthonormal matrix and the vector $\vecs{\sigma} \in \R^{\dimpp}$ contains non-negative entries in decreasing order. Note that in the case where $\rank(\Cov{\Chan}) = \dimpp' < \dimpp$, the last $\dimpp - \rank(\Cov{\Chan})$ elements of the vector $\vecs{\sigma}$ are zero.

Let us now consider the singular value decomposition (SVD) of the $\dimpp \times \dimp$ precoder matrix $\Prec = \mat{U}_\Prec \mb{Diag}\big( \sqrt{\vecs{\lambda}} \big) \mat{V}_\Prec^\T$. For the case where $\dimp \geq \dimpp$, we have that $\mat{U}_\Prec \in \R^{\dimpp \times \dimpp}$ is an orthonormal matrix, the vector $\vecs{\lambda}$ is $\dimpp$-dimensional, and the matrix $\mat{V}_\Prec \in \R^{\dimp \times \dimpp}$ contains orthonormal columns such that $\mat{V}_\Prec^\T \mat{V}_\Prec = \mat{I}_\dimpp$. For the case $\dimp < \dimpp$, the matrix $\mat{U}_\Prec \in \R^{\dimpp \times \dimp}$ contains orthonormal columns such that $\mat{U}_\Prec^\T \mat{U}_\Prec = \mat{I}_\dimp$, the vector $\vecs{\lambda}$ is $\dimp$-dimensional, and $\mat{V}_\Prec \in \R^{\dimp \times \dimp}$ is an orthonormal matrix.

In the following theorem we assume $\dimp \geq \dimpp$ for the sake of simplicity, and we characterize the concavity properties of the mutual information with respect to the entries of the squared singular values vector $\vecs{\lambda}$ for the particular case where the left singular vectors of the precoder coincide with the eigenvectors of the channel covariance matrix, $\mat{U}_\Prec = \mat{U}_\Chan$. The result for the case $\dimp < \dimpp$ is stated after the following theorem, and is left without proof because it follows similar steps.
\begin{thm}[Mutual information Hessian with respect to the squared singular values of the precoder] \label{thm:Diagonal_Concav}
Consider $\rvec{y} = \Chan \Prec \rvec{s} + \rvec{z}$, where all the terms are defined as in Theorem \ref{thm:Jacob_arb}, for the
particular case where the eigenvectors of the channel covariance matrix $\Cov{\Chan}$ and the left singular vectors of the precoder $\Prec \in \R^{\dimpp \times \dimp}$ coincide, \ie, $\mat{U}_\Prec = \mat{U}_\Chan$, and where we have $\dimp \geq \dimpp$. Then, the Hessian of the mutual information with respect to the squared singular values of the precoder $\vecs{\lambda}$ is:
\begin{gather} \label{eq:HlambdaI}
\Hess_{\vecs{\lambda}} \I(\rvec{s}; \rvec{y}) = - \frac{1}{2} \mb{Diag}(\vecs{\sigma}) \Esp[1]{\CMSE{\mat{V}_\Prec^\T \rvec{s}}{y} \circ \CMSE{\mat{V}_\Prec^\T\rvec{s}}{y}} \mb{Diag}(\vecs{\sigma}),
\end{gather}
where we recall that $\mat{A}\circ\mat{B}$ denotes the Schur (or Hadamard) product. Moreover, the Hessian matrix $\Hess_{\vecs{\lambda}} \I(\rvec{s}; \rvec{y})$ is negative semidefinite, which implies that the mutual information is a concave function of the squared singular values of the precoder.
\end{thm}
\begin{IEEEproof}
The Hessian of the mutual information $\Hess_{\vecs{\lambda}} \I\left(\rvec{s}; \rvec{y}\right)$ can be obtained from the Hessian chain rule in Lemma \ref{lem:ChainRules} as
\begin{gather} \label{eq:gen_Hlambda}
\Hess_{\vecs{\lambda}} \I\left(\rvec{s}; \rvec{y}\right) = \Jacob_{\vecs{\lambda}}^\T \Prec \: \Hess_{\Prec} \I\left(\rvec{s}; \rvec{y}\right) \Jacob_{\vecs{\lambda}} \Prec + (\Jacob_{\Prec} \I\left(\rvec{s}; \rvec{y}\right) \otimes \mat{I}_{\dimpp}) \, \Hess_{\vecs{\lambda}} \Prec.
\end{gather}

Now we need to calculate $\Jacob_{\vecs{\lambda}} \Prec$ and $\Hess_{\vecs{\lambda}} \Prec$. The expression for $\Jacob_{\vecs{\lambda}} \Prec$ follows as
\begin{align}
\Jacob_{\vecs{\lambda}} \Prec &= \Jacob_{\vecs{\lambda}} \vecop \big( \mat{U}_\Chan \mb{Diag}\big(\sqrt{\vecs{\lambda}} \big) \mat{V}_\Prec^\T \big) \\ &= (\mat{V}_\Prec \otimes \mat{U}_\Chan) \Jacob_{\vecs{\lambda}} \vecop
\mb{Diag}\big(\sqrt{\vecs{\lambda}} \big) \label{eq:pas_Hlambda} \\ &= \frac{1}{2} (\mat{V}_\Prec \otimes \mat{U}_\Chan) \Rdx{\dimpp} \big(\mb{Diag}\big(\sqrt{\vecs{\lambda}} \big)\big)^{-1},
\label{eq:pas2_lambda}
\end{align}
where, in \req{eq:pas_Hlambda}, we have used Lemmas \ref{lem:vecABC}
and \ref{lem:Jacob_chain_examples}.\ref{lem:id1.5} and where
the last step follows from
\begin{gather}
[\Jacob_{\vecs{\lambda}} \vecop \mb{Diag}\big(\sqrt{\vecs{\lambda}} \big)]_{i + (j-1)\dimpp, k} = \frac{\partial}{\partial \lambda_k}
\big(\sqrt{\lambda_i} \delta_{ij}\big) = \frac{1}{2\sqrt{\lambda_k}} \delta_{ij} \delta_{ik}, \quad \{i,j,k\} \in [1, \dimpp],
\end{gather}
and recalling the definition of the reduction matrix $\Rdx{\dimpp}$ in
\req{eq:redux}, $[\Rdx{\dimpp}]_{i + (j-1)\dimpp, k} = \delta_{ij} \delta_{ik}$.

Following steps similar to the derivation of $\Jacob_{\vecs{\lambda}} \Prec$, the Hessian matrix $\Hess_{\vecs{\lambda}} \Prec$ is obtained according to
\begin{align}
\Hess_{\vecs{\lambda}} \Prec = \Jacob_{\vecs{\lambda}} (\Jacob_{\vecs{\lambda}}^\T \Prec) &= \frac{1}{2} \Jacob_{\vecs{\lambda}} \big( \big(\mb{Diag}\big(\sqrt{\vecs{\lambda}} \big)\big)^{-1} \Rdx{\dimpp}^\T (\mat{V}_\Prec^\T \otimes \mat{U}_\Chan^\T)\big) \\ &= \frac{1}{2} ((\mat{V}_\Prec \otimes \mat{U}_\Chan) \Rdx{\dimpp} \otimes \mat{I}_\dimpp ) \Jacob_{\vecs{\lambda}} \big(\mb{Diag}\big(\sqrt{\vecs{\lambda}} \big)\big)^{-1} \\ &= - \frac{1}{4} ((\mat{V}_\Prec \otimes \mat{U}_\Chan) \Rdx{\dimpp} \otimes \mat{I}_{\dimpp} ) \Rdx{\dimpp} \big(\mb{Diag}\big(\sqrt{\vecs{\lambda}} \big)\big)^{-3}. \label{eq:pas3_lambda}
\end{align}

Plugging \req{eq:pas2_lambda} and \req{eq:pas3_lambda} in \req{eq:gen_Hlambda} and operating together with the expressions for the Jacobian matrix $\Jacob_{\Prec} \I\left(\rvec{s}; \rvec{y}\right)$ and the Hessian matrix $\Hess_{\Prec} \I\left(\rvec{s}; \rvec{y}\right)$ given in Remark \ref{rem:Jacob_mi_ent_Signal} of Lemma \ref{lem:Entropy_Jacobians} and in Remark \ref{rem:Hess_MI} of Theorem \ref{thm:entropy_Hessians}, respectively, we obtain
\begin{gather} \label{eq:HlambdaI_mig}
\begin{split}
\Hess_{\vecs{\lambda}} \I(\rvec{s};\rvec{y}) = & \frac{1}{4} \big(\mb{Diag}\big(\sqrt{\vecs{\lambda}} \big)\big)^{-1} \Rdx{\dimpp}^\T\Big( \EM_{\mat{V}_\Prec^\T \rvec{s}} \otimes \mb{Diag}(\vecs{\sigma}) - 2\big(\mat{I}_{\dimpp} \otimes
\mb{Diag}\big(\vecs{\sigma}\circ \sqrt{\vecs{\lambda}}\big) \big) \Sym{\dimpp} \\ & \Esp[1]{ \mat{V}_\Prec^\T \CMSE{s}{y} \mat{V}_\Prec
\otimes \mat{V}_\Prec^\T \CMSE{s}{y} \mat{V}_\Prec}\big(\mat{I}_{\dimpp} \otimes
\mb{Diag}\big(\vecs{\sigma}\circ \sqrt{\vecs{\lambda}}\big) \big) \Big) \Rdx{\dimpp} \big(\mb{Diag}\big(\sqrt{\vecs{\lambda}} \big)\big)^{-1} \\ &-\frac{1}{4} \big( \vecop^\T \big( \mb{Diag}\big(\vecs{\sigma}\circ \sqrt{\vecs{\lambda}}\big) \EM_{\mat{V}_\Prec^\T \rvec{s}} \big)\Rdx{\dimpp} \otimes \mat{I}_\dimpp  \big)\Rdx{\dimpp} \big(\mb{Diag}\big(\sqrt{\vecs{\lambda}} \big)\big)^{-3},
\end{split}
\end{gather}
where it can be noted that the dependence of $\Hess_{\vecs{\lambda}}
\I(\rvec{s};\rvec{y})$ on $\mat{U}_\Chan$ has disappeared.

Now, applying Lemma \ref{lem:redux} the first term in last equation becomes
\begin{align}
\big(\mb{Diag}\big(\sqrt{\vecs{\lambda}} \big)\big)^{-1}
\Rdx{\dimpp}^\T\big( \EM_{\mat{V}_\Prec^\T \rvec{s}} \otimes
\mb{Diag}(\vecs{\sigma}) \big)& \Rdx{\dimpp}
\big(\mb{Diag}\big(\sqrt{\vecs{\lambda}} \big)\big)^{-1} \nonumber \\
&= \big(\mb{Diag}\big(\sqrt{\vecs{\lambda}} \big)\big)^{-1} \big(
\EM_{\mat{V}_\Prec^\T \rvec{s}} \circ \mb{Diag}(\vecs{\sigma}) \big)
\big(\mb{Diag}\big(\sqrt{\vecs{\lambda}} \big)\big)^{-1} \\ &=
\EM_{\mat{V}_\Prec^\T \rvec{s}} \circ \mb{Diag} ( \vecs{\sigma}
\circ (1/\vecs{\lambda})), \label{eq:term1}
\end{align}
whereas the third term in \req{eq:HlambdaI_mig} can be expressed as
\begin{align}
\big( \vecop^\T \big( \mb{Diag}\big(\vecs{\sigma}\circ
\sqrt{\vecs{\lambda}}\big) \EM_{\mat{V}_\Prec^\T \rvec{s}}
\big)\Rdx{\dimpp} \otimes \mat{I}_\dimp  \big)&\Rdx{\dimpp}
\big(\mb{Diag}\big(\sqrt{\vecs{\lambda}} \big)\big)^{-3} \nonumber \\
&= \mb{Diag} \big( \mb{Diag}\big(\vecs{\sigma}\circ
\sqrt{\vecs{\lambda}}\big) \EM_{\mat{V}_\Prec^\T \rvec{s}} \big)
\big(\mb{Diag}\big(\sqrt{\vecs{\lambda}} \big)\big)^{-3}
\label{eq:pas_Rdx} \\ &= \EM_{\mat{V}_\Prec^\T \rvec{s}} \circ
\mb{Diag} ( \vecs{\sigma} \circ (1/\vecs{\lambda})),
\label{eq:term3}
\end{align}
where in \req{eq:pas_Rdx} we have used that, for any square matrix
$\mat{A} \in \R^{\dimpp\times \dimpp}$,
\begin{align}
[\vecop^\T (\mat{A}) \Rdx{\dimpp}]_k &= \sum_{i,j=1}^{\dimpp}
\mat{A}_{ij} \delta_{ij} \delta_{ik} = \mat{A}_{kk} \\
[(\mb{diag}(\mat{A})^\T \otimes \mat{I}_{\dimpp})\Rdx{\dimpp}]_{kl}
&= \sum_{i,j}^{\dimpp} \mat{A}_{jj} \delta_{ki} \delta_{ij}
\delta_{il} = \mat{A}_{ll} \delta_{kl}.
\end{align}

Now, from \req{eq:term1} and \req{eq:term3} we see that the first and third terms in \req{eq:HlambdaI_mig} cancel out and, recalling that $\mat{V}_\Prec^\T \CMSE{s}{y} \mat{V}_\Prec = \CMSE{\mat{V}_\Prec^\T s}{y}$, the expression for the Hessian matrix $\Hess_{\vecs{\lambda}} \I(\rvec{s};\rvec{y})$ simplifies to
\begin{gather} \label{eq:HlambdaI_casi}
\begin{split}
\Hess_{\vecs{\lambda}} \I(\rvec{s}; \rvec{y}) =& -\frac{1}{4} \big(\mb{Diag}\big(\sqrt{\vecs{\lambda}} \big)\big)^{-1} \EspOp\big\{
\CMSE{\mat{V}_\Prec^\T s}{y} \circ \mb{Diag}\big(\vecs{\sigma}\circ \sqrt{\vecs{\lambda}}\big) \CMSE{\mat{V}_\Prec^\T s}{y} \mb{Diag}\big(\vecs{\sigma}\circ \sqrt{\vecs{\lambda}}\big)
\\ &+ \mb{Diag}\big(\vecs{\sigma}\circ \sqrt{\vecs{\lambda}}\big) \CMSE{\mat{V}_\Prec^\T s}{y} \circ \CMSE{\mat{V}_\Prec^\T s}{y} \mb{Diag}\big(\vecs{\sigma}\circ \sqrt{\vecs{\lambda}}\big)\big\} \big(\mb{Diag}\big(\sqrt{\vecs{\lambda}} \big)\big)^{-1},
\end{split}
\end{gather}
where we have applied Lemma \ref{lem:redux} and have taken into account that
\begin{gather}
2\big(\mat{I}_\dimpp \otimes
\mb{Diag}\big(\vecs{\sigma}\circ \sqrt{\vecs{\lambda}}\big) \big)
\Sym{\dimpp} =
\big(\mat{I}_\dimpp \otimes
\mb{Diag}\big(\vecs{\sigma}\circ \sqrt{\vecs{\lambda}}\big) \big) +
\Com{\dimpp}\big(\mb{Diag}\big(\vecs{\sigma}\circ
\sqrt{\vecs{\lambda}}\big) \otimes
\mat{I}_\dimpp\big) \big),
\end{gather}
together with $\Rdx{\dimpp}^\T \Com{\dimpp} = \Rdx{\dimpp}^\T$. Now, from simple inspection of the expression in \req{eq:HlambdaI_casi} and recalling the properties of the Schur product, the desired result follows.
\end{IEEEproof}
\begin{rem}
Observe from the expression for the Hessian in \req{eq:HlambdaI} that for the case where the channel covariance matrix $\Cov{\Chan}$ is rank deficient, $\rank(\Cov{\Chan}) = \dimpp' < \dimpp$, the last $\dimpp - \dimpp'$ entries of the vector $\vecs{\sigma}$ are zero, which implies that the last $\dimpp - \dimpp'$ rows and columns of the Hessian matrix are also zero.
\end{rem}
\begin{rem} \label{rem:dimpleqdimpp}
For the case where $\dimp < \dimpp$, note that the matrix $\mat{U}_\Prec \in \R^{\dimpp \times \dimp}$ with the left singular vectors of the precoder $\Prec$ is not square. We thus consider that it contains the $\dimp$ eigenvectors in $\mat{U}_\Chan$ associated with the $\dimp$ largest eigenvalues of $\Cov{\Chan}$. In this case, the Hessian matrix of the mutual information with respect to the squared singular values $\vecs{\lambda} \in \R^\dimp$ is also negative semidefinite and its expression becomes
\begin{gather}
\Hess_{\vecs{\lambda}} \I(\rvec{s}; \rvec{y}) = - \frac{1}{2} \mb{Diag}(\vecs{\tilde{\sigma}}) \Esp[1]{\CMSE{\mat{V}_\Prec^\T \rvec{s}}{y} \circ \CMSE{\mat{V}_\Prec^\T\rvec{s}}{y}} \mb{Diag}(\vecs{\tilde{\sigma}}),
\end{gather}
where we have defined $\vecs{\tilde{\sigma}} = (\sigma_1 \sigma_2 \ldots \sigma_\dimp)^\T$ and where we recall that, in this case, $\mat{V}_\Prec \in \R^{\dimp\times \dimp}$.
\end{rem}
We now recover a result obtained in \cite{lozano:06} were it was proved that the mutual information is concave in the power allocation for the case of parallel channels. Note, however, that \cite{lozano:06} considered independence of the elements in the signaling vector $\rvec{s}$, whereas the following result shows that it is not necessary.
\begin{cor}[Mutual information concavity with respect to the power allocation in parallel channels]
Par\-ti\-cu\-lar\-i\-z\-i\-n\-g Theorem \ref{thm:Diagonal_Concav} for the case where the channel $\Chan$, the precoder $\Prec$, and the noise covariance $\Cov{z}$ are diagonal matrices, which implies that $\mat{U}_\Prec = \mat{U}_\Chan = \mat{I}_\dimpp$, it follows that the mutual information is a concave function with respect to the power allocation for parallel non-interacting channels for an arbitrary distribution of the signaling vector $\rvec{s}$.
\end{cor}

\subsection{General negative results}
\label{ssec:general_negative}

In the previous section we have proved that the mutual information is a concave function of the squared singular values of the precoder matrix $\Prec$ for the case where the left singular vectors of the precoder $\Prec$ coincide with the eigenvectors of the channel correlation matrix, $\Cov{\Chan}$. For the general case where these vectors do not coincide, the mutual information is not a concave function of the squared singular values of the precoder. This fact is formally established in the following theorem through a counterexample.
\begin{thm}[General non-concavity of the mutual information] \label{thm:CounterEx1}
Consider $\rvec{y} = \Chan \Prec \rvec{s} + \rvec{z}$, where all the terms are defined as in Theorem \ref{thm:Jacob_arb}. It then follows that, in general, the mutual information is not a concave function with respect to the squared singular values of the precoder $\vecs{\lambda}$.
\end{thm}
\begin{IEEEproof}
We present a simple two-dimensional counterexample. Assume that the
noise is white $\Cov{z} = \mat{I}_2$ and consider the following
channel matrix and precoder structure
\begin{gather}
\Chan_{ce} = \left(
\begin{array}{cc}
1 & \beta \\ \beta & 1
\end{array}
\right), \quad \Prec_{ce} = \mb{Diag}\big(\sqrt{\vecs{\lambda}}\big) = \left(
\begin{array}{cc}
\sqrt{\lambda_{1}} & 0 \\ 0 & \sqrt{\lambda_{2}}
\end{array}
\right),
\end{gather}
where $\beta \in (0,1]$ and assume that the distribution for the signal vector $\rvec{s}$ has two equally likely mass points at the following positions
\begin{gather}
\rvecr{s}^{(1)} = \left(
\begin{array}{c}
2 \\ 0
\end{array} \right), \quad \rvecr{s}^{(2)} = \left(
\begin{array}{c}
0 \\ 2
\end{array} \right).
\end{gather}
%which implies that $\Cov{s} = \mat{I}_2$.

Accordingly, we define the noiseless received vector as $\rvecr{r}^{(k)} = \Chan_{ce} \Prec_{ce} \rvecr{s}^{(k)}$, for $k=\{1,2\}$, which yields
\begin{gather}
\rvecr{r}^{(1)} = 2\left(
\begin{array}{c}
\sqrt{\lambda_{1}} \\ \beta \sqrt{\lambda_{1}}
\end{array} \right), \quad \rvecr{r}^{(2)} = 2\left(
\begin{array}{c}
\beta \sqrt{\lambda_{2}} \\ \sqrt{\lambda_{2}}
\end{array} \right).
\end{gather}

We now define the mutual information for this counterexample as
\begin{gather}
\I_{ce}(\lambda_{1}, \lambda_{2}, \beta) = \I(\rvec{s}; \Chan_{ce}\Prec_{ce}\rvec{s} + \rvec{w} ).
\end{gather}
Since there are only two possible signals to be transmitted, $\rvecr{s}^{(1)}$ and $\rvecr{s}^{(2)}$, it is clear that $0 \leq \I_{ce} \leq \log 2$. Moreover we will use the fact %reported in \cite{}
that, as $\Cov{z} = \mat{I}_2$, the mutual information is an increasing function of the squared distance of the two only possible received vectors $\d^2(\lambda_{1}, \lambda_{2}, \beta) = \Vert \rvecr{r}^{(1)} - \rvecr{r}^{(2)} \Vert^2$, which is denoted by $\I_{ce}(\lambda_{1}, \lambda_{2}, \beta) = f\big( \d^2(\lambda_{1}, \lambda_{2}, \beta)\big)$, where $f$ is an increasing  function.

For a fixed $\beta$, we want to study the concavity of $\I_{ce}(\lambda_{1}, \lambda_{2}, \beta)$ with respect to $(\lambda_{1}, \lambda_{2})$. In order to do so, we restrict ourselves to the study of concavity along straight lines of the type $\lambda_{1} + \lambda_{2} = \rho$, with $\rho > 0$, which is sufficient to disprove the concavity.

Given three aligned points, such that the point in between is at the same distance of the other two, if a function is concave in $(\lambda_{1}, \lambda_{2})$ it means that the average of the function evaluated at the two extreme points is smaller than or equal to the function evaluated at the midpoint. Consequently, concavity can be disproved by finding three aligned points, such that the aforementioned concavity property is violated.

Our three aligned points will be $(\rho, 0)$, $(0, \rho)$, and $(\rho/2, \rho/2)$ and instead of working with the mutual information we will work with the squared distance among the received points because closed form expressions are available.

Operating with the received vectors and recalling that $\beta \in (0, 1]$, we can easily obtain
\begin{align}
\d^2(\rho, 0, \beta) &= \d^2(0, \rho, \beta) = 4\rho ( 1 + \beta^2) > 4\rho \\ \d^2(\rho/2, \rho/2, \beta) &= 4\rho (1 - \beta )^2 < 4\rho.
\end{align}
The first equality means that the mutual information evaluated at the extreme points has the same quantitative value and is always above a certain threshold, $f(4\rho)$, independently of the value of $\beta$. Consequently the mean of the mutual information evaluated at the two extreme points is equal to the value on any of the extreme points. The second equality means that the function evaluated at the point in between is always below this same threshold.

Now it is clear that, given any $\rho > 0$ we can always find $\beta$ such that $0< \beta \leq 1$ and that
\begin{gather}
\I_{ce} (\rho, 0, \beta) = \I_{ce}(0, \rho, \beta) > \I_{ce} (\rho/2, \rho/2, \beta),
\end{gather}
which contradicts the concavity hypothesis.
\end{IEEEproof}

\begin{figure}[t]
\centering
\includegraphics[width=0.8\textwidth]{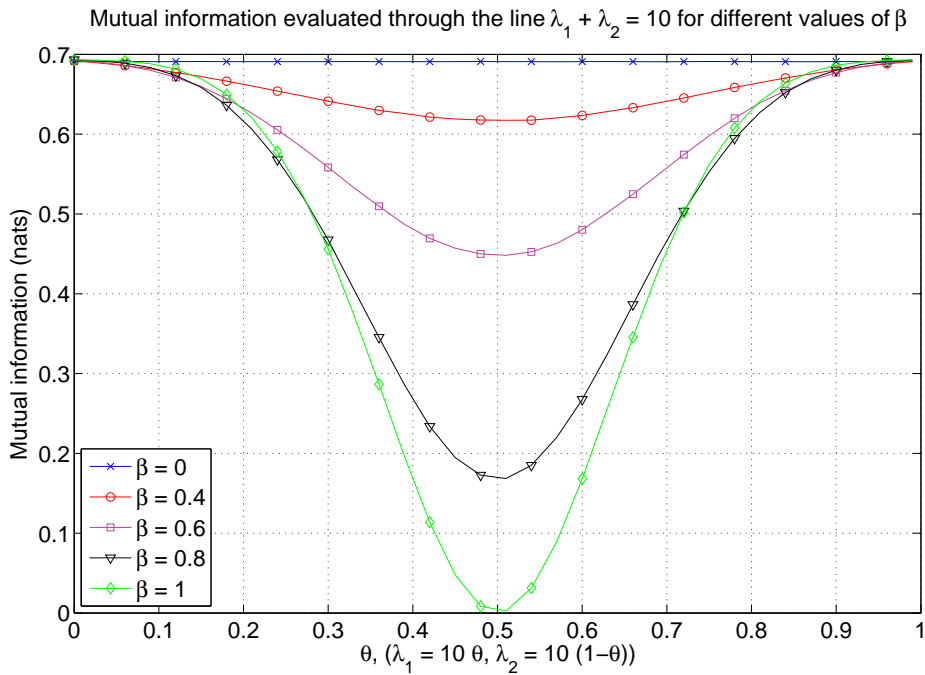}
\caption{Graphical representation of the mutual information $\I_{ce}(\lambda_1, \lambda_2, \beta)$ in the counterexample for different values of the channel parameter $\beta$ along the line $\lambda_1 + \lambda_2 = \rho = 10$. It can be readily seen that, except for the case $\beta = 0$, the function is not concave.}
\label{fig:counterexample}
\end{figure}

For illustrative purposes, in Fig.~\ref{fig:counterexample} we have depicted the mutual information for different values of the channel parameter $\beta$ for the counterexample in the proof of Theorem \ref{thm:CounterEx1}. Note that the function is only concave (and, in fact, linear) for the case where the channel is diagonal, $\beta = 0$, which agrees with the results in Theorem \ref{thm:Diagonal_Concav}.

\subsection{Concavity results summary}

At the beginning of Section \ref{sec:mi_concavity} we have argued that the mutual information is concave with respect to the full transmitted signal covariance matrix $\mat{Q}$ for the case where the signaling is Gaussian and also for the low SNR regime. Next we have discussed that this result cannot be generalized for arbitrary signaling distributions because, in the general case, the mutual information is not well defined as a function of $\mat{Q}$ alone.

In Sections \ref{ssec:scalar_conc} and \ref{ssec:commute}, we have encountered two particular cases where the mutual information is a concave function. In the first case, we have seen that the mutual information is concave with respect to the SNR and, in the second, that the mutual information is a concave function of the squared singular values of the precoder, provided that the eigenvectors of the channel covariance $\Cov{\Chan}$ and the left singular vectors of the precoder $\Prec$ coincide. For the general case where these vectors do not coincide in general, we have shown in Section \ref{ssec:general_negative} that the mutual information is not concave in the squared singular values.

A summary of the different concavity results for the mutual information as a function of the configuration of the linear vector Gaussian channel can be found in Table \ref{table:concavity}.
\begin{table}[t]
\caption{Summary of the concavity type of the mutual
information. \newline ($\checkmark$ indicates concavity, $\times$ indicates non-concavity, and $-$ indicates that it does not apply)}
\centering
\begin{tabular}{| l || c | c | c |}
  \hline & \textbf{Scalar} & \textbf{Vector} & \textbf{Matrix} \\  \cline{2-4} & Power, $\snr$, & Squared singular values, $\vecs{\lambda}$, & Transmit covariance, $\mat{Q}$, \\ \textbf{Cases} &  $\Prec = \sqrt{\snr} \mat{I}$ & $\Prec = \mat{U}_\Prec \mb{Diag}\big(\sqrt{\vecs{\lambda}}) \mat{V}_\Prec^\T$ & $\mat{Q} = \Prec \Cov{s} \Prec^\T$ \\ \hline \hline \multirow{2}{*}{General case: $\rvec{y} = \Chan \Prec \rvec{s} + \rvec{z}$} & $\checkmark$ \cite{guo:05} \cite{costa:85} & \multirow{2}{*}{$\times$ (Section \ref{ssec:general_negative})} & \multirow{2}{*}{$-$ (Section \ref{ssec:mi_not_fQ})} \\ & (Section \ref{ssec:scalar_conc}) & & \\ \hline
  Channel covariance $\Cov{\Chan}$ and precoder $\Prec$ & \multirow{2}{*}{$\checkmark$} & \multirow{2}{*}{$\checkmark$ (Section \ref{ssec:commute})} &  \multirow{2}{*}{$-$ (Section \ref{ssec:mi_not_fQ})} \\ share singular/eigenvectors: $\mat{U}_\Prec = \mat{U}_\Chan$ & & & \\ \hline Independent parallel communication: & \multirow{2}{*}{$\checkmark$} & \multirow{2}{*}{$\checkmark$ \cite{lozano:06}} & $\checkmark$ \cite{lozano:06} \\ $\Cov{\Chan} = \mat{U}_\Prec = \mat{V}_\Prec = \mat{I}$, $P_{\rvec{s}} = \prod_i P_{\rveci{s}{i}}$ & & & (Note that $\mat{Q}$ is diagonal) \\ \hline Low SNR regime: $\Cov{z} = N_0\mat{I}_\dim$, $N_0 \gg 1$ & $\checkmark$ & $\checkmark$ & $\checkmark$ \cite{prelov:04} \\ \hline Gaussian signaling: $\rvec{s} = \rvec{s}_{\mc{G}}$ & $\checkmark$ & $\checkmark$ & $\checkmark$ \\ \hline
\end{tabular}
\label{table:concavity}
\end{table}

\section{Multivariate extension of Costa's entropy power inequality}
\label{sec:Costa_EPI}

Having proved that the mutual information and, hence, the
differential entropy are concave functions of the squared singular
values $\vecs{\lambda}$ of the precoder $\Prec$ for the case where
the left singular vectors of the precoder coincide with the
eigenvectors of the channel covariance $\Cov{\Chan}$,
$\Hess_{\vecs{\lambda}} \I(\rvec{s}; \rvec{y}) =
\Hess_{\vecs{\lambda}} \Ent(\rvec{y}) \leq \vecs{0}$, we want to
study if this last result can be strengthened by proving the
concavity in $\vecs{\lambda}$ of the entropy power.

The entropy power of the random vector $\rvec{Y} \in \R^{\dim}$ was
first introduced by Shannon in his seminal work \cite{shannon:48}
and is, since then, defined as
\begin{gather} \label{eq:EP}
\EP(\rvec{Y}) = \frac{1}{2 \pi \nume} \exp \left(
\frac{2}{\dim} \Ent(\rvec{Y}) \right),
\end{gather}
where $\Ent(\rvec{Y})$ represents the differential entropy as
defined in \req{eq:diff_entropy}. The entropy power of a random
vector $\rvec{Y}$ represents the variance (or power) of a standard
Gaussian random vector $\rvec{Y}_\mc{G} \sim
\NZ{\sigma^2\mat{I}_\dim}$ such that both $\rvec{Y}$ and
$\rvec{Y}_\mc{G}$ have identical differential entropy,
$\Ent(\rvec{Y}_\mc{G}) = \Ent(\rvec{Y})$.

Costa proved in \cite{costa:85} that, provided that the random
vector $\rvec{w}$ is white Gaussian distributed, then
\begin{gather} \label{eq:CostaEPIO}
\EP(\rvec{x} + \sqrt{\CostaT}\rvec{w}) \geq (1-\CostaT)\EP(\rvec{x}) + \CostaT\EP(\rvec{x} + \rvec{w}),
\end{gather}
where $t\in[0,1]$. As Costa noted, the above entropy power inequality (EPI) is equivalent to the concavity of the entropy power function $\EP(\rvec{x} +
\sqrt{\CostaT} \rvec{w})$ with respect to the parameter $\CostaT$,
or, formally\footnote{The equivalence between equations
\req{eq:CostaEPI} and \req{eq:CostaEPIO} is due to the fact that the
function $\EP(\rvec{x} + \sqrt{\CostaT} \rvec{w})$ is twice
differentiable almost everywhere thanks to the smoothing properties
of the added Gaussian noise.}
\begin{gather} \label{eq:CostaEPI}
\frac{\d^2}{\d \CostaT^2} \EP\big(\rvec{x} + \sqrt{\CostaT}
\rvec{w}\big) = \Hess_\CostaT \EP\big(\rvec{x} + \sqrt{\CostaT}
\rvec{w}\big) \leq 0.
\end{gather}
Due to its inherent interest and to the fact that the proof by Costa
was rather involved, simplified proofs of his result have been
subsequently given in \cite{dembo:89, villani:00, guo:06, rioul:07}.

Additionally, in his paper Costa presented two extensions of his
main result in \req{eq:CostaEPI}. Precisely, he showed that the EPI
is also valid when the Gaussian vector $\rvec{w}$ is not white, and
also for the case where the $\CostaT$ parameter is multiplying the
arbitrarily distributed random vector $\rvec{X}$ instead:
\begin{gather} \label{eq:CostaEPI2}
\Hess_\CostaT \EP\big(\sqrt{t}\rvec{x} + \rvec{w}\big) \leq 0.
\end{gather}
Observe that $\sqrt{t}$ in \req{eq:CostaEPI2} plays the role of a scalar precoder. We next consider an extension of \req{eq:CostaEPI2} to the case where the scalar precoder $\sqrt{t}$ is replaced by a multivariate precoder $\Prec \in \R^{\dimpp\times \dimp}$ and a channel $\Chan \in \R^{\dim\times \dimpp}$ for the particular case where the precoder left singular vectors coincide with the channel covariance eigenvectors. Similarly as in Section \ref{ssec:commute} we assume that $\dimp \geq \dimpp$. The case $\dimp < \dimpp$ is presented after the proof of the following theorem.
\begin{thm}[Costa's multivariate EPI] \label{thm:MVCostaEPI}
Consider $\rvec{y} = \Chan \Prec \rvec{s} + \rvec{z}$, where all the terms are defined as in Theorem \ref{thm:Jacob_arb}, for the
particular case where the eigenvectors of the channel covariance matrix $\Cov{\Chan}$ coincide with the left singular vectors of the precoder $\Prec \in \R^{\dimpp\times \dimp}$ and where we assume that $\dimp \geq \dimpp$. It then follows that the entropy power $\EP(\rvec{Y})$ is a concave function of $\vecs{\lambda}$, \ie,
\begin{gather} \nonumber
\Hess_{\bs{\lambda}} \EP(\rvec{Y}) \leq \mats{0}.
\end{gather}
Moreover, the Hessian matrix of the entropy power function
$\EP(\rvec{y})$ with respect to $\bs{\lambda}$ is given by
\begin{gather} \label{eq:N}
\Hess_{\bs{\lambda}} \EP(\rvec{Y}) = \frac{\EP(\rvec{Y})}{\dim} \mb{Diag}(\vecs{\sigma})
\left( \frac{\mb{diag}(\MSE{\mat{V}_\Prec^\T s})
\mb{diag}(\MSE{\mat{V}_\Prec^\T s})^\T}{\dim} -
\Esp{\CMSE{\mat{V}_\Prec^\T s}{y} \circ \CMSE{\mat{V}_\Prec^\T
s}{y}}\right)\mb{Diag}(\vecs{\sigma}),
\end{gather}
where we recall that $\mb{diag}(\MSE{\mat{V}_\Prec^\T s})$ is a
column vector with the diagonal entries of the matrix
$\MSE{\mat{V}_\Prec^\T s}$.
\end{thm}
\begin{IEEEproof}
First, let us prove \req{eq:N}. From the definition of the entropy
power in \req{eq:EP} and applying the chain rule for Hessians in \req{eq:Hess_chain_rule} we obtain
\begin{gather} \label{eq:Nintermig}
\begin{split}
\Hess_{\vecs{\lambda}} \EP(\rvec{Y}) &= \Jacob_{\vecs{\lambda}}^\T
\Ent(\rvec{y}) \cdot \Hess_{\Ent(\rvec{y})} \EP(\rvec{y}) \cdot
\Jacob_{\vecs{\lambda}} \Ent(\rvec{y}) + \Jacob_{\Ent(\rvec{y})}
\EP(\rvec{y}) \cdot \Hess_{\vecs{\lambda}} \Ent(\rvec{y}) \\ &=
\frac{2\EP(\rvec{y})}{\dim} \left( \frac{2\Jacob_{\vecs{\lambda}}^\T
\Ent(\rvec{y}) \Jacob_{\vecs{\lambda}} \Ent(\rvec{y})}{\dim} +
\Hess_{\vecs{\lambda}} \Ent(\rvec{y}) \right).
\end{split}
\end{gather}
Now, recalling from \cite[Eq. (61)]{guo:05} that
$\Jacob_{\vecs{\lambda}}^\T\Ent(\rvec{Y}) =
(1/2)\mb{Diag}(\vecs{\sigma})\mb{diag}(\MSE{\mat{V}_\Prec^\T s})$ and incorporating the
expression for $\Hess_{\vecs{\lambda}} \Ent(\rvec{Y})$ calculated in
Theorem \ref{thm:Diagonal_Concav}, the result in \req{eq:N} follows.

Now that a explicit expression for the Hessian matrix has been
obtained, we wish to prove that it is negative semidefinite. Note
from \req{eq:Nintermig} that, except for the positive factor
$2\EP(\rvec{y})/\dim$, the Hessian matrix $\Hess_{\vecs{\lambda}}
\EP(\rvec{Y})$ is the sum of a rank one positive semidefinite matrix
and the Hessian matrix of the differential entropy, which is
negative semidefinite according to Theorem
\ref{thm:Diagonal_Concav}. Consequently, the definiteness of
$\Hess_{\vecs{\lambda}} \EP(\rvec{Y})$ is, a priori, undetermined,
and some further developments are needed to determine it, which is
what we do next.

First consider the positive semidefinite matrix $\mat{A}(\rvecr{y}) \in
\PSD{\dimpp'}$, which is obtained by selecting the first $\dimpp' = \rank(\Cov{\Chan})$ columns and rows of the positive semidefinite matrix $\mb{Diag}\big( \sqrt{\vecs{\sigma}}\big) \CMSEr{\mat{V}_\Prec^\T
s}{y} \mb{Diag}\big( \sqrt{\vecs{\sigma}}\big)$,
\begin{gather}
\left[ \mat{A}(\rvecr{y}) \right]_{ij} = \big[ \mb{Diag}\big( \sqrt{\vecs{\sigma}}\big) \CMSE{\mat{V}_\Prec^\T
s}{y} \mb{Diag}\big( \sqrt{\vecs{\sigma}}\big) \big]_{ij}, \quad \{ i, j \} = 1, \ldots, \dimpp'.
\end{gather}
With this definition, it is now easy to see that the expression
\begin{gather} \label{eq:eqN}
\frac{\Esp{\mb{diag}(\mat{A}(\rvecr{y}))}\Esp[1]{\mb{diag}(\mat{A}(\rvecr{y}))^\T}}{\dim} - \Esp[1]{\mat{A}(\rvecr{y}) \circ \mat{A}(\rvecr{y})}
\end{gather}
coincides (up to the factor $2\EP(\rvec{y})/\dim$) with the first $\dimpp'$ rows and columns of the Hessian matrix $\Hess_{\bs{\lambda}} \EP(\rvec{Y})$ in \req{eq:N}. Recalling that the remaining elements of the Hessian matrix $\Hess_{\bs{\lambda}} \EP(\rvec{Y})$ are zero due to the presence of the matrix $\mb{Diag}(\vecs{\sigma})$, it is sufficient to show that the expression in \req{eq:eqN} is negative semidefinite to prove the negative semidefiniteness of $\Hess_{\bs{\lambda}} \EP(\rvec{Y})$.

Now, we apply Pro\-po\-si\-tion \ref{prp:XXones} to $\mat{A}(\rvecr{y})$, yielding
\begin{gather} \label{eq:Av}
\mat{A}(\rvecr{y}) \circ \mat{A}(\rvecr{y}) \geq
\frac{\diag{\mat{A}(\rvecr{y})}\diag{\mat{A}(\rvecr{y})}^\T}{\dimpp'}.
\end{gather}
Taking the expectation in both sides of \req{eq:Av}, we have
\begin{gather} \label{eq:EspAv}
\Esp{\mat{A}(\rvec{y}) \circ \mat{A}(\rvec{y})} \geq
\frac{\Esp{\diag{\mat{A}(\rvec{y})}\diag{\mat{A}(\rvec{y})}^\T}}{\dimpp'},
\end{gather}
From Lemma \ref{lem:psd} we know that
\begin{gather} \nonumber
\Esp[1]{\diag{\mat{A}(\rvec{y})}\diag{\mat{A}(\rvec{y})}^\T} \geq
\Esp[1]{\diag{\mat{A}(\rvec{y})}}\Esp[1]{\diag{\mat{A}(\rvec{y})}^\T},
\end{gather}
from which it follows that
\begin{gather} \nonumber
\Esp{\mat{A}(\rvec{y}) \circ \mat{A}(\rvec{y})} \geq
\frac{\Esp{\diag{\mat{A}(\rvec{y})}}\Esp{\diag{\mat{A}(\rvec{y})}^\T}}{\dimpp'}.
\end{gather}
Since the operators $\diag{\mat{A}}$ and expectation commute we
finally obtain the desired result as
\begin{gather} \nonumber
\Esp{\mat{A}(\rvec{y}) \circ \mat{A}(\rvec{y})} \geq
\frac{\diag{\Esp{\mat{A}(\rvec{y})}}\diag{\Esp{\mat{A}(\rvec{y})}}^\T}{\dimpp'} \geq \frac{\diag{\Esp{\mat{A}(\rvec{y})}}\diag{\Esp{\mat{A}(\rvec{y})}}^\T}{\dim},
\end{gather}
where in last inequality we have used that $\dimpp' = \rank(\Cov{\Chan}) \leq \min\{\dim, \dimpp \} \leq \dim$, as $\Cov{\Chan} = \Chan^\T \Cov{z}^{-1} \Chan$ and $\Chan \in \R^{\dim\times\dimpp}$.
\end{IEEEproof}
\begin{rem}
For the case where $\dimp < \dimpp$, we assume that the matrix $\mat{U}_\Prec \in \R^{\dimpp \times \dimp}$ contains the $\dimp$ eigenvectors in $\mat{U}_\Chan$ associated with the $\dimp$ largest eigenvalues of $\Cov{\Chan}$. It then follows that the Hessian matrix $\Hess_{\vecs{\lambda}} \EP(\rvec{Y})$ is also negative semidefinite and its expression is the same given in \req{eq:N} by simply replacing $\vecs{\sigma}$ by $\vecs{\tilde{\sigma}} = (\sigma_1 \ldots \sigma_\dimp)^\T$.
\end{rem}
\begin{rem}
For the case where $\Cov{\Chan} = \mat{I}_\dimpp$ and $\dimpp = \dimp$ we recover our results in \cite{payaro:08a}.
\end{rem}

Another possibility of multivariate generalization of Costa's EPI
would be to study the concavity of $\EP(\rvec{x} + \rvec{z})$ with
respect to the covariance of the noise vector $\Cov{z}$. Numerical
computations seem to indicate that the entropy power is indeed
concave in $\Cov{z}$. However, a proof has been elusive, mainly due to the fact that, differently from the conditional MSE, $\CMSEr{s}{y}$, the
conditional Fisher information matrix $\CJr{y}$, which appears when differentiating with respect to $\Cov{z}$, is not a positive
definite function $\forall \rvecr{y}$.

\section{Extensions to the complex field}
\label{sec:complex}

So far, the presented results hold for the case where all the
variables and parameters take values from the field of real numbers.
Due to the simplicity of working with baseband equivalent
models, it is a common practice when studying communication systems
to model the parameters and random variables in the
complex field, and work with the following complex linear vector Gaussian channel:
\begin{gather} \label{eq:MIMOio_cpx}
\rvec{y}_c = \preS_c \rvec{s}_c + \rvec{z}_c,
\end{gather}
where $\preS_c \in \C^{\dim\times \dimp}$ and all the other
dimensions are defined accordingly and the noise $\rvec{z}_c$ is a zero
mean circularly symmetric (or proper \cite{neeser:93}) complex
Gaussian random vector with covariance
$\Esp{\rvec{z}_c\rvec{z}_c^\H} = {\Cov{z}}_c$. The complex model in \req{eq:MIMOio_cpx} can be equivalently rewritten by defining an extended double-dimensional real model of \req{eq:MIMOio_cpx}. We consider the extended vectors and matrices
\begin{gather}
\rvec{y}_r = \left[
\begin{array}{c}
\real \rvec{y}_c \\ \imag \rvec{y}_c
\end{array}
\right], \: \preS_r = \left[
\begin{array}{cc}
\real \preS_c & - \imag \preS_c \\  \imag \preS_c & \real \preS_c
\end{array}
\right], \: \rvec{s}_r = \left[
\begin{array}{c}
\real \rvec{s}_c \\ \imag \rvec{s}_c
\end{array}
\right], \: \rvec{z}_r = \left[
\begin{array}{c}
\real \rvec{z}_c \\ \imag \rvec{z}_c
\end{array}
\right], \label{eq:ext_model}
\end{gather}
and then rewrite the input-output relation in \req{eq:MIMOio_cpx} according to the real model
\begin{gather} \label{eq:MIMOio_real_ex}
\rvec{y}_r = \preS_r \rvec{s}_r + \rvec{z}_r.
\end{gather}
With these definitions, we have that, for example, $\Ent(\rvec{y}_c) = \Ent(\rvec{y}_r)$ or $\I(\rvec{s}_c; \rvec{y}_c) = \I(\rvec{s}_r; \rvec{y}_r)$ \cite{neeser:93}.

Working with the real model in \req{eq:MIMOio_real_ex}, it is possible to calculate, for example, the Jacobian of the mutual information with respect to the complex precoder $\preS_c$ by using the results for the real case and the chain rule as
\begin{align}
\Jacob_{\preS_c} \I(\rvec{s}_c; \rvec{y}_c) = \Jacob_{\preS_c} \I(\rvec{s}_r; \rvec{y}_r) & \triangleq \frac{1}{2} \big(\Jacob_{\real \preS_c} \I(\rvec{s}_r; \rvec{y}_r) - \textrm{j} \Jacob_{\imag \preS_c} \I(\rvec{s}_r; \rvec{y}_r)\big) \\ & = \frac{1}{2} \big(\Jacob_{\preS_r} \I(\rvec{s}_r; \rvec{y}_r) \Jacob_{\real \preS_c} \preS_r - \textrm{j}\Jacob_{\preS_r} \I(\rvec{s}_r; \rvec{y}_r) \Jacob_{\imag \preS_c} \preS_r \big), \label{eq:indirect}
\end{align}
where we have used the convention for the complex derivative defined in \cite{brandwood:83} and where the Jacobians $\Jacob_{\real \preS_c} \preS_r$ and $\Jacob_{\imag \preS_c} \preS_r$ can be found using the definition in \req{eq:ext_model} and the results in \cite[Chapter 9]{magnus:88}. Similarly, expressions for $\Hess_{\preS_c} \I(\rvec{s}_c; \rvec{y}_c)$ or $\Hess_{\preS_c^*} \I(\rvec{s}_c; \rvec{y}_c)$ can also be obtained by successive application of the complex derivative definition and the chain rule.

In the following we present a simplified complex counterpart of the Hessian result in Theorem \ref{thm:Diagonal_Concav} for the real case, which, despite its simplicity, illustrates the particularities of the complex case.
\begin{thm}[Mutual information Hessian in the complex case]
Consider the complex signal model $\rvec{y}_c = \mb{Diag}\big(
\sqrt{\vecs{\lambda}_c} \big) \rvec{s}_c + \rvec{w}_c$, where
$\mb{Diag}\big( \sqrt{\vecs{\lambda}_c} \big) \in \PSD{\dim}$ is an
arbitrary deterministic diagonal matrix ($\vecs{\lambda}_c \in
\R^{\dim}$), the signaling $\rvec{s}_c \in \C^{\dim}$ is arbitrarily
distributed, and the noise vector $\rvec{w}_c \in \C^{\dim}$ follows
a white Gaussian proper distribution and is independent of the input
$\rvec{s}_c$. Then, the differential entropy of the output vector
$\rvec{y}_c$, $\Ent(\rvec{y})$, satisfies
\begin{gather}
 \Hess_{\vecs{\lambda}_c} \Ent(\rvec{y}_c) = - \Esp{ \CM_{\rvec{s}_c}(\rvec{y}) \circ \CM^*_{\rvec{s}_c}(\rvec{y}) + \overline{\CM}_{\rvec{s}_c}(\rvec{y}) \circ \overline{\CM}^*
 _{\rvec{s}_c}(\rvec{y}) },
\end{gather}
where we have defined
\begin{align}
\CM_{\rvec{s}_c}(\rvecr{y}) &= \CEsp{ \left(\rvec{s}_c - \CEsp{\rvec{s}_c}{\rvecr{y}}\right) \left( \rvec{s}_c - \CEsp{\rvec{s}_c}{\rvecr{y}} \right)^\H }{\rvecr{y}} \label{eq:CMSE_c} \\ \overline{\CM}_{\rvec{s}_c}(\rvecr{y}) &= \CEsp{ \left(\rvec{s}_c - \CEsp{\rvec{s}_c}{\rvecr{y}}\right) \left( \rvec{s}_c - \CEsp{\rvec{s}_c}{\rvecr{y}} \right)^\T }{\rvecr{y}}. \label{eq:CPMSE_c}
\end{align}
\end{thm}
\begin{IEEEproof}
The real extended model of $\rvec{y}_c = \mb{Diag}\big( \sqrt{\vecs{\lambda}_c} \big) \rvec{s}_c + \rvec{w}_c$ is readily obtained as
\begin{gather}
\rvec{y}_r = \mb{Diag}\big( \sqrt{\vecs{\lambda}_r} \big) \rvec{s}_r
+ \rvec{n}_r = \left[
\begin{array}{cc}
\mb{Diag}\big( \sqrt{\vecs{\lambda}_c} \big) & \vecs{0} \\ \vecs{0} & \mb{Diag}\big( \sqrt{\vecs{\lambda}_c} \big)
\end{array}
\right] \left[
\begin{array}{c}
\real \rvec{s}_c \\ \imag \rvec{s}_c
\end{array}
\right] + \frac{1}{\sqrt{2}} \rvec{w}_r,
\end{gather}
where now we have $\Esp[1]{\rvec{w}_r \rvec{w}_r^\T} =
\mat{I}_{2\dim}$.

Now, applying the chain rule for $\vecs{\lambda}_r^\T =
\big[\vecs{\lambda}_c^\T \: \vecs{\lambda}_c^\T \big]$ the elements
of the Hessian matrix read as
\begin{gather}
\left[ \Hess_{\vecs{\lambda}_c} \Ent(\rvec{y}_c) \right]_{ij} =
\frac{\partial^2 \Ent(\rvec{y}_c)}{\partial \lambda_{c,i} \partial
\lambda_{c,j}} = \frac{\partial^2 \Ent(\rvec{y}_r)}{\partial
\lambda_{r,i} \partial \lambda_{r,j}} + \frac{\partial^2
\Ent(\rvec{y}_r)}{\partial \lambda_{r,i+\dim} \partial
\lambda_{r,j}} + \frac{\partial^2 \Ent(\rvec{y}_r)}{\partial
\lambda_{r,i} \partial \lambda_{r,j+\dim}} + \frac{\partial^2
\Ent(\rvec{y}_r)}{\partial \lambda_{r,i+\dim} \partial \lambda_{r,j
+ \dim}}. \nonumber
\end{gather}
The four terms in the complex Hessian can be identified with the elements of the Hessian for the real case, which thanks to Theorem \ref{thm:Diagonal_Concav} can be written as
\begin{gather}
\frac{\partial^2 \Ent(\rvec{y}_r)}{\partial \lambda_{r,i} \partial
\lambda_{r,j}} = -2 \Esp{\left( \CEsp{ \left(\rveci{s}{r,i} -
\CEsp{\rveci{s}{r,i}}{\rvec{y}} \right)\left(\rveci{s}{r,j} -
\CEsp{\rveci{s}{r,j}}{\rvec{y}} \right)}{\rvec{y}} \right)^2}.
\end{gather}
Noting that $\rveci{s}{r,i} = \real \rveci{s}{c,i}$ and
$\rveci{s}{r,i + \dim} = \imag \rveci{s}{c,i}$, we can finally write
\begin{align}
\left[ \Hess_{\vecs{\lambda}_c} \Ent(\rvec{y}_c) \right]_{ij} = & -2
\Esp{\left( \CEsp{ \real\left(\rveci{s}{c,i} -
\CEsp{\rveci{s}{c,i}}{\rvec{y}} \right)\real\left(\rveci{s}{c,j} -
\CEsp{\rveci{s}{c,j}}{\rvec{y}} \right)}{\rvec{y}} \right)^2} \\ &
-2 \Esp{\left( \CEsp{ \real\left(\rveci{s}{c,i} -
\CEsp{\rveci{s}{c,i}}{\rvec{y}} \right)\imag\left(\rveci{s}{c,j} -
\CEsp{\rveci{s}{c,j}}{\rvec{y}} \right)}{\rvec{y}} \right)^2} \\ &
-2 \Esp{\left( \CEsp{ \imag\left(\rveci{s}{c,i} -
\CEsp{\rveci{s}{c,i}}{\rvec{y}} \right)\real\left(\rveci{s}{c,j} -
\CEsp{\rveci{s}{c,j}}{\rvec{y}} \right)}{\rvec{y}} \right)^2} \\ &
-2 \Esp{\left( \CEsp{ \imag\left(\rveci{s}{c,i} -
\CEsp{\rveci{s}{c,i}}{\rvec{y}} \right)\imag\left(\rveci{s}{c,j} -
\CEsp{\rveci{s}{c,j}}{\rvec{y}} \right)}{\rvec{y}} \right)^2}.
\end{align}
Now, with the definitions in \req{eq:CPMSE_c} and \req{eq:CMSE_c} and a slight amount of algebra, the result follows.
\end{IEEEproof}

It is important to highlight that, whereas in the real case the
conditional MMSE matrix $\CMSEr{s}{y}$ was enough to compute the
Hessian, in the complex case, in addition to the conditional MMSE
matrix (as defined in \req{eq:CMSE_c}) there is an extra matrix
$\overline{\CM}_{\rvec{s}_c}(\rvecr{y})$ defined as in
\req{eq:CPMSE_c}, and which is referred to as the conditional
pseudo-MMSE matrix.

\appendix

\subsection{The commutation $\Com{q, r}$, symmetrization $\Sym{q}$, duplication $\Dup{q}$, and reduction $\Rdx{q}$ matrices.}
\label{ap:special_matrices}

In this appendix we present four matrices that are very important
when calculating Hessian matrices. The definitions of the
commutation $\Com{q, r}$, symmetrization $\Sym{q}$, and duplication
$\Dup{q}$ matrices have been taken from \cite{magnus:88} and the
reduction matrix $\Rdx{q}$ has been defined by the authors of the
present work.

Given any matrix $\mat{A} \in \R^{q \times r}$, the two vectors $\vecop \mat{A}$ and $\vecop \mat{A}^{\! \T}$ contain the same elements but arranged in a different order. Consequently, there exists a unique permutation matrix $\Com{q, r} \in \R^{qr \times qr}$ independent of $\mat{A}$, which is called commutation matrix, that satisfies
\begin{gather}
\vecop \mat{A}^{\! \T} = \Com{q, r} \vecop \mat{A}, \quad \textrm{and} \quad
\Com{q, r}^\T = \Com{q, r}^{-1} = \Com{r, q}.
\end{gather}
It is easy to verify that the entries of the commutation matrix satisfy
\begin{gather} \label{eq:Comm_delta}
\left[ \Com{q, r} \right]_{i + (j-1)r, i' + (j'-1)q} = \delta_{i'j} \delta_{j'i}, \quad \{i',j\}\in[1, q], \quad \{i,j'\} \in [1, r].
\end{gather}

The main reason why we have introduced the commutation matrix is due to the property from which it obtains its name, as it enables us to commute the two matrices of a Kronecker product \cite[Theorem 3.9]{magnus:88},
\begin{gather} \label{eq:ComAkronB}
\Com{s, q} (\mat{A}\otimes\mat{B}) = (\mat{B} \otimes \mat{A}) \Com{t, r},
\end{gather}
where we have considered $\mat{A} \in \R^{q \times r}$ and $\mat{B} \in \R^{s \times t}$.

We also define $\Com{q} = \Com{q, q}$ for the case where the commutation matrix is square. An important property of the square matrix $\Com{q}$ is given in the following lemma.
\begin{lemap} \label{lem:Kron_comm}
Let $\mat{A} \in \R^{q\times r}$ and $\mat{B} \in \R^{q\times t}$. Then,
\begin{gather}
\begin{split}
\left[ \mat{A} \otimes \mat{B} \right]_{i + (j-1)q, k + (l-1)t} &= \mat{A}_{jl} \mat{B}_{ik}, \\ \left[ \Com{q} (\mat{A} \otimes \mat{B}) \right]_{i + (j-1)q, k + (l-1)t} &= \mat{A}_{il} \mat{B}_{jk},
\end{split}
\quad \quad \{i,j\} \in [1, q], \quad k \in [1, t], \quad l \in [1, r],
\end{gather}
\end{lemap}
\begin{IEEEproof}
The equality for the entries of the product $\mat{A} \otimes \mat{B}$ follows straightforwardly from the definition \cite[Section 4.2]{horn:91}. Concerning the entries of $\Com{q} (\mat{A} \otimes \mat{B})$, we have
\begin{align}
\left[ \Com{q} (\mat{A} \otimes \mat{B}) \right]_{i + (j-1)q, k + (l-1)t} &= \sum_{i' = 1}^{q} \sum_{j'=1}^{q} \left[ \Com{q} \right]_{i + (j-1)q, i' + (j'-1)q} \left[ \mat{A} \otimes \mat{B} \right]_{i' + (j'-1)q, k + (l-1)t} \\ &= \sum_{i' = 1}^{q} \sum_{j' = 1}^{q} \delta_{i'j} \delta_{j'i} \mat{A}_{j'l} \mat{B}_{i'k} \\ &= \mat{A}_{il} \mat{B}_{jk},
\end{align}
where we have used the expression for the elements of $\Com{q}$ in \req{eq:Comm_delta}.
\end{IEEEproof}

When calculating Jacobian and Hessian matrices, the form $\mat{I}_q + \Com{q}$ is usually encountered. Hence, we define the symmetrization  matrix $\Sym{q} = \tfrac{1}{2}(\mat{I}_q + \Com{q})$, which is singular and has the following properties
\begin{align}
\Sym{q} &= \Sym{q}^\T = \Sym{q}^{2} \\ \Sym{q} &= \Sym{q} \Com{q}  = \Com{q} \Sym{q}.
\end{align}
The name of the symmetrization matrix comes from the fact that given any square matrix $\mat{A} \in \R^{q\times q}$, then
\begin{gather} \label{eq:symm_vecA}
\Sym{q}\vecop \mat{A} = \tfrac{1}{2}\big(\vecop \mat{A} + \vecop\mat{A}^{\! \T}\big) = \tfrac{1}{2}\vecop \big( \mat{A} + \mat{A}^{\! \T}\big).
\end{gather}
The last important property of the symmetrization matrix is
\begin{gather} \label{eq:symm_AkronA}
\Sym{q}(\mat{A}\otimes\mat{A}) = (\mat{A} \otimes \mat{A})\Sym{q},
\end{gather}
which follows from the definition $\Sym{q} = \tfrac{1}{2}(\mat{I}_q + \Com{q})$ together with \req{eq:ComAkronB}.

Another important matrix related to the calculation of Jacobian and Hessian matrices, specially when symmetric matrices are involved, is the duplication matrix $\Dup{q}$. Given any symmetric matrix $\mat{R} \in \SymM{q}$, we denote by $\vechop\mat{R}$ the $\tfrac{1}{2}q(q+1)$-dimensional vector that is obtained from $\vecop\mat{R}$ by eliminating all the repeated elements that lie strictly above the diagonal of $\mat{R}$. Then, the duplication matrix $\Dup{q} \in \R^{q^2\times \tfrac{1}{2}q(q+1)}$ fulfills \cite[Section 3.8]{magnus:88}
\begin{gather} \label{eq:duplication}
\vecop \mat{R} = \Dup{q} \vechop \mat{R},
\end{gather}
for any $q$-dimensional symmetric matrix $\mat{R}$. The duplication matrix takes its name from the fact that it duplicates the entries of $\vechop \mat{R}$ which correspond to off-diagonal elements of $\mat{R}$ to produce the elements of $\vecop \mat{R}$.

Since $\Dup{q}$ has full column rank, it is possible to invert the transformation in \req{eq:duplication} to obtain
\begin{gather} \label{eq:inv_duplication}
\vechop \mat{R} = \Dup{q}^\pinv \vecop \mat{R} = \big( \Dup{q}^\T \Dup{q} \big)^{-1} \Dup{q}^\T \vecop \mat{R}.
\end{gather}
The most important properties of the duplication matrix are \cite[Theorem 3.12]{magnus:88}
\begin{gather} \label{eq:DupProp}
\Com{q} \Dup{q} = \Dup{q}, \quad \Sym{q}\Dup{q} = \Dup{q}, \quad \Dup{q} \Dup{q}^\pinv = \Sym{q}, \quad \Dup{q}^\pinv \Sym{q} = \Dup{q}^\pinv.
\end{gather}

The last one of the matrices introduced in this appendix is the reduction matrix $\Rdx{q} \in \R^{q^2\times q}$. The entries of the reduction matrix are defined as
\begin{gather} \label{eq:redux}
[\Rdx{q}]_{i + (j-1)q, k} = \delta_{ij}\delta_{ik} = \delta_{ijk}, \quad \{ i, j, k \} \in [1, q]
\end{gather}
from which it is easy to verify that the reduction matrix fulfills
\begin{gather} \label{eq:rdx_prop}
\Com{q} \Rdx{q} = \Rdx{q}, \quad \Sym{q} \Rdx{q} = \Rdx{q}.
\end{gather}
However, the most important property of the reduction matrix is that it can be used to reduce the Kronecker product of two matrices to their Schur product as it is detailed in the next lemma.
\begin{lemap} \label{lem:redux}
Let $\mat{A} \in \R^{q\times r}$, $\mat{B} \in \R^{q\times r}$. Then,
\begin{gather}
\Rdx{q}^\T (\mat{A} \otimes \mat{B}) \Rdx{r} = \Rdx{q}^\T (\mat{B} \otimes \mat{A}) \Rdx{r} = \mat{A} \circ \mat{B}.
\end{gather}
\end{lemap}
\begin{IEEEproof}
From the expression for the elements of the Kronecker product in Lemma \ref{lem:Kron_comm} and the expression for the elements of the reduction matrix we have that, for any $i \in [1, q]$ and $j \in [1, r]$,
\begin{align}
\big[ \Rdx{q}^\T (\mat{A} \otimes \mat{B}) \Rdx{r} \big]_{i, j} &= \sum_{k,l}^q \sum_{k',l'}^r [\Rdx{q}]_{k + (l-1)q, i} [\mat{A} \otimes \mat{B}]_{k + (l-1)q, k' + (l'-1)r} [\Rdx{r}]_{k' + (l'-1)r, j} \\ &= \sum_{k,l}^q \sum_{k',l'}^r \mat{A}_{ll'} \mat{B}_{kk'} \delta_{kli} \delta_{k'l'j} \\ &= \mat{A}_{ij} \mat{B}_{ij},
\end{align}
from which the result in the lemma follows.
\end{IEEEproof}

Finally, to conclude this appendix, we present two basic lemmas concerning the Kronecker product and the $\vecop$ operator.%, followed by another lemma which also involves the commutation matrix.
\begin{lemap} \label{lem:ACkronBD}
Let $\mat{A}$, $\mat{B}$, $\mat{F}$, and $\mat{T}$ be four matrices such that the products $\mat{A}\mat{B}$ and $\mat{F}\mat{T}$ are defined. Then, $
(\mat{A} \otimes \mat{F})(\mat{B} \otimes \mat{T}) = \mat{A}\mat{B} \otimes \mat{F}\mat{T}$.
\end{lemap}
\begin{IEEEproof}
See \cite[Chapter 2]{magnus:88}.
\end{IEEEproof}
\begin{lemap} \label{lem:vecABC}
Let $\mat{A}$, $\mat{T}$, and $\mat{B}$ be three matrices such that the product $\mat{A} \mat{T} \mat{B}$ is defined. Then,
\begin{gather}
\vecop (\mat{A} \mat{T} \mat{B}) = \big( \mat{B}^\T \otimes \mat{A} \big) \vecop \mat{T}.
\end{gather}
\end{lemap}
\begin{IEEEproof}
See \cite[Theorem 2.2]{magnus:88} or \cite[Proposition 7.1.6]{bernstein:05}.
\end{IEEEproof}

\subsection{Conventions used for Jacobian and Hessian matrices}
\label{ap:diff_conventions}

In this work we make extensive use of differentiation of matrix
functions $\mats{\Psi}$ with respect to a matrix argument $\mat{T}$.
From the many possibilities of displaying the partial derivatives
$\partial^r \mats{\Psi}_{st}/\partial \mat{T}_{ij} \cdots \partial
\mat{T}_{kl}$, we will stick to the ``good notation'' introduced by
Magnus and Neudecker in \cite[Section 9.4]{magnus:88} which is
briefly reproduced next for the sake of completeness.
\begin{dfnap}
Let $\mats{\Psi}$ be a differentiable $q \times t$ real matrix function of an $r \times s$ matrix of real variables $\mat{T}$. The Jacobian matrix of $\mats{\Psi}$ at $\mat{T} = \mat{T}_0$ is the $qt \times rs$ matrix
\begin{gather} \label{eq:Jacobian}
\Jacob_{\mat{T}} \mats{\Psi}(\mat{T}_0) = \left.\frac{\partial \, \vecop \, \mats{\Psi}(\mat{T})}{\partial (\vecop \mat{T})^\T}\right|_{\mat{T} = \mat{T}_0}.
\end{gather}
\end{dfnap}

\begin{remap}
To properly deal with the case where $\mats{\Psi}$ is a symmetric
matrix, the $\vecop$ operator in the numerator in \req{eq:Jacobian}
has to be replaced by a $\vechop$ operator to avoid obtaining
repeated elements. Similarly, $\vechop$ has to replace $\vecop$ in
the denominator in \req{eq:Jacobian} for the case where $\mat{T}$ is
a symmetric matrix. For practical purposes, it is enough to
calculate the Jacobian without taking into account any symmetry
properties and then add a left factor $\Dup{q}^\pinv$ to the
obtained Jacobian when $\mats{\Psi}$ is symmetric and/or a right factor
$\Dup{r}$ when $\mat{T}$ is symmetric. This proceeding will become
more clear in the examples given below.
\end{remap}

\begin{dfnap}
Let $\mats{\Psi}$ be a twice differentiable $q \times t$ real matrix function of an $r \times s$ matrix of real variables $\mat{T}$. The Hessian matrix of $\mats{\Psi}$ at $\mat{T} = \mat{T}_0$ is the $qtrs \times rs$ matrix
\begin{gather} \label{eq:Hessian}
\Hess_{\mat{T}} \mats{\Psi}(\mat{T}_0) = \left. \Jacob_{\mat{T}}\big(\Jacob_{\mat{T}}^\T \mats{\Psi}(\mat{T})\big) \right|_{\mat{T} = \mat{T}_0} = \left. \frac{\partial}{\partial (\vecop \mat{T})^\T} \vecop\left( \frac{\partial \, \vecop \, \mats{\Psi}(\mat{T})}{\partial (\vecop \mat{T})^\T} \right)^{\!\!\T} \right|_{\mat{T} = \mat{T}_0}.
\end{gather}
One can verify that the obtained Hessian matrix for the matrix function $\mats{\Psi}$ is the stacking of the $qt$ Hessian matrices corresponding to each individual element of vector $\vecop \mats{\Psi}$.
\end{dfnap}
\begin{remap}
Similarly to the Jacobian case, when $\mats{\Psi}$ or $\mat{T}$ are symmetric matrices, the $\vechop$ operator has to replace the $\vecop$ operator where appropriate in \req{eq:Hessian}.
\end{remap}

One of the major advantages of using the notation of \cite{magnus:88} is that a simple chain rule can be applied for both the Jacobian and Hessian matrices, as detailed in the following lemma.
\begin{lemap}[\mbox{\cite[Theorems 5.8 and 6.9]{magnus:88}}] \label{lem:ChainRules}
Let $\mats{\Upsilon}$ be a twice differentiable $u\times v$ real matrix function of a $q\times t$ real matrix argument. Let $\mats{\Psi}$ be a twice differentiable $q \times t$ real matrix function of an $r \times s$ matrix of real variables $\mat{T}$. Define $\mats{\Omega}(\mat{T}) = \mats{\Upsilon}(\mats{\Psi}(\mat{T}))$. The Jacobian and Hessian matrices of $\mats{\Omega}(\mat{T})$ at $\mat{T} = \mat{T}_0$ are:
\begin{align} \label{eq:Jacob_chain_rule}
\Jacob_{\mat{T}} \mats{\Omega}(\mat{T}_0) &= (\Jacob_{\mats{\Psi}} \mats{\Upsilon}(\mats{\Psi}_0)) (\Jacob_{\mat{T}} \mats{\Psi}(\mat{T}_0)) \\ \Hess_{\mat{T}} \mats{\Omega}(\mat{T}_0) &= (\mat{I}_{uv} \otimes \Jacob_{\mat{T}} \mats{\Psi}(\mat{T}_0))^\T \Hess_{\mats{\Psi}} \mats{\Upsilon}(\mats{\Psi}_0) (\Jacob_{\mat{T}} \mats{\Psi}(\mat{T}_0)) + (\Jacob_{\mats{\Psi}} \mats{\Upsilon}(\mats{\Psi}_0) \otimes \mat{I}_{rs}) \Hess_{\mat{T}} \mats{\Psi}(\mat{T}_0), \label{eq:Hess_chain_rule}
\end{align}
where $\mats{\Psi}_0 = \mats{\Psi}(\mat{T}_0)$.
\end{lemap}

The notation introduced above unifies the study of scalar $(q=t=1)$, vector $(t=1)$, and matrix functions of scalar $(r=s=1)$, vector $(s=1)$, or matrix arguments into the study of vector functions of vector arguments through the use of the $\vecop$ and $\vechop$ operators. However, the idea of arranging the partial derivatives of a scalar function of a matrix argument $\psi(\mat{T})$ into a matrix rather than a vector is quite appealing and sometimes useful, so we will also make use of the notation described next.
\begin{dfnap}
Let $\psi$ be differentiable scalar function of an $r \times s$ matrix of real variables $\mat{T}$. The gradient of $\psi$ at $\mat{T} = \mat{T}_0$ is the $r \times s$ matrix
\begin{gather} \label{eq:Gradient}
\nabla_{\mat{T}} \psi(\mat{T}_0) = \left. \frac{\partial \psi}{\partial \mat{T}}\right|_{\mat{T} = \mat{T}_0}.
\end{gather}
It is easy to verify that $\Jacob_{\mat{T}} \psi(\mat{T}_0) = \vecop^\T \nabla_{\mat{T}} \psi(\mat{T}_0)$.
\end{dfnap}

We now give expressions for the most common Jacobian and Hessian matrices encountered during our developments.
\begin{lemap} \label{lem:Jacob_chain_examples}
Consider $\mat{A} \in \R^{q\times r}$, $\mat{T} \in \R^{r\times s}$, $\mat{B} \in \R^{s\times t}$, $\mat{R} \in \PSD{s}$, and $\vec{f} \in \R^{r\times 1}$, such that $\vec{f}$ is a function of $\mat{T}$. Then, the following holds:
\begin{enumerate}
\item If $\mats{\Psi} = \mat{A} \mat{T} \mat{B}$, we have $\Jacob_{\mat{T}} \mats{\Psi} = \big( \mat{B}^\T \otimes \mat{A} \big)$. If, in addition, $\mat{B}$ is a function of $\mat{T}$, then we have $\Jacob_{\mat{T}} \mats{\Psi} = \big( \mat{B}^\T \otimes \mat{A} \big) + (\mat{I}_t \otimes \mat{A}\mat{T}) \Jacob_{\mat{T}} \mat{B}$. \label{lem:id1}

\item If $\mats{\Psi} = \mat{A} \vec{f}$, we have $\Jacob_{\mat{T}} \mats{\Psi} = \mat{A} \Jacob_{\mat{T}} \vec{f}$. \label{lem:id1.5}

\item If $\mats{\Psi} = \mat{A} \mat{T} \mat{A}^{\! \T}$, with $\mat{T}$ being a symmetric matrix, we have $\Jacob_{\mat{T}} \mats{\Psi} = \Dup{q}^\pinv \big( \mat{A} \otimes \mat{A} \big) \Dup{r}$. \label{lem:id2}

\item If $\mats{\Psi} = \mat{B}^\T \mat{T}^\T \mat{A}^{\! \T}$, we have $\Jacob_{\mat{T}} \mats{\Psi} = \big( \mat{A} \otimes \mat{B}^\T \big) \Com{r,s}$. \label{lem:id3}

\item If $\mats{\Psi} = (\mat{T} \otimes \mat{A})$, we have
$\Jacob_{\mat{T}} \mats{\Psi} = (\mat{I}_s \otimes \Com{t, r}
\otimes \mat{I}_q) (\mat{I}_{rs} \otimes \vecop \mat{A})$ and if
$\mats{\Psi} = (\mat{A} \otimes \mat{T})$, we have $\Jacob_{\mat{T}}
\mats{\Psi} = (\mat{I}_t \otimes \Com{s, q} \otimes \mat{I}_r)
(\vecop \mat{A} \otimes \mat{I}_{rs})$, where in this case we have assumed that $\mat{A} \in \R^{q\times t}$. \label{lem:id4}

\item If $\mats{\Psi} = \mat{T}^{-1}$, we have $\Jacob_{\mat{T}} \mats{\Psi} = - \big( \mat{T}^\T \otimes \mat{T}\big)^{-1}$, where $\mat{T}$ is a square invertible matrix. \label{lem:id4.5}

\item If $\mats{\Psi} = \mat{A} \mat{T} \mat{R} \mat{T}^\T \mat{A}^{\! \T}$, we have $\Jacob_{\mat{T}} \mats{\Psi} = 2 \Dup{q}^\pinv ( \mat{A} \mat{T} \mat{R} \otimes \mat{A} )$. \label{lem:id5}

\item If $\mats{\Psi} = \mat{A} \mat{T} \mat{R} \mat{T}^\T \mat{A}^{\! \T}$, we have $\Hess_{\mat{T}} \mats{\Psi} = 2 (\Dup{q}^\pinv \otimes \mat{I}_{rs})(\mat{I}_q \otimes \Com{q,s} \otimes \mat{I}_r)\big( \mat{A}\otimes \mat{R} \otimes \vecop \mat{A}^{\!\T} \big) \Com{r, s}$. \label{lem:id6}
\end{enumerate}
\end{lemap}
\begin{IEEEproof}
The identities from \ref{lem:id1}) to \ref{lem:id5}) can be found in \cite[Chapter 9]{magnus:88}. Concerning identity \ref{lem:id6}), it can be calculated through the definition of the Hessian as
\begin{gather}
\Hess_{\mat{T}} \big( \mat{A} \mat{T} \mat{R} \mat{T}^\T \mat{A}^{\! \T}\big) = 2 \Jacob_{\mat{T}} \big( \Dup{q}^\pinv ( \mat{A} \mat{T} \mat{R} \otimes \mat{A} ) \big)^\T = 2 \Jacob_{\mat{T}} \big(\big( \mat{R} \mat{T}^\T \mat{A}^{\! \T} \otimes \mat{A}^{\! \T} \big) \Dup{q}^{\pinv \T} \big).
\end{gather}
Now, we define $\mats{\Upsilon} = \mat{R} \mat{T}^\T \mat{A}^{\! \T}$ and $\mats{\Omega} = \mats{\Upsilon} \otimes \mat{A}^{\! \T}$ and apply the chain rule twice to obtain
\begin{gather}
\Jacob_{\mat{T}} \big(\big( \mat{R} \mat{T}^\T \mat{A}^{\! \T} \otimes \mat{A}^{\! \T} \big) \Dup{q}^{\pinv \T} \big) = \Jacob_{\mats{\Omega}} \big(\mats{\Omega} \Dup{q}^{\pinv \T}\big) \Jacob_{\mats{\Upsilon}} \mats{\Omega} \Jacob_{\mat{T}} \mats{\Upsilon},
\end{gather}
from which the result in \ref{lem:id6}) follows by application of identities \ref{lem:id1}), \ref{lem:id4}), and \ref{lem:id3}) and from Lemma \ref{lem:ACkronBD}.
\end{IEEEproof}

\subsection{Differential properties of the quantities $\pdfvec{y}$, $\pdfvec{y|s}$, and $\CEsp{\rvec{s}}{\rvecr{y}}$.}
\label{ap:partials_Py}

In this appendix we present a number of lemmas which are used in the proofs of Theorems \ref{thm:DcJ} and \ref{thm:DbE} in Appendices \ref{ap:DcJ_proof} and \ref{ap:DbE_proof}.

In the proofs of the following lemmas we interchange the order of differentiation and expectation, which can be justified following similar steps as in \cite[Appendix B]{palomar:06}.

\begin{lemap} \label{lem:DCP}
Let $\rvec{y} = \rvec{x} + \preN\rvec{n}$, where $\rvec{x}$ is arbitrarily distributed and where $\rvec{n}$ is a zero-mean Gaussian random variable with covariance matrix $\Cov{n}$ and independent of $\rvec{x}$. Then, the probability density function $\pdfvec{y}$ satisfies
\begin{gather} \label{eq:heat}
\nabla_{\preN} \pdfvec{y} = \Hess_{\rvecr{y}} \pdfvec{y} \preN \Cov{n} .
\end{gather}
\end{lemap}
\begin{IEEEproof}
%= \Esp{ \pdf{\rvec{z}|\rvec{x}}{\rvecr{z}|\rvec{x}}\left( \Rcn^{-1} (\rvecr{z} - \rvec{x})(\rvecr{z} - \rvec{x})^\T \Rcn^{-1} - \Rcn^{-1}\right)} \preN \Rn \\ &=
First, we recall that $\pdfvec{y} = \Esp{\pdf{\rvec{y}|\rvec{x}}{\rvecr{y}|\rvec{x}}}$. Thus the matrix gradient of the density $\pdfvec{y}$ with respect to $\preN$ is $\nabla_{\preN} \pdfvec{y} = \Esp{\nabla_{\preN} \pdf{\rvec{y}|\rvec{x}}{\rvecr{y}|\rvec{x}}}$. The computation of the inner the gradient $\nabla_{\preN} \pdf{\rvec{y}|\rvec{x}}{\rvecr{y}|\rvec{x}}$ can be performed by replacing $\preS \rvecr{s}$ by $\rvecr{x}$ in \req{eq:pdfYcondS} together with
\begin{align}
\nabla_\preN \: \vec{a}^\T \big( \preN \Cov{n} \preN^\T \big)^{-1} \vec{a} &= - 2 \big( \preN \Cov{n} \preN^\T \big)^{-1} \vec{aa}^\T \big( \preN \Cov{n} \preN^\T \big)^{-1} \preN \Cov{n} \label{eq:dTraceInv} \\ \nabla_\preN \det\big( \preN \Cov{n} \preN^\T \big) &= 2 \det \big( \preN \Cov{n} \preN^\T \big) \big( \preN \Cov{n} \preN^\T \big)^{-1} \preN \Cov{n}, \label{eq:gdet}
\end{align}
where $\vec{a}$ is a fixed vector of the appropriate dimension and where we have used \cite[p. 178, Exercise 9.9.3]{magnus:88} and the chain rule in Lemma \ref{lem:ChainRules} in \req{eq:dTraceInv} and, \cite[p. 180, Exercise 9.10.2]{magnus:88} in \req{eq:gdet}. With these expressions at hand and recalling that $\Cov{z} = \preN \Cov{n} \preN^\T$, the expression for the gradient $\nabla_{\preN} \pdfvec{y}$ is equal to
\begin{gather} \label{eq:gradCPz}
\nabla_{\preN} \pdfvec{y} = \Esp{ \pdf{\rvec{y}|\rvec{x}}{\rvecr{y}|\rvec{x}}\left( \Cov{z}^{-1} (\rvecr{y} - \rvec{x})(\rvecr{y} - \rvec{x})^\T \Cov{z}^{-1} - \Cov{z}^{-1}\right)} \preN \Cov{n}.
\end{gather}

To complete the proof, we need to calculate the Hessian matrix, $\Hess_{\rvecr{y}} \pdfvec{y}$. First consider the following two Jacobians
\begin{align} \label{eq:DxTx}
\Jacob_{\rvecr{y}} (\rvecr{y} - \rvecr{x})^\T
\Cov{z}^{-1} (\rvecr{y} - \rvecr{x}) &= 2 (\rvecr{y} - \rvecr{x})^\T
\Cov{z}^{-1} \\ \Jacob_{\rvecr{y}} \Cov{z}^{-1} (\rvecr{y} - \rvecr{x}) &=
\Cov{z}^{-1} \label{eq:DTx},
\end{align}
which follow directly from \cite[Section 9.9, Table 3]{magnus:88} and \cite[Section 9.12]{magnus:88}. Now, from \req{eq:DxTx}, we can first obtain the Jacobian row vector $\Jacob_{\rvecr{y}} \pdfvec{y}$ as
\begin{gather} \label{eq:DyPy}
\Jacob_{\rvecr{y}} \pdfvec{y} = - \Esp{\pdf{\rvec{y}|\rvec{x}}{\rvecr{y}|\rvec{x}} (\rvecr{y} - \rvec{x})^\T \Cov{z}^{-1}}.
\end{gather}
Recalling the expression in \req{eq:DTx} and that $\Hess_{\rvecr{y}} \pdfvec{y} = \Jacob_{\rvecr{y}} \left( \Jacob_{\rvecr{y}}^\T \pdfvec{y} \right)$ the Hessian matrix becomes
\begin{gather} \label{eq:HessPz}
\Hess_{\rvecr{y}} \pdfvec{y} = \Esp{ \pdf{\rvec{y}|\rvec{x}}{\rvecr{y}|\rvec{x}}\left( \Cov{z}^{-1} (\rvecr{y} - \rvec{x})(\rvecr{y} - \rvec{x})^\T \Cov{z}^{-1} - \Cov{z}^{-1}\right)}.
\end{gather}
By simple inspection from \req{eq:gradCPz} and \req{eq:HessPz} the result in \req{eq:heat} follows.
\end{IEEEproof}

\begin{lemap} \label{lem:DBP}
Let $\rvec{y} = \preS\rvec{s} + \rvec{z}$, where $\rvec{s}$ is arbitrarily distributed and where $\rvec{z}$ is a zero-mean Gaussian random variable with covariance matrix $\Cov{z}$  and independent of $\rvec{s}$. Then, $\pdfvec{y}$ satisfies
\begin{gather} \label{eq:heatB}
\nabla_{\preS} \pdfvec{y} = - \Esp{\Jacob_{\rvecr{y}}^\T
\pdf{\rvec{y}|\rvec{s}}{\rvecr{y}|\rvec{s}}\rvec{s}^\T}.
\end{gather}
\end{lemap}
\begin{IEEEproof}
First we write
\begin{gather} \label{eq:JyPys}
\Jacob_{\rvecr{y}} \pdf{\rvec{y}|\rvec{s}}{\rvecr{y}|\rvecr{s}} = - \pdf{\rvec{y}|\rvec{s}}{\rvecr{y}|\rvecr{s}} (\rvecr{y} - \preS\rvecr{s})^\T
\Cov{z}^{-1},
\end{gather}
where we have used \req{eq:DxTx}. Now, we simply need to notice that
\begin{gather} \label{eq:DBPys}
\nabla_\preS \pdf{\rvec{y}|\rvec{s}}{\rvecr{y}|\rvecr{s}} = \pdf{\rvec{y}|\rvec{s}}{\rvecr{y}|\rvecr{s}} \Cov{z}^{-1}(\rvecr{y} - \preS\rvecr{s})\rvecr{s}^\T = -\Jacob_{\rvecr{y}}^\T \pdf{\rvec{y}|\rvec{s}}{\rvecr{y}|\rvecr{s}}\rvecr{s}^\T,
\end{gather}
where we have used $\nabla_\preS (\rvecr{y} - \preS\rvecr{s})^\T
\Cov{z}^{-1} (\rvecr{y} - \preS\rvecr{s}) = - 2
\Cov{z}^{-1}(\rvecr{y} - \preS\rvecr{s})\rvecr{s}^\T$, which follows
from \cite[Section 9.9, Table 4]{magnus:88}. Recalling that
$\nabla_\preS \pdfvec{y} = \Esp{\nabla_\preS
\pdf{\rvec{y}|\rvec{s}}{\rvecr{y}|\rvec{s}}}$ the result follows.
%The existence of the right hand side of \req{eq:heatB} is proved in \cite[App. B]{palomar:06}.
\end{IEEEproof}

\begin{lemap} \label{lem:DyCEsp}
Let $\rvec{y} = \preS\rvec{s} + \rvec{z}$, where $\rvec{s}$ is arbitrarily distributed and where $\rvec{z}$ is a zero-mean Gaussian random variable with covariance matrix $\Cov{z}$ and independent of $\rvec{s}$. Then,
\begin{gather}
\Jacob_{\rvecr{y}} \CEsp{\rvec{s}}{\rvecr{y}} = \CMSEr{s}{y} \preS \Cov{z}^{-1}.
\end{gather}
\end{lemap}
\begin{IEEEproof}
\begin{align}
\Jacob_{\rvecr{y}} \CEsp{\rvec{s}}{\rvecr{y}} &= \Jacob_{\rvecr{y}} \Esp{\rvec{s} \frac{\pdf{\rvec{y}|\rvec{s}}{\rvecr{y}|\rvec{s}}}{\pdfvec{y}}} \\ &= \Esp{\rvec{s} \frac{\pdfvec{y} \Jacob_{\rvecr{y}} \pdf{\rvec{y}|\rvec{s}}{\rvecr{y}|\rvec{s}} - \pdf{\rvec{y}|\rvec{s}}{\rvecr{y}|\rvec{s}} \Jacob_{\rvecr{y}} \pdfvec{y} }{\pdfvec{y}^2}} \\ &= \Esp{\rvec{s} \frac{ - \pdf{\rvec{y}|\rvec{s}}{\rvecr{y}|\rvec{s}} (\rvecr{y} - \preS\rvec{s})^\T \Cov{z}^{-1} + \pdf{\rvec{y}|\rvec{s}}{\rvecr{y}|\rvec{s}} (\rvecr{y} - \preS\CEsp{\rvec{s}}{\rvecr{y}})^\T \Cov{z}^{-1}  }{\pdfvec{y}}} \\ &= \big( \CEsp[1]{\rvec{s}\rvec{s}^\T}{\rvecr{y}} - \CEsp{\rvec{s}}{\rvecr{y}} \CEsp[1]{\rvec{s}^\T}{\rvecr{y}} \big)\preS^\T \Cov{z}^{-1}.
\end{align}
Now, from the definition in \req{eq:CondE} the result in the lemma follows. Note that we have used the expression in \req{eq:JyPys} for $\Jacob_{\rvecr{y}} \pdf{\rvec{y}|\rvec{s}}{\rvecr{y}|\rvec{s}}$ and that from \req{eq:DyPy} we can write
\begin{gather}
\Jacob_{\rvecr{y}} \pdfvec{y} = -
\Esp{\pdf{\rvec{y}|\rvec{x}}{\rvecr{y}|\rvec{x}} (\rvecr{y} -
\rvec{x})^\T \Cov{z}^{-1}} = - \pdfvec{y} (\rvecr{y} - \preS
\CEsp{\rvec{s}}{\rvecr{y}})^\T \Cov{z}^{-1}.
\end{gather}
\end{IEEEproof}

\begin{lemap} \label{lem:DlogP}
Let $\rvec{y} = \preS\rvec{s} + \rvec{z}$, where $\rvec{s}$ is arbitrarily distributed and where $\rvec{z}$ is a zero-mean Gaussian random variable with covariance matrix $\Cov{z}$ and independent of $\rvec{s}$. Then, the Jacobian and Hessian of $\log \pdfvec{y}$ satisfy
\begin{align}
\Jacob_{\rvecr{y}} \log \pdfvec{y} &= \left(\CEsp{\rvec{x}}{\rvecr{y}} - \rvecr{y}\right)^\T \Cov{z}^{-1} \\ \Hess_{\rvecr{y}} \log \pdfvec{y} &= \Cov{z}^{-1} \CMSEr{x}{y} \Cov{z}^{-1} - \Cov{z}^{-1}. \label{eq:HesslogPz}
\end{align}
\end{lemap}
\begin{IEEEproof}
Recalling the expression in \req{eq:DyPy} we can write
\begin{align}
\Jacob_{\rvecr{y}} \log \pdfvec{y} &= \frac{1}{\pdfvec{y}} \Jacob_{\rvecr{y}} \pdfvec{y} \\ &= - \frac{1}{\pdfvec{y}} \Esp[1]{\pdf{\rvec{y}|\rvec{x}}{\rvecr{y}|\rvec{x}} (\rvecr{y} - \rvec{x})^\T \Cov{z}^{-1}} \\ &= (\CEsp{\rvec{x}}{\rvecr{y}} - \rvecr{y})^\T \Cov{z}^{-1}.
\end{align}
From the Jacobian expression, the Hessian can be computed as
\begin{align}
\Hess_{\rvecr{y}} \log \pdfvec{y} &= \Jacob_{\rvecr{y}} \Cov{z}^{-1} (\CEsp{\rvec{x}}{\rvecr{y}} - \rvecr{y}) \\ &= \Cov{z}^{-1}(\preS \Jacob_{\rvecr{y}} \CEsp{\rvec{s}}{\rvecr{y}} - \mat{I}_\dim) \\ &= \Cov{z}^{-1}(\preS \CMSEr{s}{y}\preS^\T \Cov{z}^{-1} - \mat{I}_\dim),
\end{align}
where the expression for $\Jacob_{\rvecr{y}} \CEsp{\rvec{s}}{\rvecr{y}}$ follows from Lemma \ref{lem:DyCEsp}.
\end{IEEEproof}

\begin{lemap} \label{lem:DbCEsp}
Let $\rvec{y} = \preS\rvec{s} + \rvec{z}$, where $\rvec{s}$ is arbitrarily distributed (with $i$-th element denoted by $\rveci{s}{i}$) and where $\rvec{z}$ is a zero-mean Gaussian random variable with covariance matrix $\Cov{z}$ and independent of $\rvec{s}$. Then,
\begin{gather}
\nabla_\preS \CEsp{\rveci{s}{i}}{\rvecr{y}} = \frac{1}{\pdfvec{y}} \left( \CEsp{\rveci{s}{i}}{\rvecr{y}}\Esp[1]{\Jacob_{\rvecr{y}}^\T \pdf{\rvec{y}|\rvec{s}}{\rvecr{y}|\rvec{s}}\rvec{s}^\T} - \Esp[1]{\rveci{s}{i}\Jacob_{\rvecr{y}}^\T \pdf{\rvec{y}|\rvec{s}}{\rvecr{y}|\rvec{s}}\rvec{s}^\T} \right). \nonumber
\end{gather}
\end{lemap}
\begin{IEEEproof}
The proof follows from this chain of equalities
\begin{align}
\nabla_\preS \CEsp{\rveci{s}{i}}{\rvecr{y}} &= \nabla_\preS \Esp{\rveci{s}{i} \frac{ \pdf{\rvec{y}|\rvec{s}}{\rvecr{y}|\rvec{s}}}{\pdfvec{y}}} \nonumber \\ &= \Esp{\rveci{s}{i} \frac{ \pdfvec{y} \nabla_\preS \pdf{\rvec{y}|\rvec{s}}{\rvecr{y}|\rvec{s}} - \pdf{\rvec{y}|\rvec{s}}{\rvecr{y}|\rvec{s}} \nabla_\preS \pdfvec{y} }{\pdfvec{y}^2}} \nonumber \\ &= \Esp{\rveci{s}{i} \left( - \frac{\Jacob_{\rvecr{y}}^\T \pdf{\rvec{y}|\rvec{s}}{\rvecr{y}|\rvec{s}}\rvec{s}^\T}{\pdfvec{y}} + \frac{\pdf{\rvec{y}|\rvec{s}}{\rvecr{y}|\rvec{s}} \Esp{\Jacob_{\rvecr{y}}^\T \pdf{\rvec{y}|\rvec{s}}{\rvecr{y}|\rvec{s}}\rvec{s}^\T}}{\pdfvec{y}^2} \right)} \label{eq:DbCEsp} \\ &= \frac{1}{\pdfvec{y}}\left( - \Esp[1]{\rveci{s}{i}\Jacob_{\rvecr{y}}^\T \pdf{\rvec{y}|\rvec{s}}{\rvecr{y}|\rvec{s}}\rvec{s}^\T} + \Esp{\rveci{s}{i}\frac{\pdf{\rvec{y}|\rvec{s}}{\rvecr{y}|\rvec{s}}}{\pdfvec{y}}} \Esp[1]{\Jacob_{\rvecr{y}}^\T \pdf{\rvec{y}|\rvec{s}}{\rvecr{y}|\rvec{s}}\rvec{s}^\T} \right), \nonumber
\end{align}
where \req{eq:DbCEsp} follows from Lemma \ref{lem:DBP} and from \req{eq:DBPys}.
\end{IEEEproof}

\subsection{Proof of Theorem \ref{thm:DcJ}}
\label{ap:DcJ_proof}

Let us begin by considering the expression for the entries of the Jacobian of the vector $\vecop\J{y}$, which is $[\Jacob_{\preN} \vecop \J{y}]_{i + (j-1)\dim, k + (l-1)\dim} = \Jacob_{\preN_{kl}} \left[\J{y}\right]_{ij}$, where throughout this proof $\{i, j, k\} \in [1, \dim]$ and $l \in [1, \dim']$. From \req{eq:FIM_Reg} and \req{eq:CondJ} we have that the entries of the Fisher information matrix are given by
\begin{gather}
\left[\J{y}\right]_{ij} = \Esp[1]{\left[\CJ{y}\right]_{ij}} = - \int \pdfvec{y} \dijlogPz{y}{i}{j} \d \rvecr{y}.
\end{gather}
We now differentiate the expression above with respect to the
entries of the matrix $\preN$ and we get
\begin{align}
\Jacob_{\preN_{kl}} \left[\J{y}\right]_{ij} &= - \frac{\partial }{\partial \preN_{kl}} \int \pdfvec{y} \dijlogPz{y}{i}{j} \d \rvecr{y} \\ &= - \int \frac{\partial \pdfvec{y}}{\partial \preN_{kl}} \dijlogPz{y}{i}{j} \d \rvecr{y} - \int \pdfvec{y} \frac{\partial^2}{\partial \rvecri{y}{i} \partial \rvecri{y}{j}}\left( \frac{1}{\pdfvec{y}} \frac{\partial \pdfvec{y}}{\partial \preN_{kl}} \right) \d \rvecr{y},
\end{align}
where the interchange of the order of integration and differentiation can be justified from the Dominated Convergence Theorem following similar steps as in \cite[Appendix B]{palomar:06}. Now, using Lemma \ref{lem:DCP} we transform the partial derivatives with respect to $\preN$ into derivatives with respect to the entries of vector $\rvecr{y}$, yielding
\begin{multline}
\Jacob_{\preN_{kl}} \left[\J{y}\right]_{ij} = - \int [\Hess_{\rvecr{y}} \pdfvec{y} \preN \Cov{n}]_{kl} \dijlogPz{y}{i}{j} \d \rvecr{y} \\
- \int \pdfvec{y} \frac{\partial^2}{\partial \rvecri{y}{i} \partial \rvecri{y}{j}}\left( \frac{1}{\pdfvec{y}} [\Hess_{\rvecr{y}} \pdfvec{y} \preN \Cov{n}]_{kl} \right) \d \rvecr{y}.
\end{multline}
Expressing the elements of $[\Hess_{\rvecr{y}} \pdfvec{y} \preN \Cov{n}]_{kl}$ as the sum of the product of the elements of $\Hess_{\rvecr{y}} \pdfvec{y}$ and $\preN \Cov{n}$ we get
\begin{multline} \label{eq:intermig}
\Jacob_{\preN_{kl}} \left[\J{y}\right]_{ij} = - \sum_{r=1}^\dim [\preN \Cov{n}]_{rl} \left( \int \dijPz{y}{k}{r} \dijlogPz{y}{i}{j} \d \rvecr{y} \right. \\
\left. + \int \pdfvec{y} \frac{\partial^2}{\partial \rvecri{y}{i} \partial \rvecri{y}{j}}\left( \frac{1}{\pdfvec{y}} \dijPz{y}{k}{r} \right) \d \rvecr{y} \right).
\end{multline}
We can now combine the integral identities \req{eq:int_id1} and \req{eq:int_id2} derived in Proposition \ref{prp:int_id} to rewrite the first term in the last equation as
\begin{gather}
\int \dijPz{y}{k}{r} \dijlogPz{y}{i}{j} \d \rvecr{y} = \int \pdfvec{y} \dijkllogPz{y}{i}{j}{k}{r}\d \rvecr{y}.
\end{gather}

Now, applying a scalar version of the logarithm identity in \req{eq:log_id},
\begin{gather}
\frac{1}{\pdfvec{y}} \dijPz{y}{k}{r} = \dijlogPz{y}{k}{r} + \dilogPz{y}{k} \dilogPz{y}{r},
\end{gather}
the second term in the right hand side of \req{eq:intermig} becomes
\begin{multline}
\int \pdfvec{y} \frac{\partial^2}{\partial \rvecri{y}{i} \partial \rvecri{y}{j}}\left( \frac{1}{\pdfvec{y}} \dijPz{y}{k}{r} \right) \d \rvecr{y} = \int \pdfvec{y} \dijkllogPz{y}{i}{j}{k}{r}\d \rvecr{y} \\ + \int \pdfvec{y} \left( \dijlogPz{y}{j}{r} \dijlogPz{y}{i}{k} + \dijlogPz{y}{i}{r} \dijlogPz{y}{j}{k} \right) \d \rvecr{y} \\ + \int \pdfvec{y} \left( \dijklogPz{y}{i}{j}{k} \dilogPz{y}{r} + \dijklogPz{y}{i}{j}{r} \dilogPz{y}{k} \right) \d \rvecr{y}.
\end{multline}

Using the regularity condition \req{eq:int_symmetry} in Corollary \ref{cor:Espijkl}, we finally obtain the desired result
\begin{multline}
\Jacob_{\preN_{kl}} \left[\J{y}\right]_{ij} = - \int \pdfvec{y} \left( \sum_{r=1}^\dim \dijlogPz{y}{j}{r} [\preN \Cov{n}]_{rl}\right) \dijlogPz{y}{i}{k} \d \rvecr{y} \\ - \int \pdfvec{y} \left( \sum_{r=1}^\dim \dijlogPz{y}{i}{r} [\preN \Cov{n}]_{rl}\right) \dijlogPz{y}{j}{k} \d \rvecr{y}. \label{eq:Dani}
\end{multline}
Now, recalling that $\CJr{y} = \Hess_{\rvecr{y}} \log \pdfvec{y}$ and identifying the elements of the two matrices $\CJr{y} \preN \Cov{n}$ and $\CJr{y}$ with the terms in \req{eq:Dani}, we obtain
\begin{gather} \label{eq:DcJijkl}
\Jacob_{\preN_{kl}} \left[\J{y}\right]_{ij} = - \Esp{ [\CJ{y} \preN
\Cov{n}]_{jl} [\CJ{y}]_{ik} + [\CJ{y} \preN \Cov{n}]_{il}
[\CJ{y}]_{jk}}.
\end{gather}
Finally, taking into account that $[\Jacob_{\preN} \vecop \J{y}]_{i + (j-1)\dim, k + (l-1)\dim} = \Jacob_{\preN_{kl}} \left[\J{y}\right]_{ij}$ and applying Lemma \ref{lem:Kron_comm} with $\mat{A} = \CJ{y} \preN \Cov{n}$ and $\mat{B} = \CJ{y}$ it can be easily shown that
\begin{align}
\Jacob_{\preN} \vecop \J{y} &= - \Esp{\CJ{y} \preN \Cov{n} \otimes \CJ{y} + \Com{\dim} ( \CJ{y} \preN \Cov{n} \otimes \CJ{y})} \\ &= -2\Sym{\dim} \Esp{\CJ{y} \preN \Cov{n} \otimes \CJ{y}}.
\end{align}

\subsection{Proof of Theorem \ref{thm:DbE}} \label{ap:DbE_proof}

Throughout this proof we assume that $\{i, j, l\} \in [1, \dimp]$ and $k \in [1, \dim]$. Now, let us begin by considering the expression for the entries of the matrix $\MSE{s}$:
\begin{gather}
\left[ \MSE{s} \right]_{ij} = \Esp{(\rveci{s}{i} - \CEsp{\rveci{s}{i}}{\rvec{y}})(\rveci{s}{j} - \CEsp{\rveci{s}{j}}{\rvec{y}})} = \Esp{\rveci{s}{i}\rveci{s}{j}} - \Esp{\CEsp{\rveci{s}{i}}{\rvec{y}}\CEsp{\rveci{s}{j}}{\rvec{y}}}.
\end{gather}
Since the first term in last expression does not depend on $\preS$, we have that
\begin{multline}
\Jacob_{\preS_{kl}} \left[ \MSE{s} \right]_{ij} = - \Jacob_{\preS_{kl}} \Esp{\CEsp{\rveci{s}{i}}{\rvec{y}}\CEsp{\rveci{s}{j}}{\rvec{y}}} = - \frac{\partial}{\partial\preS_{kl}} \int \pdfvec{y} \CEsp{\rveci{s}{i}}{\rvecr{y}}\CEsp{\rveci{s}{j}}{\rvecr{y}} \d \rvecr{y} \\ = - \int \frac{\partial\pdfvec{y}}{\partial\preS_{kl}} \CEsp{\rveci{s}{i}}{\rvecr{y}}\CEsp{\rveci{s}{j}}{\rvecr{y}} \d \rvecr{y} - \int \pdfvec{y} \frac{\partial \CEsp{\rveci{s}{i}}{\rvecr{y}}}{\partial\preS_{kl}} \CEsp{\rveci{s}{j}}{\rvecr{y}} \d \rvecr{y} \\ - \int \pdfvec{y} \CEsp{\rveci{s}{i}}{\rvecr{y}} \frac{\partial \CEsp{\rveci{s}{j}}{\rvecr{y}}}{\partial\preS_{kl}} \d \rvecr{y}, \label{eq:JbCEsp}
\end{multline}
where, as in Appendix \ref{ap:DcJ_proof}, the justification of this interchange of the order of derivation and integration and two other interchanges below follow similar steps as in \cite[Appendix B]{palomar:06}.

Note that the second and third terms in \req{eq:JbCEsp} have the same structure and, thus, we will deal with them jointly. The first term in \req{eq:JbCEsp} can be rewritten as
\begin{gather} \label{eq:1sttermE}
- \int \frac{\partial\pdfvec{y}}{\partial\preS_{kl}} \CEsp{\rveci{s}{i}}{\rvecr{y}}\CEsp{\rveci{s}{j}}{\rvecr{y}} \d \rvecr{y} = \int \Esp{\rveci{s}{l}\frac{\partial \pdf{\rvec{y}|\rvec{s}}{\rvecr{y}|\rvec{s}}}{\partial \rvecri{y}{k}}} \CEsp{\rveci{s}{i}}{\rvecr{y}}\CEsp{\rveci{s}{j}}{\rvecr{y}} \d \rvecr{y},
\end{gather}
where we have used Lemma \ref{lem:DBP} to transform the derivative with respect to $\preS$ into a derivative with respect to $\rvecr{y}$. Using Lemma \ref{lem:DbCEsp}, the second term in \req{eq:JbCEsp} can be computed as
\begin{multline} \label{eq:2ndtermE}
- \int \pdfvec{y} \frac{\partial \CEsp{\rveci{s}{i}}{\rvecr{y}}}{\partial\preS_{kl}} \CEsp{\rveci{s}{j}}{\rvecr{y}} \d \rvecr{y} = - \int \CEsp{\rveci{s}{i}}{\rvecr{y}} \Esp{\rveci{s}{l} \frac{\partial \pdf{\rvec{y}|\rvec{s}}{\rvecr{y}|\rvec{s}} }{\partial \rvecri{y}{k}} } \CEsp{\rveci{s}{j}}{\rvecr{y}}  \d \rvecr{y}
\\ + \int \Esp{\rveci{s}{i} \rveci{s}{l} \frac{\partial \pdf{\rvec{y}|\rvec{s}}{\rvecr{y}|\rvec{s}} }{\partial \rvecri{y}{k}} } \CEsp{\rveci{s}{j}}{\rvecr{y}}  \d \rvecr{y}.
\end{multline}
Note that the third term in \req{eq:JbCEsp} can be obtained by interchanging the roles of $i$ and $j$ in last equation. Plugging the expressions \req{eq:1sttermE} and \req{eq:2ndtermE} into \req{eq:JbCEsp} we can write
\begin{multline} \label{eq:intermigE}
\Jacob_{\preS_{kl}} \left[ \MSE{s} \right]_{ij} = - \int \CEsp{\rveci{s}{i}}{\rvecr{y}} \Esp{\rveci{s}{l} \frac{\partial \pdf{\rvec{y}|\rvec{s}}{\rvecr{y}|\rvec{s}} }{\partial \rvecri{y}{k}} } \CEsp{\rveci{s}{j}}{\rvecr{y}}  \d \rvecr{y} \\ + \int \Esp{\rveci{s}{i} \rveci{s}{l} \frac{\partial \pdf{\rvec{y}|\rvec{s}}{\rvecr{y}|\rvec{s}} }{\partial \rvecri{y}{k}} } \CEsp{\rveci{s}{j}}{\rvecr{y}}  \d \rvecr{y} + \int \Esp{\rveci{s}{j} \rveci{s}{l} \frac{\partial \pdf{\rvec{y}|\rvec{s}}{\rvecr{y}|\rvec{s}} }{\partial \rvecri{y}{k}} } \CEsp{\rveci{s}{i}}{\rvecr{y}}  \d \rvecr{y}.
\end{multline}

We now simplify the obtained expression. The first term can be reformulated as
\begin{align}
-\int \Esp{\rveci{s}{l} \frac{\partial \pdf{\rvec{y}|\rvec{s}}{\rvecr{y}|\rvec{s}}}{\partial \rvecri{y}{k}}} \CEsp{\rveci{s}{i}}{\rvecr{y}}\CEsp{\rveci{s}{j}}{\rvecr{y}} \d \rvecr{y} &= -\int \frac{\partial \Esp{\rveci{s}{l} \pdf{\rvec{y}|\rvec{s}}{\rvecr{y}|\rvec{s}}}}{\partial \rvecri{y}{k}} \CEsp{\rveci{s}{i}}{\rvecr{y}}\CEsp{\rveci{s}{j}}{\rvecr{y}} \d \rvecr{y} \nonumber \\ &= -\int \frac{\partial \pdfvec{y}  \CEsp{\rveci{s}{l}}{\rvecr{y}}}{\partial \rvecri{y}{k}} \CEsp{\rveci{s}{i}}{\rvecr{y}}\CEsp{\rveci{s}{j}}{\rvecr{y}} \d \rvecr{y} \nonumber \\ &= \int \pdfvec{y} \CEsp{\rveci{s}{l}}{\rvecr{y}} \frac{\partial \CEsp{\rveci{s}{i}}{\rvecr{y}}\CEsp{\rveci{s}{j}}{\rvecr{y}}}{\partial \rvecri{y}{k}} \d \rvecr{y}, \nonumber
\end{align}
where in the last step we have integrated by parts as detailed in Proposition \ref{prp:int_id2}. We now make use of Lemma \ref{lem:DyCEsp} to simplify the derivative inside the integration sign in last equation to obtain
\begin{multline} \label{eq:terme1}
\int \pdfvec{y} \CEsp{\rveci{s}{l}}{\rvecr{y}} \frac{\partial \CEsp{\rveci{s}{i}}{\rvecr{y}}\CEsp{\rveci{s}{j}}{\rvecr{y}}} {\partial \rvecri{y}{k}} \d \rvecr{y} \\ = \int \pdfvec{y} \CEsp{\rveci{s}{l}}{\rvecr{y}} \Big( \CEsp{\rveci{s}{j}}{\rvecr{y}} \big[ \CMSEr{s}{y} \preS^\T \Cov{z}^{-1} \big]_{ik} +  \CEsp{\rveci{s}{i}}{\rvecr{y}} \big[ \CMSEr{s}{y} \preS^\T \Cov{z}^{-1} \big]_{jk} \Big) \d \rvecr{y}.
\end{multline}

We now proceed to the computation of the second and third terms in \req{eq:intermigE} (note that they are in fact the same term with the roles of $i$ and $j$ interchanged). We have
\begin{align}
\int \Esp{\rveci{s}{i} \rveci{s}{l} \frac{\partial \pdf{\rvec{y}|\rvec{s}}{\rvecr{y}|\rvec{s}} }{\partial \rvecri{y}{k}} } \CEsp{\rveci{s}{j}}{\rvecr{y}}  \d \rvecr{y} &= \int \frac{\partial \Esp{ \rveci{s}{i} \rveci{s}{l} \pdf{\rvec{y}|\rvec{s}}{\rvecr{y}|\rvec{s}} }}{\partial \rvecri{y}{k}} \CEsp{\rveci{s}{j}}{\rvecr{y}} \d \rvecr{y} \\ &= \int \frac{\partial \pdfvec{y} \CEsp{ \rveci{s}{i} \rveci{s}{l}}{\rvecr{y}}}{\partial \rvecri{y}{k}} \CEsp{\rveci{s}{j}}{\rvecr{y}} \d \rvecr{y} \\ &= - \int \pdfvec{y} \CEsp{\rveci{s}{i} \rveci{s}{l}}{\rvecr{y}} \frac{\partial \CEsp{\rveci{s}{j}}{\rvecr{y}}}{\partial \rvecri{y}{k}} \d \rvecr{y},
\end{align}
where last equality follows by integrating by parts as Proposition \ref{prp:int_id2}. We are now ready to apply Lemma \ref{lem:DyCEsp} again to obtain
\begin{gather} \label{eq:terme2}
- \int \pdfvec{y} \CEsp{\rveci{s}{i} \rveci{s}{l}}{\rvecr{y}} \frac{\partial \CEsp{\rveci{s}{j}}{\rvecr{y}}}{\partial \rvecri{y}{k}} \d \rvecr{y} = - \int \pdfvec{y} \CEsp{\rveci{s}{i}\rveci{s}{l}}{\rvecr{y}} \big[ \CMSEr{s}{y} \preS^\T \Cov{z}^{-1} \big]_{jk} \d \rvecr{y}.
\end{gather}

Plugging \req{eq:terme1} and \req{eq:terme2} into \req{eq:intermigE} and recalling that $[\CMSEr{s}{y}]_{jl} = \CEsp{\rveci{s}{j} \rveci{s}{l}}{\rvecr{y}} - \CEsp{\rveci{s}{j}}{\rvecr{y}}\CEsp{\rveci{s}{l}}{\rvecr{y}}$, we finally have
\begin{gather}
\Jacob_{\preS_{kl}} \left[ \MSE{s} \right]_{ij} = - \int \pdfvec{y} \Big( [ \CMSEr{s}{y} ]_{jl} \big[ \CMSEr{s}{y} \preS^\T \Cov{z}^{-1} \big]_{ik} + [ \CMSEr{s}{y} ]_{il} \big[ \CMSEr{s}{y} \preS^\T \Cov{z}^{-1} \big]_{jk} \Big) \d \rvecr{y}.
\end{gather}
Taking into account that $\Jacob_{\preS_{kl}} \left[ \MSE{s} \right]_{ij} = [ \Jacob_{\preS} \vecop \MSE{s} ]_{i + (j-1) \dim , k + (l-1)\dimp }$ and applying Lemma \ref{lem:Kron_comm} with $\mat{A} = \CMSEr{s}{y}$ and $\mat{B} = \CMSEr{s}{y} \preS^\T \Cov{z}^{-1}$ we obtain
\begin{align}
\Jacob_{\preS} \vecop \MSE{s} &= - \Esp[1]{ \CMSE{s}{y} \otimes  \CMSE{s}{y} \preS^\T \Cov{z}^{-1} + \Com{\dimp}\CMSE{s}{y} \otimes  \CMSE{s}{y} \preS^\T \Cov{z}^{-1}} \\ &= -2\Sym{\dimp} \Esp[1]{ \CMSE{s}{y} \otimes  \CMSE{s}{y} \preS^\T \Cov{z}^{-1}}.
\end{align}

\subsection{Proof of Theorem \ref{thm:entropy_Hessians}} \label{ap:entropy_Hessians}

The developments leading to the expressions for the Hessian matrices $\Hess_{\Prec} \Ent(\rvec{y})$, $\Hess_{\Chan} \Ent(\rvec{y})$, and $\Hess_{\preN} \Ent(\rvec{y})$ follow a very similar pattern. Consequently, we will present only one of them here.

Consider the Hessian $\Hess_{\Prec} \Ent(\rvec{y})$, from the expression for the Jacobian $\Jacob_{\Prec} \Ent(\rvec{y})$ in \req{eq:JPh} it follows that
\begin{align}
\Hess_{\Prec} \Ent(\rvec{y}) = \Jacob_{\Prec}(\Jacob_{\Prec}^\T \Ent(\rvec{y})) &= \Jacob_{\Prec} \vecop \Chan^\T \Cov{z}^{-1} \Chan \Prec \MSE{s} \\ &= \big(\MSE{s} \otimes \Chan^\T \Cov{z}^{-1} \Chan \big) + (\mat{I}_\dimp \otimes \Chan^\T \Cov{z}^{-1} \Chan \Prec) \Dup{\dimp} \: \Jacob_{\Prec} \MSE{s}, \label{eq:proofHph}
\end{align}
where in \req{eq:proofHph} we have used Lemma \ref{lem:Jacob_chain_examples}.\ref{lem:id1} adding the matrix $\Dup{\dimp}$ because $\MSE{s}$ is a symmetric matrix. The final expression for $\Hess_{\Prec} \Ent(\rvec{y})$ is obtained by plugging in \req{eq:proofHph} the expression for $\Jacob_{\Prec} \MSE{s}$ obtained in Theorem \ref{thm:Jacob_arb} and recalling that $\Dup{\dimp} \Dup{\dimp}^\pinv = \Sym{\dimp}$.

The calculation of the Hessian matrix $\Hess_{\Cov{z}} \Ent(\rvec{y})$ from its Jacobian $\Jacob_{\Cov{z}} \Ent(\rvec{y})$ in \req{eq:JRzh} follows:
\begin{gather} \label{eq:proof_HRzh}
\Hess_{\Cov{z}} \Ent(\rvec{y}) = \frac{1}{2} \Jacob_{\Cov{z}} \Dup{\dim}^\T \vecop \J{y} = \frac{1}{2} \Jacob_{\Cov{z}} \Dup{\dim}^\T \Dup{\dim} \vechop \J{y} = \frac{1}{2} \Dup{\dim}^\T \Dup{\dim} \Jacob_{\Cov{z}} \J{y},
\end{gather}
where, in last equality, we have used Lemma \ref{lem:Jacob_chain_examples}.\ref{lem:id1.5}. Now, we only need to plug in the expression for $\Jacob_{\Cov{z}} \J{y}$, which can be found in Theorem \ref{thm:Jacob_arb} and note that $\Dup{\dim}^\T \Dup{\dim} \Dup{\dim}^\pinv = \Dup{\dim}^\T \Sym{\dim} = \Dup{\dim}^\T$.

Finally, the Hessian matrix $\Hess_{\Cov{n}} \Ent(\rvec{y})$ can be computed from its Jacobian $\Jacob_{\Cov{n}} \Ent(\rvec{y})$ in \req{eq:JRnh} as
\begin{align}
\Hess_{\Cov{n}} \Ent(\rvec{y}) &= \frac{1}{2} \Jacob_{\Cov{n}} \Dup{\dim'}^\T \vecop (\preN^\T \J{y} \preN) \\ &= \frac{1}{2} \Dup{\dim'}^\T (\preN^\T \otimes \preN^\T) \Dup{\dim} \Jacob_{\Cov{n}} \J{y},
\end{align}
where we have used Lemmas \ref{lem:vecABC} and \ref{lem:Jacob_chain_examples}.\ref{lem:id1.5} and also that $\Jacob_{\Cov{n}} \vecop \J{y} = \Dup{\dim} \Jacob_{\Cov{n}} \vechop \J{y} = \Dup{\dim} \Jacob_{\Cov{n}} \J{y}$ similarly as in \req{eq:proof_HRzh}. Recalling the expression for $\Jacob_{\Cov{n}} \J{y}$ in Theorem \ref{thm:Jacob_arb}, the result follows.

\subsection{Matrix algebra results for the proof of the multidimensional EPI in Theorem \ref{thm:MVCostaEPI}}  \label{ap:matrix_algebra}

In this appendix we present a number of lemmas and propositions that
are used in the proof of our multidimensional EPI in Section
\ref{sec:Costa_EPI}.

\begin{lemap}[Bhatia {\cite[p.~15]{bhatia:07}}] \label{lem:XXXX}
Let $\mat{R} \in \PSD{s}$ be a positive
semidefinite matrix, $\mat{R} \geq \mats{0}$. Then,
\begin{gather} \nonumber
\left[
\begin{array}{cc}
\mat{R} & \mat{R} \\
\mat{R} & \mat{R}
\end{array}
\right] \geq \mats{0}.
\end{gather}
\end{lemap}
\begin{IEEEproof}
Since $\mat{R} \geq \mats{0}$, consider $\mat{R} =
\mat{A}\mat{A}^\T$ and write
\begin{gather} \nonumber
\left[
\begin{array}{cc}
\mat{R} & \mat{R} \\
\mat{R} & \mat{R}
\end{array}
\right] = \left[
\begin{array}{c}
\mat{A} \\
\mat{A}
\end{array}
\right]\left[
\begin{array}{cc}
\mat{A}^\T & \mat{A}^\T
\end{array}
\right].
\end{gather}
\end{IEEEproof}

\begin{lemap}[Bhatia {\cite[Exercise 1.3.10]{bhatia:07}}] \label{lem:XIIinvX}
Let $\mat{R} \in \PD{s}$ be a positive definite matrix, $\mat{R}
> \mats{0}$. Then,
\begin{gather} \label{eq:XIIinvX}
\left[
\begin{array}{cc}
\mat{R} & \mat{I}_s \\
\mat{I}_s & \mat{R}^{-1}
\end{array}
\right] \geq \mats{0}.
\end{gather}
\end{lemap}
\begin{IEEEproof}
Consider again $\mat{R} = \mat{A}\mat{A}^\T$, then we have
$\mat{R}^{-1} = \mat{A}^{-\T}\mat{A}^{-1}$. Now, simply write
\req{eq:XIIinvX} as
\begin{gather} \nonumber
\left[
\begin{array}{cc}
\mat{R} & \mat{I}_s \\
\mat{I}_s & \mat{R}^{-1}
\end{array}
\right] %\\
= \left[
\begin{array}{cc}
\mat{A} & \mats{0} \\
\mats{0} & \mat{A}^{-\T}
\end{array}
\right]\left[
\begin{array}{cc}
\mat{I}_s & \mat{I}_s \\
\mat{I}_s & \mat{I}_s
\end{array}
\right]\left[
\begin{array}{cc}
\mat{A}^{\T} & \mats{0} \\
\mats{0} & \mat{A}^{-1}
\end{array}
\right], \nonumber
\end{gather}
which, from Sylvester's law of inertia for congruent matrices
\cite[p.~5]{bhatia:07} and Lemma \ref{lem:XXXX}, is positive
semidefinite.
\end{IEEEproof}

\begin{lemap} \label{lem:kronecker_th}
If the matrices $\mat{R}$ and $\mat{T}$ are positive (semi)definite,
then so is the product $\mat{R} \otimes \mat{T}$. In other words, the class of positive (semi)definite matrices is closed under the Kronecker product.
\end{lemap}
\begin{IEEEproof}
See \cite[p. 254, Fact 7.4.15]{bernstein:05}
\end{IEEEproof}
\begin{corap}[Schur Theorem] \label{lem:schur_th}
The class of positive (semi)definite matrices is also
closed under the Schur matrix product, $\mat{R} \circ \mat{T}$.
\end{corap}
\begin{IEEEproof}
The proof follows from Lemma \ref{lem:kronecker_th} by noting that the Schur product $\mat{R} \circ \mat{T}$ is a principal submatrix of the Kronecker product $\mat{R} \otimes \mat{T}$ as in \cite[Proposition 7.3.1]{bernstein:05} and that any principal submatrix of a positive (semi)definite matrix is also positive (semi)definite, \cite[Propositions 8.2.6 and 8.2.7]{bernstein:05}. Alternatively, see \cite[Theorem 7.5.3]{horn:85} or \cite[Theorem 5.2.1]{horn:91} for a completely different proof.
\end{IEEEproof}

\begin{lemap}[Schur complement] \label{lem:schur_compl}
Let the matrices $\mat{R}\in \PD{s}$ and $\mat{T}\in \PD{q}$
be positive definite, $\mat{R}>\mats{0}$ and $\mat{T}>\mats{0}$, and
not necessarily of the same dimension. Then the following statements
are equivalent
\begin{enumerate}
\item
$\left[
\begin{array}{cc}
\mat{R} & \mat{A} \\
\mat{A}^\T & \mat{T}
\end{array}
\right] \geq 0,$

\item $\mat{T} \geq \mat{A}^\T \mat{R}^{-1} \mat{A}$,

\item $\mat{R} \geq \mat{A} \mat{T}^{-1} \mat{A}^\T$,
\end{enumerate}
where $\mat{A}\in\R^{s \times r}$ is any arbitrary matrix.
\end{lemap}
\begin{IEEEproof}
See \cite[Theorem 7.7.6]{horn:85} and the second exercise following it
or \cite[Propostition 8.2.3]{bernstein:05}.
\end{IEEEproof}

With the above lemmas at hand, we are now ready to prove the
following proposition:
\begin{prpap} \label{prp:XoinvZ}
Consider two positive definite matrices $\mat{R} \in \PD{s}$ and
$\mat{T} \in \PD{s}$ of the same dimension. Then it follows that
\begin{gather} \label{eq:AoinvB}
\mat{R}\circ\mat{T}^{-1} \geq \Diag{\mat{R}}
\left(\mat{R}\circ\mat{T}\right)^{-1} \Diag{\mat{R}}.
\end{gather}
\end{prpap}
\begin{IEEEproof}
From Lemmas \ref{lem:XXXX}, \ref{lem:XIIinvX}, and
\ref{lem:schur_th}, it follows that
\begin{gather} \nonumber
\left[
\begin{array}{cc}
\mat{R} & \mat{R} \\
\mat{R} & \mat{R}
\end{array}
\right] \circ \left[
\begin{array}{cc}
\mat{T} & \mat{I}_s \\
\mat{I}_s & \mat{T}^{-1}
\end{array}
\right] = \left[
\begin{array}{cc}
\mat{R}\circ\mat{T} & \Diag{\mat{R}} \\
\Diag{\mat{R}} & \mat{R}\circ\mat{T}^{-1}
\end{array}
\right] \geq 0.
\end{gather}
Now, from Lemma \ref{lem:schur_compl}, the result follows directly.
\end{IEEEproof}

\begin{corap} \label{cor:dinvXXd}
Let $\mat{R} \in \PD{s}$ be a positive definite matrix. Then,
\begin{gather}
\diag{\mat{R}}^\T \left(\mat{R}\circ\mat{R}\right)^{-1}
\diag{\mat{R}} \leq s. \label{eq:dinvXXd}
\end{gather}
\end{corap}
\begin{IEEEproof}
Particularizing the result in Proposition \ref{prp:XoinvZ} with
$\mat{T} = \mat{R}$ and pre- and post-multiplying it by
$\mats{1}^\T$ and $\mats{1}$ we obtain
\begin{gather} \nonumber
\mats{1}^\T\left(\mat{R}\circ\mat{R}^{-1}\right)\mats{1} \geq
\mats{1}^\T \Diag{\mat{R}} \left(\mat{R}\circ\mat{R}\right)^{-1}
\Diag{\mat{R}} \mats{1}.
\end{gather}
The result in \req{eq:dinvXXd} now follows straightforwardly from
the fact $\vecs{1}^\T \left( \mat{R}\circ\mat{R}^{-\T}
\right)\vecs{1} = s$, \cite{johnson:86} (see also \cite[Fact
7.6.10]{bernstein:05}, \cite[Lemma 5.4.2(a)]{horn:91}). Note that
$\mat{R}$ is symmetric and thus $\mat{R}^{\T} = \mat{R}$ and
$\mat{R}^{-\T} = \mat{R}^{-1}$.
\end{IEEEproof}
\begin{remap}
Note that the proof of Corollary \ref{cor:dinvXXd} is based on the
result of Proposition \ref{prp:XoinvZ} in \req{eq:AoinvB}. An
alternative proof could follow similarly from a different inequality
by Styan in \cite{styan:73}
\begin{gather} \nonumber %\label{eq:ineq_sim}
\mat{R}\circ\mat{R}^{-1} + \mat{I}_s \geq
2\left(\mat{R}\circ\mat{R}\right)^{-1},
\end{gather}
where, in this case, $\mat{R}$ is constrained to have ones in its main diagonal, \ie, $\mat{R}\circ\mat{I}_s = \mat{I}_s$.
\end{remap}

\begin{prpap} \label{prp:XXones}
Consider now the positive semidefinite matrix $\mat{R} \in
\PSD{s}$. Then,
\begin{gather} \nonumber
\mat{R}\circ\mat{R} \geq
\frac{\diag{\mat{R}}\diag{\mat{R}}^\T}{s}.
\end{gather}
\end{prpap}
\begin{IEEEproof}
For the case where $\mat{R} \in \PD{s}$ is positive definite,
from \req{eq:dinvXXd} in Corollary \ref{cor:dinvXXd} and Lemma
\ref{lem:schur_compl}, it follows that
\begin{gather} \nonumber
\left[
\begin{array}{cc}
\mat{R}\circ\mat{R} & \diag{\mat{R}} \\
\diag{\mat{R}}^\T & s
\end{array}
\right] \geq 0.
\end{gather}
Applying again Lemma \ref{lem:schur_compl}, we get
\begin{gather} \label{eq:XXones_proof}
\mat{R}\circ\mat{R} \geq
\frac{\diag{\mat{R}}\diag{\mat{R}}^\T}{s}.
\end{gather}
Now, assume that $\mat{R} \in \PSD{s}$ is positive semidefinite.
We thus define $\epsilon > 0$ and consider the positive definite
matrix $\mat{R} + \epsilon\mat{I}_s$. From \req{eq:XXones_proof}, we
know that
\begin{gather} \nonumber
\left(\mat{R}+ \epsilon\mat{I}_s \right) \circ \left( \mat{R} +
\epsilon\mat{I}_s \right) \geq \frac{\diag{\mat{R}+
\epsilon\mat{I}_s }\diag{\mat{R}+ \epsilon\mat{I}_s }^\T}{s}.
\end{gather}
Taking the limit as $\epsilon$ tends to 0, the validity of \req{eq:XXones_proof} for positive semidefinite matrices follows from continuity.
\end{IEEEproof}

The last lemma in this section follows.
\begin{lemap} \label{lem:psd}
For a given random vector $\rvec{x}$, it follows that
$\Esp{\rvec{X}\rvec{X}^\T} \geq
\Esp[1]{\rvec{X}}\Esp{\rvec{X}^\T}$.
\end{lemap}
\begin{IEEEproof}
Simply note that
\begin{gather} \nonumber
\Esp[1]{\rvec{X}\rvec{X}^\T} - \Esp[1]{\rvec{X}}\Esp[1]{\rvec{X}^\T}
= \Esp[1]{(\rvec{X} - \Esp{\rvec{X}})(\rvec{X} - \Esp{\rvec{X}})^\T}
\geq \mats{0},
\end{gather}
where last inequality follows from the fact that the expectation
preserves positive semidefiniteness.
\end{IEEEproof}

\subsection{Integral identities involving functions and derivatives of $\pdfvec{y}$.} \label{ap:key_step}

The integral identities presented in this section are derived through a sequence of lemmas which lead to the main proposition containing the identities.

First, we present a lemma, which is a straightforward generalization
for non-white Gaussian random variables of \cite[Lemma
4.1]{toscani:99}
\begin{lemap} \label{lem:toscani}
Assume $\rvec{y} = \rvec{x} + \rvec{z}$ is an
$\dim$-dimensional random vector, where $\rvec{x}$ is arbitrarily distributed and $\rvec{z}$ is distributed following a zero-mean Gaussian distribution with covariance $\Cov{z}$ and consider a non-empty set of natural numbers $\mc{I}$, whose elements range from $1$ to $\dim$. Then, given $\varphi > 1$, there exists a finite positive constant $\kappa$ not depending on $\rvecr{y}$ such that
\begin{gather}
\left| \frac{\partial^{|\mc{I}|} \pdfvec{y}}{\prod_{i\in\mc{I}} \partial \rvecri{y}{i}} \right| \leq \kappa(\dim, \mc{I}, \varphi, \Cov{z}) (\pdfvec{y})^{1/\varphi},
\end{gather}
where we use the notation $\prod_{i\in \mc{I}}\partial \rvecri{y}{i}$ to denote, \eg, $\prod_{i\in \{3, 1, 3, 5 \}}\partial \rvecri{y}{i} = \partial \rvecri{y}{1}\partial \rvecri{y}{3}^2 \partial \rvecri{y}{5}$.
\end{lemap}
\begin{IEEEproof}
This proof follows the guidelines of the proof of \cite[Lemma
4.1]{toscani:99}. For any $\Cov{z} > \mats{0}$, which implies that $\Cov{z}^{-1}$ exists, we have that $\pdfvec{y}$ is continuously differentiable in $\rvecr{y}$, and
\begin{align}
\diPz{y}{i} &= \frac{\partial}{\partial \rvecri{y}{i}} \Esp{\pdf{\rvec{y}|\rvec{x}}{\rvecr{y}|\rvec{x}}}
= \Esp{\frac{\partial}{\partial \rvecri{y}{i}}\pdf{\rvec{z}}{\rvecr{y} - \rvec{x}}}  \\
&= - \int \pdfvec{x} \pdf{\rvec{z}}{\rvecr{y} - \rvecr{x}}
\big[(\rvecr{y} - \rvecr{x})^\T \Cov{z}^{-1}\big]_i \d \rvecr{x}.
\end{align}
Now, using Hölder's inequality, for $\varphi > 1$ and $1/\varphi + 1/\psi = 1$ we
have %(TAKE CARE WITH THE FIRST INEQUALITY, MAYBE SHOULD BE EQUALITY, MAYBE NEEDS ABSOLUTE VALUE INSIDE THE INTEGRAL)
\begin{gather} \label{eq:Holder1}
\begin{split}
\left| \diPz{y}{i} \right| &\leq \int \left( \pdfvec{x}
\pdf{\rvec{z}}{\rvecr{y} - \rvecr{x}} \right)^{1/\varphi} \left(
\pdfvec{x} \pdf{\rvec{z}}{\rvecr{y} - \rvecr{x}} \big|\big[(\rvecr{y} -
\rvecr{x})^\T \Cov{z}^{-1}\big]_i\big|^\psi \right)^{1/\psi} \d \rvecr{x} \\
&\leq (\pdfvec{y})^{1/\varphi} \left( \int \pdfvec{x}
\pdf{\rvec{z}}{\rvecr{y} - \rvecr{x}} \big|\big[(\rvecr{y} -
\rvecr{x})^\T \Cov{z}^{-1}\big]_i\big|^\psi \d \rvecr{x} \right)^{1/\psi},
\end{split}
\end{gather}
from which the result for $\mc{I} = \{ i \}$, such that $|\mc{I}| = 1$, follows since
$\pdf{\rvec{z}}{\rvecr{y}} \big|\big[\rvecr{y}^\T
\Cov{z}^{-1}\big]_i\big|^\psi$ is bounded above by a constant depending
only on $\Cov{z}$ and $\psi = \varphi/(\varphi-1)$. %Note that $\left[\rvecr{y}^\T
%\Cov{z}^{-1}\right]^q_i$ is simply a polynomial of the entries of $\rvecr{y}$ of finite degree $q$.

The inequalities for $|\mc{I}| > 1$ follow in a similar fashion from
the fact that for any $\Cov{z} > \mats{0}$,
\begin{gather}
\frac{\partial^{|\mc{I}|} \pdf{\rvec{z}}{\rvecr{y}}}{\partial \rvecri{y}{i}^{|\mc{I}|}} = (-1)^{|\mc{I}|} \left(\frac{\left[ \Cov{z}^{-1}\right]_{ii}}{2}\right)^{|\mc{I}|/2} H_{|\mc{I}|} \left( \frac{\left[\Cov{z}^{-1}\rvecr{y}\right]_i}{\sqrt{2\left[\Cov{z}^{-1}\right]_{ii}}} \right) \pdf{\rvec{z}}{\rvecr{y}}
\end{gather}
where $H_{|\mc{I}|}(x)$ is the $|\mc{I}|$-th order Hermite polynomial defined following the convention in \cite[p. 817]{arfken:05}, and noting that the partial derivatives of the Hermite polynomial are other polynomials. %IMPROVE EXPLANATION.
\end{IEEEproof}

%The following lemma BLA BLA BLA\ldots
\begin{lemap} \label{lem:dmlogPz}
Assume $\rvec{y} = \rvec{x} + \rvec{z}$ is an
$\dim$-dimensional random vector, where $\rvec{x}$ is arbitrarily distributed and $\rvec{z}$ is distributed following a zero-mean Gaussian distribution with covariance $\Cov{z}$. Then, given $\varphi > 1$, there exist a set of finite positive constants $\xi$ not depending on $\rvecr{y}$ such that for all $|\mc{I}| \geq 1$
\begin{gather}
\left| \frac{\partial^{|\mc{I}|} \log \pdfvec{y}}{\prod_{i\in\mc{I}} \partial \rvecri{y}{i}} \right| \leq \sum_{\pi}  \xi(\dim, \pi, \varphi, \Cov{z}) (\pdfvec{y})^{\frac{|\pi|(1-\varphi)}{\varphi}},
\end{gather}
where the sum is over the partitions $\pi$ of the set $\mc{I}$.
\end{lemap}
\begin{IEEEproof}
We recall the Arbogast-Faà di Bruno's formula in its most general form as given in \cite[Eq. (5)]{hardy:06} for the partial derivative of a composite function,
\begin{gather} \label{eq:faadibruno}
\frac{\partial^{|\mc{I}|} f(g)}{\prod_{i\in\mc{I}} \partial \rvecri{z}{i}} = \sum_\pi \frac{\d^{|\pi|} f(g)}{\d g^{|\pi|}} \prod_{B\in \pi} \frac{\partial^{|B|}g}{\prod_{j\in B}\partial \rvecri{z}{j}},
\end{gather}
where, as explained in \cite{hardy:06}, the sum is over the partitions $\pi$ of the set $\mc{I}$, and where $B$ represents an element of the partition $\pi$.\footnote{Note that $B$ is simply a set of indices.} Thus, for each given partition $\pi$, $B$ can take $|\pi|$ different values. Consequently the order of the derivative with respect to $f$, $|\pi|$, coincides with the number of factors in the product indexed by $B$.

Particularizing \req{eq:faadibruno} for our case we obtain
\begin{gather} \label{eq:faalogPz}
\frac{\partial^{|\mc{I}|} \log \pdfvec{y}}{\prod_{i\in\mc{I}} \partial \rvecri{y}{i}} = \sum_\pi (-1)^{|\pi|}|\pi|! \left(\pdfvec{y}\right)^{-|\pi|}
\prod_{B\in \pi} \frac{\partial^{|B|} \pdfvec{y}}{\prod_{j\in B}\partial \rvecri{y}{j}}.
\end{gather}
Now, let us fix $\varphi > 1$ and $\Cov{z} > \mats{0}$ and apply the bound found in Lemma \ref{lem:toscani} to each factor $\frac{\partial^{|B|} \pdfvec{y}}{\prod_{j\in B}\partial \rvecri{y}{j}}$. Recalling that there are $|\pi|$ of these factors, the bound becomes
\begin{align}
\left| \frac{\partial^{|\mc{I}|} \log \pdfvec{y}}{\prod_{i\in\mc{I}} \partial \rvecri{y}{i}} \right| &\leq \sum_\pi |\pi|! \left(\pdfvec{y}\right)^{-|\pi|}
\prod_{B\in \pi} \kappa(\dim, B, \varphi, \Cov{z}) (\pdfvec{y})^{1/\varphi} \\ &= \sum_\pi \xi(\dim, \pi, \varphi, \Cov{z})(\pdfvec{y})^{|\pi|/\varphi-|\pi|}
\end{align}
where we have defined $\xi(\dim, \pi, \varphi, \Cov{z}) = |\pi|!
\prod_{B\in \pi} \kappa(\dim, B, \varphi, \Cov{z})$
\end{IEEEproof}

Next, we present the lemma which is key in the proof of the proposition that contains the integral identities.
\begin{lemap} \label{lem:PzqdmlogPz}
Assume $\rvec{y} = \rvec{x} + \rvec{z}$ is an
$\dim$-dimensional random vector, where $\rvec{x}$ is arbitrarily distributed and $\rvec{z}$ is distributed
following a zero-mean Gaussian distribution with covariance $\Cov{z}$. Let us consider a set $\omega$ whose elements $\mc{I}$ are sets of indices ranging from $1$ to $\dim$. Then, given $\phi > 0$, it follows that
\begin{gather} \label{eq:limPqdmlogP}
\lim_{|\rvecri{y}{k}| \rightarrow \infty} (\pdfvec{y})^\phi \prod_{\mc{I}\in\omega} \frac{\partial^{|\mc{I}|} \log \pdfvec{y}}{\prod_{j\in\mc{I}} \partial \rvecri{y}{i_j}} = 0,
\end{gather}
where $k$ is an arbitrary index for the entries of vector $\rvecr{y}$.
\end{lemap}
\begin{IEEEproof}
Applying Lemma \ref{lem:dmlogPz} to each one of the individual factors inside the product in \req{eq:limPqdmlogP}, yields, for any $\varphi > 1$, that
\begin{align} \label{eq:PzqdmlogPz}
\left| (\pdfvec{y})^\phi \prod_{\mc{I}\in\omega} \frac{\partial^{|\mc{I}|} \log \pdfvec{y}}{\prod_{j\in\mc{I}} \partial \rvecri{y}{i_j}} \right| &\leq (\pdfvec{y})^\phi \prod_{\mc{I} \in \omega} \sum_{\pi(\mc{I})} \xi(\dim, \pi(\mc{I}), \varphi, \Cov{z})(\pdfvec{y})^{|\pi(\mc{I})|/\varphi - |\pi(\mc{I})|} \\ &= \prod_{\mc{I} \in \omega} \sum_{\pi(\mc{I})} \xi(\dim, \pi(\mc{I}), \varphi, \Cov{z})(\pdfvec{y})^{|\pi(\mc{I})|/\varphi-|\pi(\mc{I})| + \phi/|\omega|}, \label{eq:PzqdmlogPz2}
\end{align}
where we have made explicit the dependence of the partition $\pi$ on the current value of the set of indices $\mc{I}$. Now we consider a generic term $\xi(\dim, \pi(\mc{I}), \varphi, \Cov{z})(\pdfvec{y})^{|\pi(\mc{I})|/\varphi - |\pi(\mc{I})| + \phi/|\omega|}$ and we note that for all $\phi > 0$, there exists a value $\varphi_{\max}(\phi, |\pi(\mc{I})|, |\omega|) > 1$ such that the exponent $|\pi(\mc{I})|/\varphi - |\pi(\mc{I})| + \phi/|\omega|$ is positive for any $\varphi$ inside the interval $\varphi \in (1, \varphi_{\max}(\phi, |\pi(\mc{I})|, |\omega|))$. Note that, if $\phi > |\pi||\omega|$, then $\varphi_{\max}(\phi, |\pi(\mc{I})|, |\omega|) = \infty$. Next, simply taking
\begin{gather}
\varphi_{\min}(\phi, |\omega|) = \min_{\pi} \varphi_{\max}(\phi, |\pi(\mc{I})|, |\omega|),
\end{gather}
which fulfills that $\varphi_{\min}(\phi, |\omega|) > 1$, we have that, for all $\varphi$ inside the non-empty interval $(1, \varphi_{\min}(\phi, |\omega|))$, all the exponents of $\pdfvec{y}$ in \req{eq:PzqdmlogPz2} are positive. Since we have that $\lim_{|\rvecri{y}{k}| \rightarrow \infty} \pdfvec{y} = 0$ and the product and sum have a finite number of factors and terms, respectively, we readily obtain the result of the lemma.
\end{IEEEproof}

With this last lemma at hand we are ready to prove the following proposition, which is the main purpose of this appendix.
\begin{prpap} \label{prp:int_id}
Assume $\rvec{y} = \rvec{x} + \rvec{z}$ is an
$\dim$-dimensional random vector, where $\rvec{x}$ is arbitrarily distributed and $\rvec{z}$ is distributed
following a zero-mean Gaussian distribution with covariance $\Cov{z}$. Then the following integral identities hold
\begin{align}
\int \dijPz{y}{k}{l} \dijlogPz{y}{i}{j}\d \rvecr{y} &= - \int \diPz{y}{l} \dijklogPz{y}{i}{j}{k}\d \rvecr{y}, \label{eq:int_id1} \\ \int \diPz{y}{l} \dijklogPz{y}{i}{j}{k}\d \rvecr{y} &= - \int
\pdfvec{y} \dijkllogPz{y}{i}{j}{k}{l}\d \rvecr{y}, \label{eq:int_id2}%\\
\end{align}
\end{prpap}
\begin{IEEEproof}
The proof is based in integrating by parts the left hand side of
\req{eq:int_id1}-\req{eq:int_id2} and showing that there is a term
that vanishes.

Integrating by parts the left hand side of \req{eq:int_id1} we obtain
\begin{gather}
\int \dijPz{y}{k}{l} \dijlogPz{y}{i}{j} \d \rvecr{y} = \left[ \diPz{y}{l} \dijlogPz{y}{i}{j} \right]_{\rvecri{y}{k} = - \infty}^{\rvecri{y}{k} = \infty} - \int \diPz{y}{l} \dijklogPz{y}{i}{j}{k}\d \rvecr{y}.
\end{gather}
Casting Lemma \ref{lem:toscani} with $\mc{I} = \{ l \}$ to bound the first factor in the term inside the evaluation limits in last equation yields
\begin{gather}
\left| \diPz{y}{l} \dijlogPz{y}{i}{j} \right| \leq \kappa(\dim, \{ l \}, \varphi, \Cov{z} ) \left| (\pdfvec{y})^{1/\varphi}  \dijlogPz{y}{i}{j} \right|.
\end{gather}
According to Lemma \ref{lem:PzqdmlogPz} with $\phi = 1/\varphi$ and $\omega = \{ \{i,j\} \}$, the right hand side of last equation vanishes in the limit as $|\rvecri{y}{k}| \rightarrow \infty$, which implies the identity in \req{eq:int_id1}.

Repeating the procedure for \req{eq:int_id2}, the resulting term when integrating by parts is
\begin{gather}
\left[ \pdfvec{y} \dijklogPz{y}{i}{j}{k} \right]_{\rvecri{y}{l}=-\infty}^{\rvecri{y}{l}=\infty},
\end{gather}
which is easily shown that it vanishes applying Lemma \ref{lem:PzqdmlogPz} with $\phi = 1$ and $\omega = \{ \{i,j, k\} \}$.
\end{IEEEproof}

A simple corollary results from Proposition \ref{prp:int_id}.
\begin{corap} \label{cor:Espijkl}
Using that $\diPz{y}{l} = \pdfvec{y}\dilogPz{y}{l}$ in the left hand side of \req{eq:int_id2}, it readily follows that
\begin{align} \label{eq:int_symmetry}
\int \pdfvec{y}\dilogPz{y}{l} \dijklogPz{y}{i}{j}{k}\d \rvecr{y} &= - \int
\pdfvec{y} \dijkllogPz{y}{i}{j}{k}{l}\d \rvecr{y}, \\
\Esp{\dilogP{y}{l} \dijklogP{y}{i}{j}{k}} &= - \Esp{\dijkllogP{y}{i}{j}{k}{l}}, \label{eq:esp_symmetry}
\end{align}
which is a higher-dimensional version of the regularity condition
\begin{gather}
\Esp{\dilogP{y}{l} \dilogP{y}{k}} = - \Esp{\dijlogP{y}{k}{l}},
\end{gather}
which is used for example in the derivation of the CRLB in \cite{kay:93}. %IMPROVE
\end{corap}

\subsection{Integral identities involving functions of $\CEsp{\rvec{s}}{\rvecr{y}}$.} \label{ap:key_step2}

Similarly as in the previous appendix, the integral identities presented in this section are derived through a lemma which leads to the main proposition containing the identities.
\begin{lemap} \label{lem:boundPyEs}
Assume $\rvec{y} = \preS\rvec{s} + \rvec{z}$ is an
$\dim$-dimensional random vector, where $\preS$ is a deterministic matrix, $\rvec{s}$ is an arbitrarily distributed random vector, and $\rvec{z}$ is distributed following a zero-mean Gaussian distribution with covariance $\Cov{z}$. Consider a set of $M$ functions $f_i(\rvec{s})$, which have polynomial dependence on the elements of $\rvec{s}$. Then, given $\varphi > 1$, there exists a finite positive constant $\kappa$ not depending on $\rvecr{y}$ such that
\begin{gather}
\left| \pdfvec{y} \prod_{i=1}^{M} \CEsp{f_i(\rvec{s})}{\rvecr{y}} \right| \leq \kappa(\dim, \{ f_i \}, \varphi, \Cov{z}) (\pdfvec{y})^{\frac{M - \varphi(M - 1)}{\varphi}},
\end{gather}
\end{lemap}
\begin{IEEEproof}
The proof follows by first noticing that
\begin{gather} \nonumber
\left| \CEsp{f_i(\rvec{s})}{\rvecr{y}} \right| \leq \frac{1}{\pdfvec{y}} \int \left| f_i(\rvecr{s}) \right| \pdf{\rvec{y}|\rvec{s}}{\rvecr{y}|\rvecr{s}} \pdfvec{s} \d \rvecr{s} = \frac{1}{\pdfvec{y}} \int \left| f_i(\rvecr{s}) \right| \pdf{\rvec{z}}{\rvecr{y} - \preS \rvecr{s}} \pdfvec{s} \d \rvecr{s},
\end{gather}
and then using H\"older's inequality with $1/\varphi + 1/\psi = 1$ in an analogous way as we have done in \req{eq:Holder1} we obtain
\begin{align}
\int \left| f_i(\rvecr{s}) \right| \pdf{\rvec{z}}{\rvecr{y} - \preS \rvecr{s}} \pdfvec{s} \d \rvecr{s} &= \int \left( \pdfvec{s}
\pdf{\rvec{z}}{\rvecr{y} - \preS\rvecr{s}} \right)^{1/\varphi} \left(
\pdfvec{s} \pdf{\rvec{z}}{\rvecr{y} - \preS\rvecr{s}} |f_i(\rvecr{s})|^\psi \right)^{1/\psi} \d \rvecr{s} \nonumber \\
&\leq (\pdfvec{y})^{1/\varphi} \left[ \int \pdfvec{s} \pdf{\rvec{z}}{\rvecr{y} - \preS\rvecr{s}} |f_i(\rvecr{s})|^\psi \d \rvecr{s} \right]^{1/\psi}, \\ &\leq  \xi(\dim, f_i, \varphi, \Cov{z}) (\pdfvec{y})^{1/\varphi},
\end{align}
where last inequality follows from the fact that $\pdf{\rvec{z}}{\rvecr{y} - \preS\rvecr{s}} |f_i(\rvecr{s})|^\psi$ is bounded above by the constant $\xi(\dim, i, \varphi, \Cov{z})$ not depending on $\rvecr{y}$ due to the fact that $f_i(\rvecr{s})$ is a polynomial on the entries of $\rvecr{s}$.

Considering the product $\left| \pdfvec{y} \prod_{i=1}^{M} \CEsp{\rveci{s}{i}}{\rvec{y}} \right|$ the result of the lemma follows by noting that the new constant becomes
$\kappa(\dim, \{ f_i \}, \varphi, \Cov{z}) = \prod_{i=1}^{M} \xi(\dim, f_i, \varphi, \Cov{z})$.
\end{IEEEproof}

\begin{prpap} \label{prp:int_id2}
Assume $\rvec{y} = \preS\rvec{s} + \rvec{z}$ is an
$\dim$-dimensional random vector, where $\preS$ is a deterministic matrix, $\rvec{s}$ is an arbitrarily distributed random vector, and $\rvec{z}$ is distributed following a zero-mean Gaussian distribution with covariance $\Cov{z}$. Then, the following integral identities hold
\begin{align}
\int \frac{\partial \pdfvec{y} \CEsp{ \rveci{s}{i} \rveci{s}{l}}{\rvecr{y}}}{\partial \rvecri{y}{k}} \CEsp{\rveci{s}{j}}{\rvecr{y}} \d \rvecr{y} &= - \int \pdfvec{y} \CEsp{\rveci{s}{i} \rveci{s}{l}}{\rvecr{y}} \frac{\partial \CEsp{\rveci{s}{j}}{\rvecr{y}}}{\partial \rvecri{y}{k}} \d \rvecr{y}
\label{eq:int_id_E} \\ -\int \frac{\partial \pdfvec{y}  \CEsp{\rveci{s}{l}}{\rvecr{y}}}{\partial \rvecri{y}{k}} \CEsp{\rveci{s}{i}}{\rvecr{y}}\CEsp{\rveci{s}{j}}{\rvecr{y}} \d \rvecr{y} \nonumber &= \int \pdfvec{y} \CEsp{\rveci{s}{l}}{\rvecr{y}} \frac{\partial \CEsp{\rveci{s}{i}}{\rvecr{y}}\CEsp{\rveci{s}{j}}{\rvecr{y}}}{\partial \rvecri{y}{k}} \d \rvecr{y}
\end{align}
\end{prpap}
\begin{IEEEproof}
Integrating by parts the left hand side of \req{eq:int_id_E} we have
\begin{multline} \label{eq:ipp_E}
\int \frac{\partial \pdfvec{y} \CEsp{ \rveci{s}{i} \rveci{s}{l}}{\rvecr{y}}}{\partial \rvecri{y}{k}} \CEsp{\rveci{s}{j}}{\rvecr{y}} \d \rvecr{y} \\ = \left[ \pdfvec{y} \CEsp{ \rveci{s}{i} \rveci{s}{l}}{\rvecr{y}} \CEsp{\rveci{s}{j}}{\rvecr{y}} \right]_{\rvecri{y}{k} = -\infty}^{\rvecri{y}{k} = \infty} - \int \pdfvec{y} \CEsp{\rveci{s}{i} \rveci{s}{l}}{\rvecr{y}} \frac{\partial \CEsp{\rveci{s}{j}}{\rvecr{y}}}{\partial \rvecri{y}{k}} \d \rvecr{y}.
\end{multline}
Using Lemma \ref{lem:boundPyEs} with $M = 2$, $f_1(\rvec{s}) = \rveci{s}{i} \rveci{s}{l}$, and $f_2(\rvec{s}) = \rveci{s}{k}$, we have that
\begin{gather}
\left| \pdfvec{y} \CEsp{ \rveci{s}{i} \rveci{s}{l}}{\rvecr{y}} \CEsp{\rveci{s}{j}}{\rvecr{y}} \right| \leq \kappa(\dim, \{ f_i \}, \varphi, \Cov{z}) (\pdfvec{y})^{(2-\varphi)/\varphi}.
\end{gather}
Now choosing $1 < \varphi < 2$ it is easy to see that the first term in the right hand side of \req{eq:ipp_E} vanishes as $\lim_{\rvecri{y}{k} \rightarrow \infty} \pdfvec{y} = 0$. Proceeding similarly with the second integral identity with $M = 3$, $f_1(\rvec{s}) = \rveci{s}{l}$, $f_1(\rvec{s}) = \rveci{s}{i}$, and $f_1(\rvec{s}) = \rveci{s}{j}$ and choosing $1 < \varphi < 3/2$ the result in the lemma follows.
\end{IEEEproof}

% Generated by IEEEtran.bst, version: 1.13 (2008/09/30)

\end{document}